\documentclass[prd,preprint,showpacs,preprintnumbers,nofootinbib,eqsecnum,superscriptaddress]{revtex4}

 \usepackage[dvips,final]{graphicx}
  \usepackage{amssymb}
   \usepackage{amsmath}
    \usepackage{amsfonts}
     \usepackage{epsfig}
      \usepackage{bm}

\usepackage{mathpazo}

\usepackage[section]{placeins}

\usepackage{multirow}
\usepackage{ctable}
\usepackage{booktabs}
\usepackage{array}
\usepackage{tabularx}
\usepackage{xcolor}
\usepackage{pstricks}

\def\lsim{\mathrel{\rlap{\lower4pt\hbox{\hskip1pt$\sim$}}
    \raise1pt\hbox{$<$}}}         
\def\gsim{\mathrel{\rlap{\lower4pt\hbox{\hskip1pt$\sim$}}
    \raise1pt\hbox{$>$}}}         

\newcommand{\bp}{{\bf{p}}}
\newcommand{\bq}{{\bf{q}}}

\begin{document}

\title{
Semiexclusive production of vector mesons in proton-proton
collisions with electromagnetic dissociation
}

\author{Anna Cisek}
\email{acisek@ur.edu.pl}
\affiliation{College of Natural Sciences, Institute of Physics, University of Rzesz\'ow,
ul. Pigonia 1, PL-35-310 Rzesz\'ow, Poland}

\author{Wolfgang Sch\"afer}
\email{Wolfgang.Schafer@ifj.edu.pl}
\affiliation{Institute of Nuclear Physics, Polish Academy of Sciences,
ul. Radzikowskiego 152, PL-31-342 Krak{\'o}w, Poland}

\author{Antoni Szczurek}
\email{Antoni.Szczurek@ifj.edu.pl}
\affiliation{Institute of Nuclear Physics, Polish Academy of Sciences,
ul. Radzikowskiego 152, PL-31-342 Krak{\'o}w, Poland}
\affiliation{College of Natural Sciences, Institute of Physics, University of Rzesz\'ow,
ul. Pigonia 1, PL-35-310 Rzesz\'ow, Poland}

\begin{abstract}
We calculate distributions of different vector mesons in
purely exclusive ($p p \to p p V$) and semi-exclusive ($p p \to p X V$) processes
with electromagnetic dissociation of a proton. The cross section for the electromagnetic dissociation
is expressed through electromagnetic structure functions of the proton.
We include the transverse momentum distribution of initial photons in the associated flux.
Contributions of the exclusive and semi-exclusive processes are compared for different
vector mesons ($V = \phi, J/\psi, \Upsilon$). We discuss how the relative contribution of the 
semi-exclusive processes depends on the mass of the vector meson as well as on different
kinematical variables of the vector meson ($y$, $p_t$). The ratio of semi-exclusive to exclusive contributions
are shown and compared for different mesons in different variables.
\end{abstract}

\pacs{13.87.Ce, 14.65.Dw}

\maketitle

\section{Introduction}

 Exclusive production of vector mesons $p p \to p p V$ is a source of knowledge about
 gluon distributions in the proton. In contrast to the collinear approach, in 
 the $k_t$-factorization approach the cross section depends not only on 
 (unintegrated) gluon distribution function (UGDF) but also
 on the quarkonium quark-antiquark wave function \cite{CSS2015}. It was shown that different UGDFs give different results. 
 In order to ``extract`` or check the collinear gluon distribution or UGDF one has to be sure that the measured cross sections are not contaminated by any other mechanism.
 
 So far both $J/\psi$ and $\Upsilon$
 \cite{Aaij:2013jxj,Aaij:2015kea} were measured in proton-proton collisions. 
 The measurements are not fully exclusive as so far the outgoing protons were not measured.
 To increase the exclusivity of the reaction a rapidity veto (no emission around rapidity of vector meson) is being imposed. How good is such an approach is not fully understood
 in our opinion. In our earlier paper on $J/\psi$ production \cite{Cisek:2016kvr} we have developed a formalism how to calculate such processes with rapdidity gaps, but including proton dissociation. To calculate electromagnetic dissociation, the method uses parametrizations of the proton structure functions which are used to derive an inelastic photon flux. We have shown in \cite{Cisek:2016kvr} that the semi-exclusive mechanism cannot be completely removed by the requirement of rapidity veto. To our surprise the electromagnetic dissociation seems the most important (the largest) in this context. Here we wish to show more systematic studies
 for production of different vector mesons and better understand the competition of the purely exclusive and the semiexclusive processes.
 
\begin{figure}[!h] 
\begin{minipage}{0.47\textwidth}
 \centerline{\includegraphics[width=1.0\textwidth]{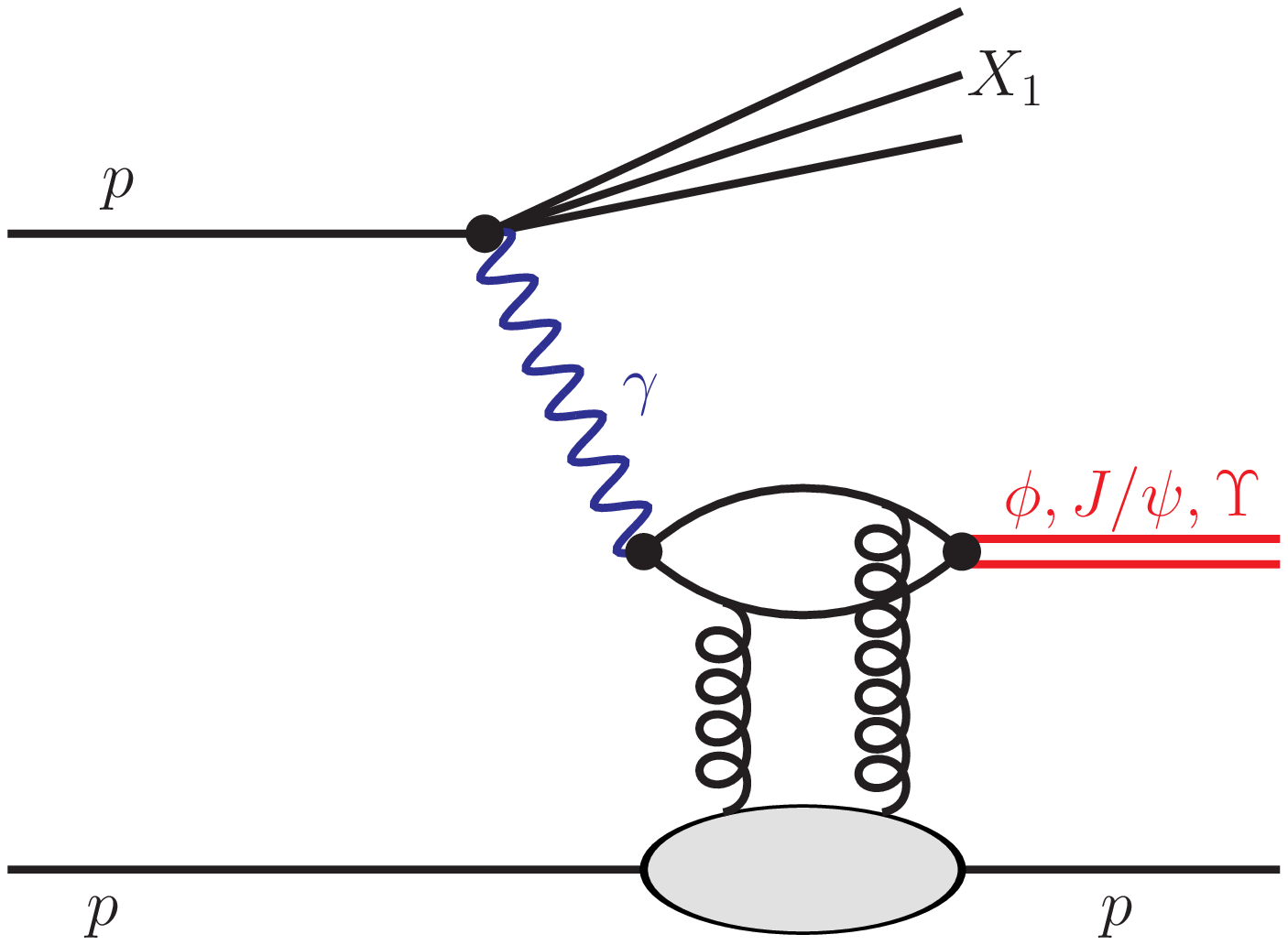}}
\end{minipage}
\begin{minipage}{0.47\textwidth}
 \centerline{\includegraphics[width=1.0\textwidth]{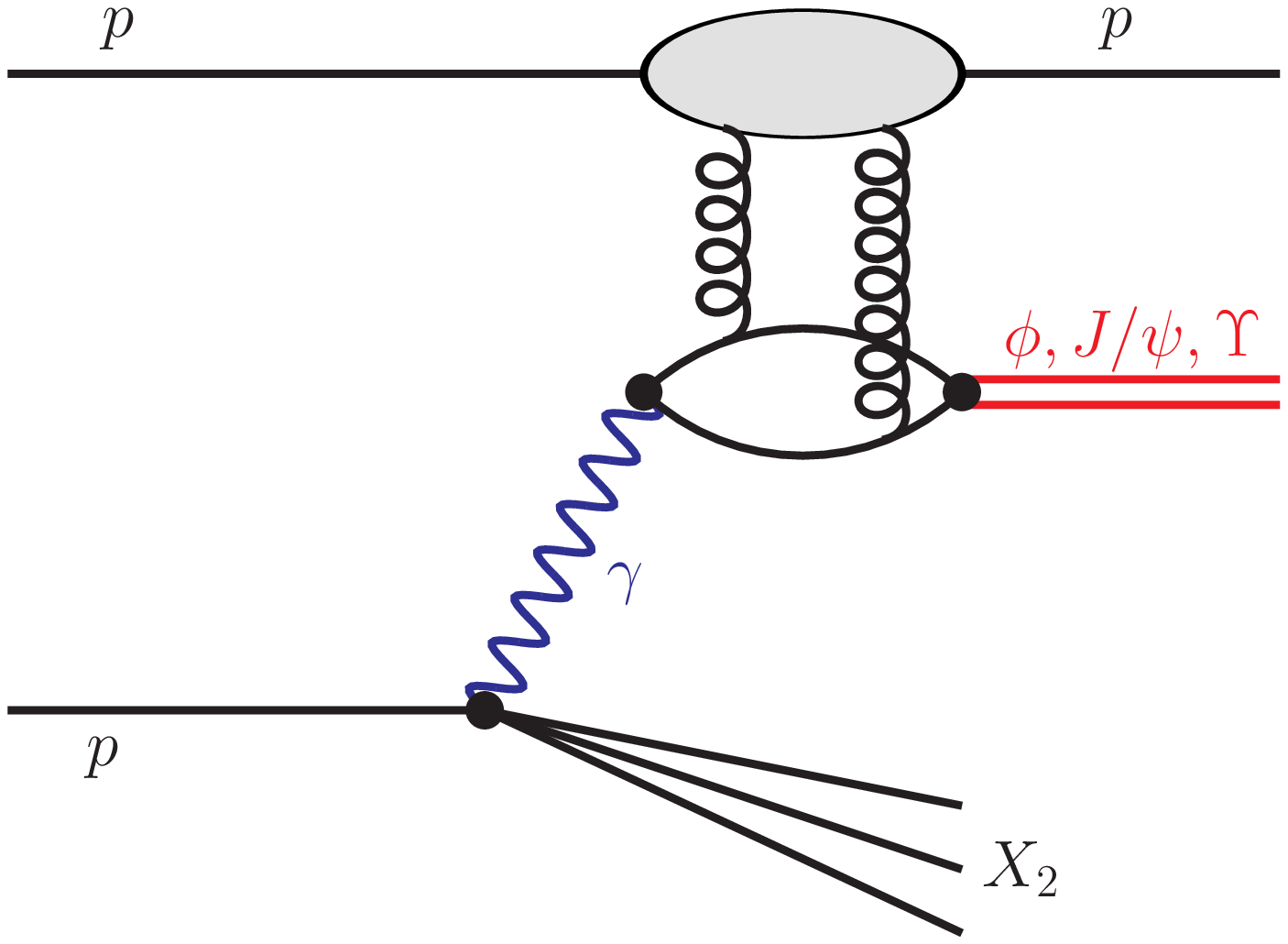}}
\end{minipage}
   \caption{
\small Schematic representation of the electromagnetic excitation
of one (left panel) or second (right panel) proton.
 }
 \label{fig:diag_EM_exc}
\end{figure}

 The semiexclusive production mechanisms are shown for illustration in Fig.\ref{fig:diag_EM_exc}.
 Here, due to the quantum numbers 
 of vector mesons, the dominant mechanism is photon-pomeron fusion. As shown in the figure the photon can be coupled to either one of the two protons.

 In this work, we will calculate different differential distributions.
 Of special interest is the ratio of semi-exclusive to exclusive cross section.
 Such a ratio may be considered as a measure of ``unwanted'' contamination of 
 exlusive processes when using the rapidity gap method.
 The predictions of cross section for the semiexclusive processes may be also
 valueable as it could be in principle measured.

 \section{Sketch of the formalism}
 
Let us concentrate on 
the events with electromagnetic dissociation of one of the protons.
The important property of these processes is that the $p \gamma^* \to X$
transition is given by the electromagnetic structure functions of the
proton, and thus to a large extent calculable ``from data''. 
The cross section for such processes can be written as:
\begin{eqnarray}
 {d \sigma (pp \to X V p; s) \over dy d^2\bp dM_X^2} &=& 
  \int {d^2\bq \over \pi \bq^2} {\cal{F}}^{(\mathrm{inel})}_{\gamma/p}(z_+,\bq^2,M_X^2) 
  {1\over \pi} {d \sigma^{\gamma^* p \to Vp} \over dt}(z_+s,t = -(\bq - \bp)^2) \nonumber \\
  &+& ( z_+ \leftrightarrow z_-),  
  \nonumber \\
\end{eqnarray}
where $z_\pm = e^{\pm y} \sqrt{(\bp^2 + m_V^2)/s}$ is the fraction of the proton's 
longitudinal momentum carried by the photon, and $M_X$ is the invariant mass of the excited system $X$,
$\bp$ is the transverse momentum of the vector meson, and $-\bq$ is the transverse momentum of the 
outgoing hadronic system $X$. Below we also use $p_t = |\bp|$ for the absolute value of the transverse momentum. 
The mass of the excited hadronic system must be above the threshold 
$M_{\rm thr} = m_\pi + m_p$. 
In the kinematics of interest the ``fully unintegrated'' flux of photons associated with the breakup of the proton
is calculable in terms of the structure function $F_2$ of a proton : 
\begin{eqnarray}
 {\cal{F}}^{(\mathrm{inel})}_{\gamma/p}(z,\bq^2,M_X^2) = {\alpha_{\mathrm{em}} \over \pi} (1 - z) \theta( M_X^2- M^2_{\mathrm{thr}}) 
 {F_2(x_{Bj},Q^2)  \over M_X^2 + Q^2 - m_p^2}  \Big[ {\bq^2 \over \bq^2 + z (M_X^2 - m_p^2) + z^2 m_p^2} \Big]^2 
\, ,
\nonumber \\
\end{eqnarray}
where we calculate the photon virtuality $Q^2$ and the Bjorken variable $x_{\rm Bj}$ from 
\begin{eqnarray}
Q^2 =  {1 \over 1 - z} \Big[ \bq^2 + z (M_X^2 -m_p^2)  + z^2 m_p^2 \Big],
\ \ \ x_{Bj} = {Q^2 \over Q^2 + M_X^2 - m_p^2} .
\end{eqnarray}
We use the following parametrizations of the proton structure function $F_2(x,Q^2)$:

\begin{enumerate}
	\item A parametrization of Ref. \cite{Fiore:2002re,Fiore:2004xb} which is fitted to the lower energy CLAS data and
	is meant to give an accurate description especially in the resonance region $M_X \lsim 2 \, \rm {GeV}$.
	In the figures it will be labeled as FFJLM. This parametrization does
	not describe data well, when it is extrapolated beyond the region of its intended use. Therefore we only
	use it when calculating observables with $M_X \lsim 2 \, \rm {GeV}$.
	
	\item The Abramowicz-Levy-Levin-Maor fit \cite{Abramowicz:1991xz,Abramowicz:1997ms} used previously also
	 in \cite{Luszczak:2015aoa}, abbreviated here ALLM.
	 
	\item A newly constructed parametrization, which at $Q^2 > 9 \, \rm{GeV}^2$ uses an NNLO calculation 
	of $F_2$ and $F_L$ from NNLO MSTW 2008 partons \cite{Martin:2009iq}. It employs a useful code by the MSTW group \cite{Martin:2009iq}
	to calculate structure functions. At $Q^2 > 9 \, \rm{GeV}^2$ this fit uses the parametrization of Bosted and Christy \cite{Bosted:2007xd}
	in the resonance region, and a version of the ALLM fit published by the HERMES Collaboration \cite{Airapetian:2011nu} for the continuum
	region. It also uses information on the longitudinal structure function from SLAC \cite{Abe:1998ym}. As the fit is constructed closely following
	the LUXqed work Ref.\cite{Manohar:2016nzj,Manohar:2017eqh}, we call this fit LUX-like.
	
	\item A Vector-Meson-Dominance model inspired fit of $F_2$ proposed in \cite{SU} at low $Q^2$, which is completed by the same 
	NNLO MSTW structure function as above at large $Q^2$. This fit is labelled SU for brevity. 
\end{enumerate}

\begin{figure}[!htb] 
\begin{center}
\includegraphics[width = .4\textwidth]{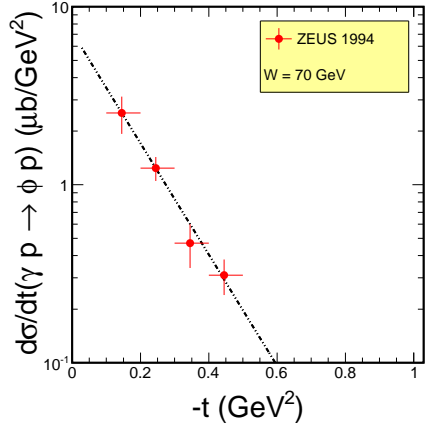}
\end{center}
\caption{Differential cross section for diffractive photoproduction of $\phi$ 
	mesons. Data are from the ZEUS collaboration at HERA \cite{ZEUS}.}

\label{fig:dsig_dt_phi}
\end{figure}

Our formalism is valid for photons which carry momentum fractions $z \ll 1$, this is an appropriate approximation for the production
of vector mesons away from the forward rapidity regions.
The differential cross section for the $\gamma^* p \to V p$ process is
\begin{eqnarray}
 {d \sigma^{\gamma^* p \to Vp} \over dt} = {d \sigma_T^{\gamma_T^* p \to Vp} \over dt} + {d \sigma_L^{\gamma_L^* p \to Vp} \over dt}
 = {d \sigma_T^{\gamma_T^* p \to Vp} \over dt} \, \Big( 1 + R_{LT}(Q^2) \Big) \, .
 \label{eq:dsig_dt}
\end{eqnarray}
We parametrize the differential cross section by a simple analytic form as in Ref. \cite{CSS2015}
for $J/\Psi$. An analogous analysis was made in the papers \cite{Ivanov:2004ax,SS2007}
and experimental data can be found in \cite{H1_new,H1_old}.
In Fig. \ref{fig:dsig_dt_phi} we show as an example the differential cross section for
$\phi$ photoproduction, comparing our simple fit with the data taken by the ZEUS collaboration
at HERA \cite{ZEUS}.

\section{Results}

\begin{figure}[!htb] 
	\begin{center}
		\includegraphics[width = .4\textwidth]{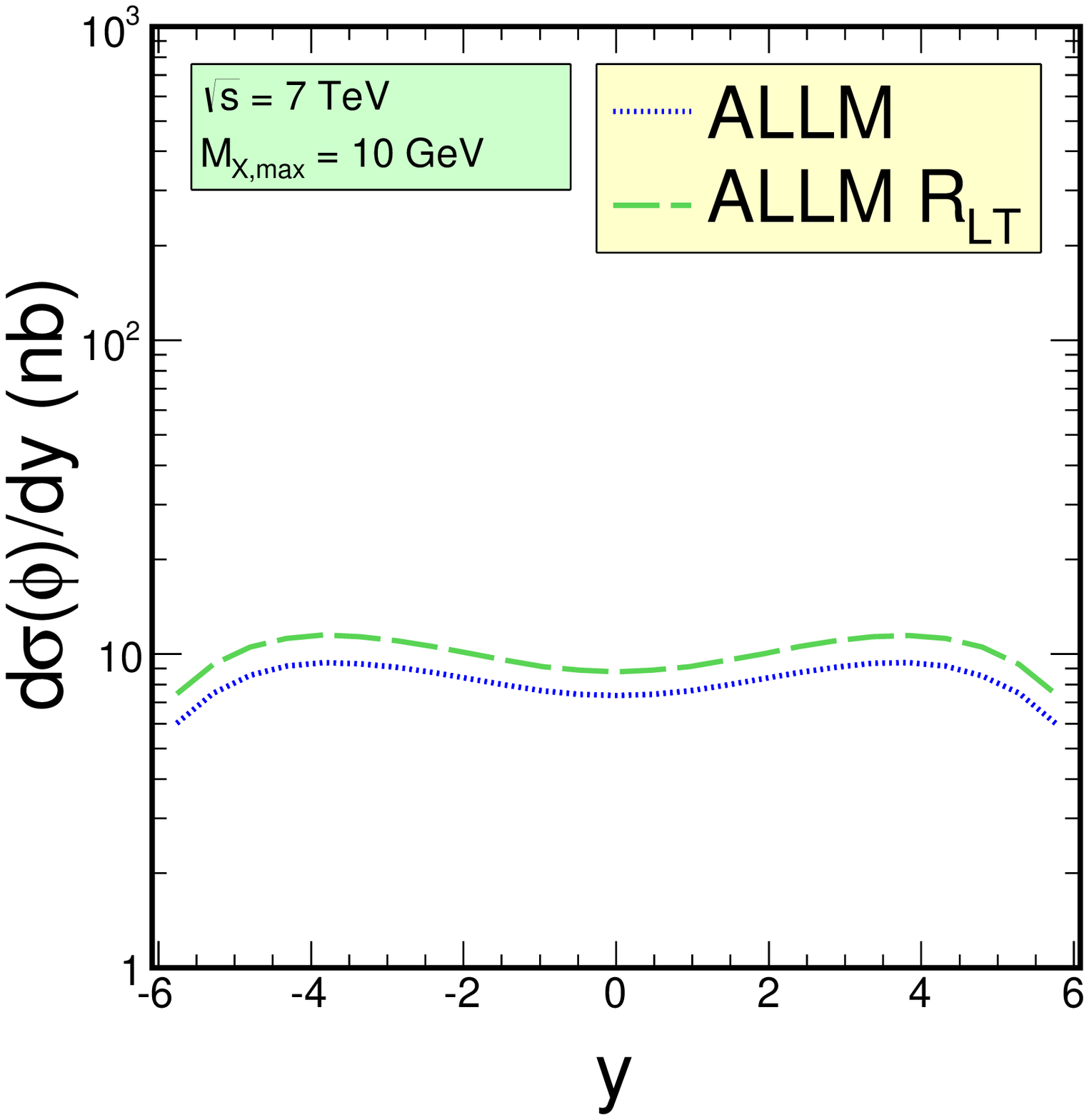}
		\includegraphics[width = .4\textwidth]{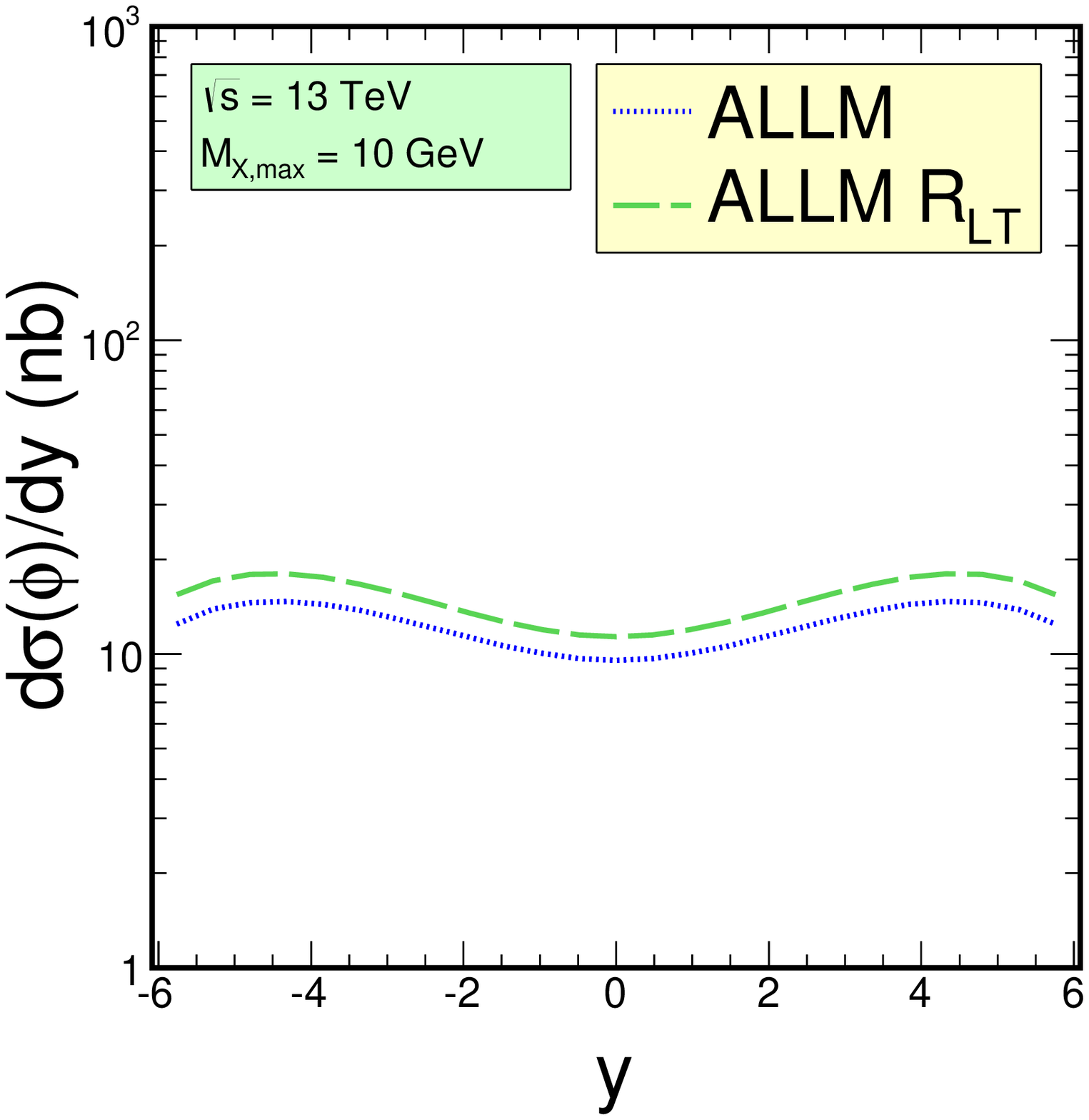}
	\end{center}
	\caption{Rapidity distribution of $\phi$-mesons at two different energies, with and without 
	the contribution from longitudinal photons.}
	\label{fig:dsig_dy_phi_RLT}
\end{figure}

\begin{figure}[!htb] 
	\begin{center}
		\includegraphics[width = .4\textwidth]{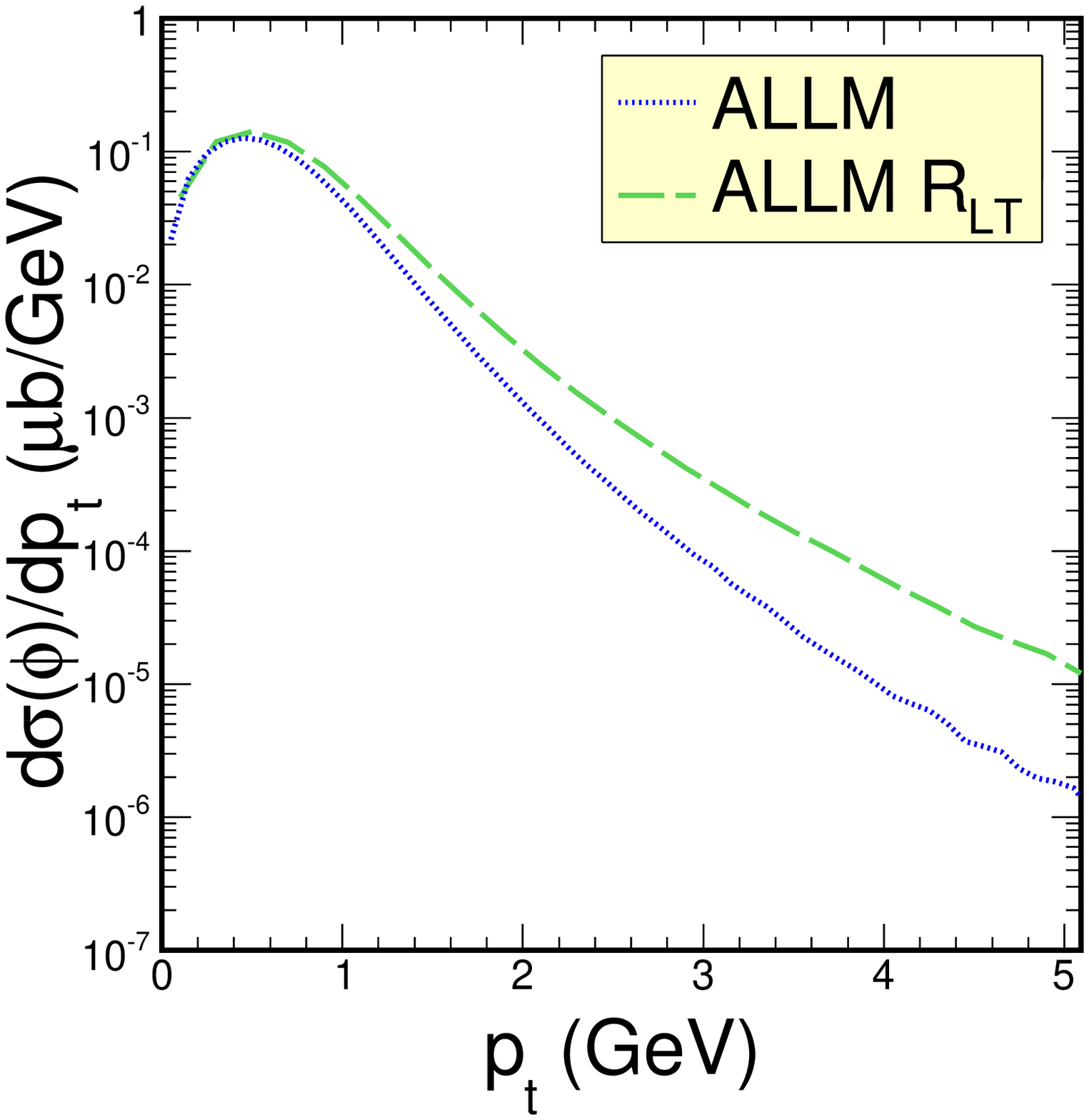}
		\includegraphics[width = .4\textwidth]{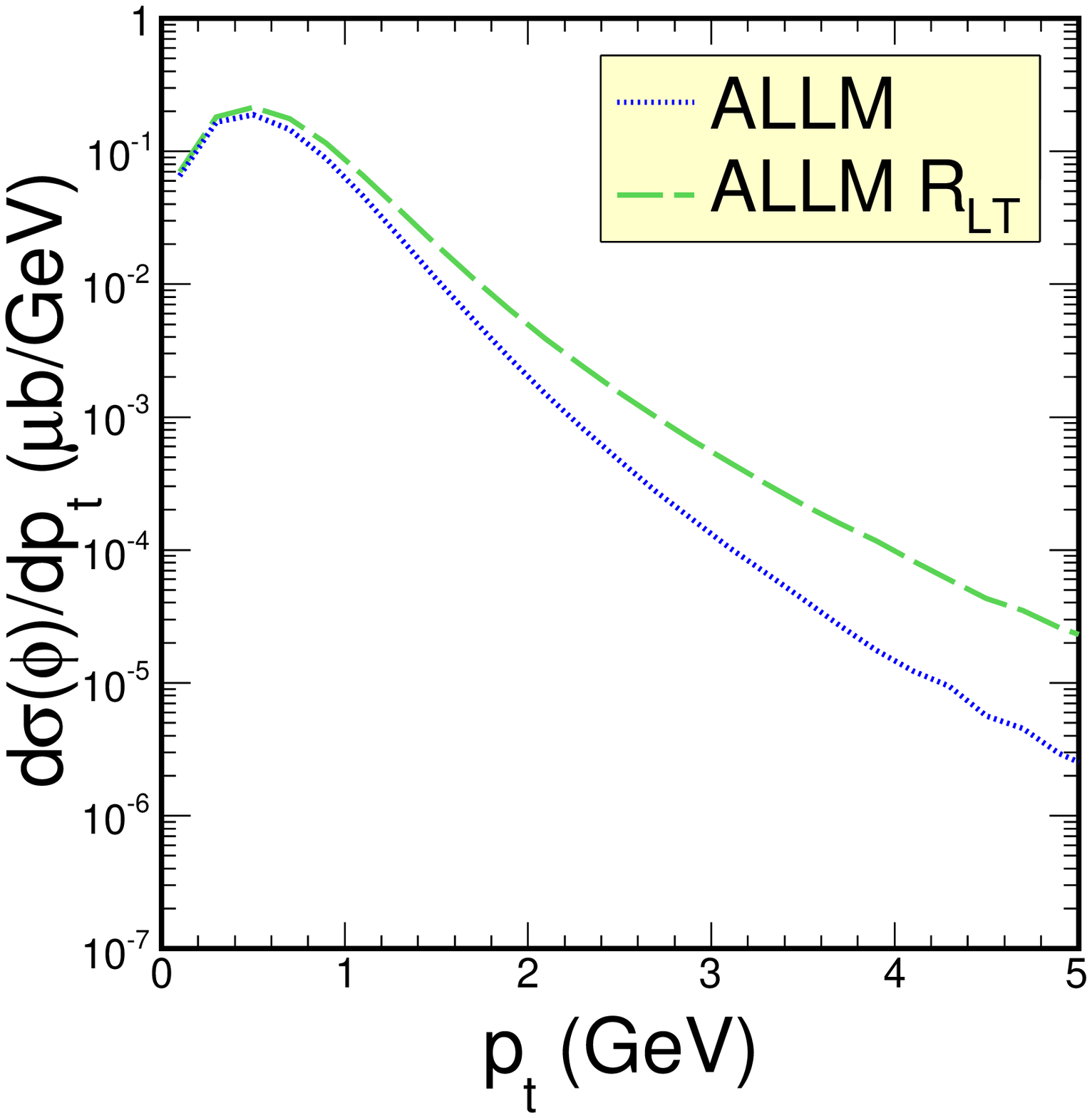}
	\end{center}
	\caption{Transverse momentum distribution of $\phi$-mesons at two different energies, with and without 
	the contribution from longitudinal photons for energy $\sqrt{s} =$ 7 TeV (left panel) and $\sqrt{s} =$
	13 TeV (right panel).}
	\label{fig:dsig_dpt_phi_RLT}
\end{figure}

We start our discussion of results with the rapidity- and transverse-momentum distributions of
semi-exclusively produced $\phi$-mesons. Here we want to discuss the effect of longitudinal
photons quantified by $R_{LT}$ in Eq.(\ref{eq:dsig_dt}). In Fig.~\ref{fig:dsig_dy_phi_RLT}
we show the rapidity dependent cross section
\begin{eqnarray}
{d \sigma (\phi) \over dy } \equiv \int_{M_{\rm thr}}^{M_{X, \rm max}} dM_X {d \sigma(pp \to \phi X;s) \over dy dM_X},
\end{eqnarray}
integrated up to masses $M_{X, \rm max} = 10 \, \rm{GeV}$ with and without the $R_{LT}$-term included.
We observe, that longitudinal photons enhance the cross section by about $\sim 20 \%$ uniformly in $y$.
As we can see from the transverse-momentum distribution of $\phi$-mesons shown in Fig.~\ref{fig:dsig_dpt_phi_RLT}, the effect of longitudinal photons
is important at large transverse momenta, $p_t > 1 \rm{GeV}$.
For heavier mesons we find small effects of longitudinal photons, as the ratio 
behaves like $R_{LT} \propto Q^2/m_V^2$, over a broad range of $Q^2$. This means
a suppression of longitudinal photons in the relevant for us range of $Q^2$.

\begin{figure}[!htb] 
\includegraphics[height=5.5cm]{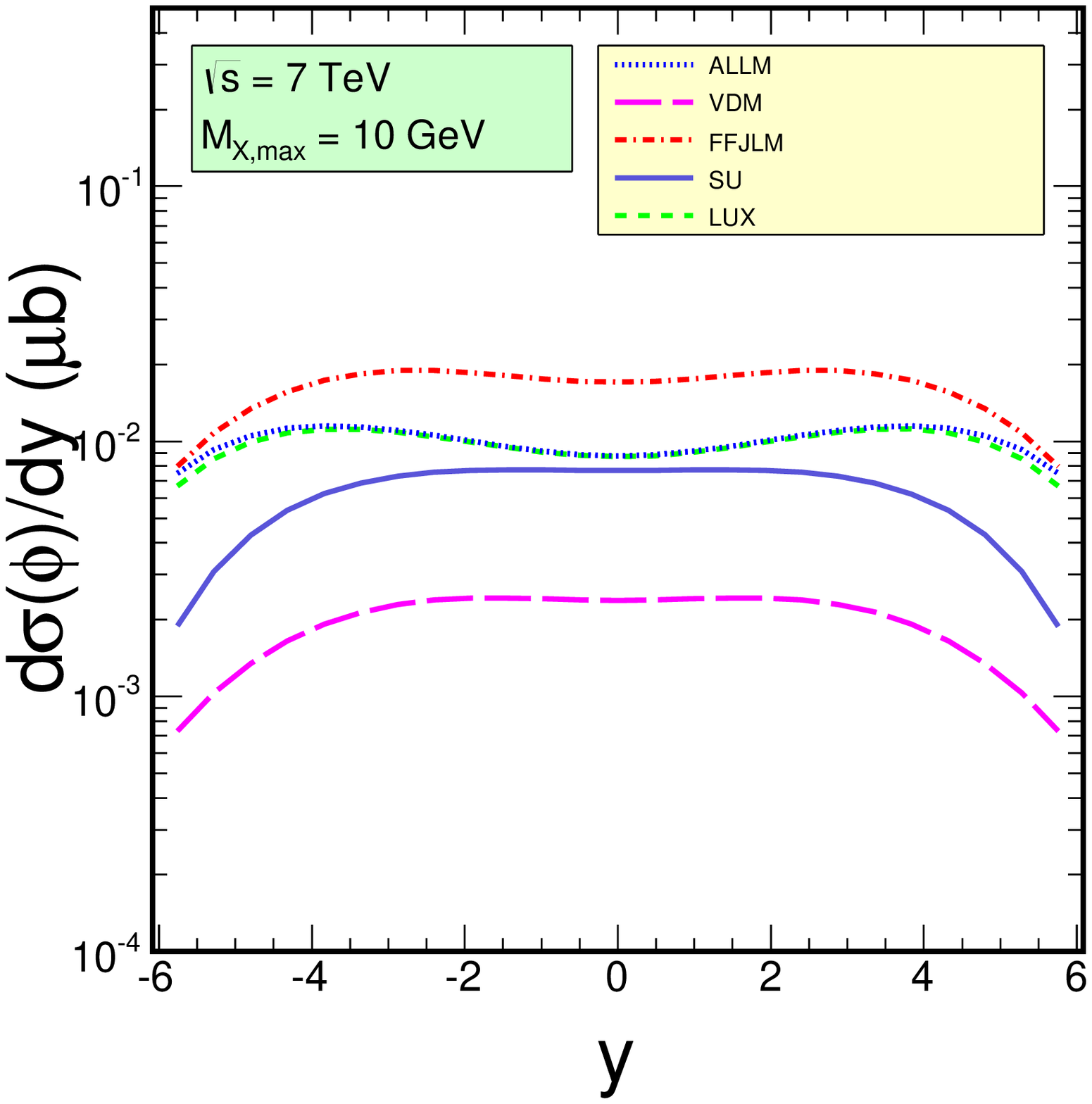}
\includegraphics[height=5.5cm]{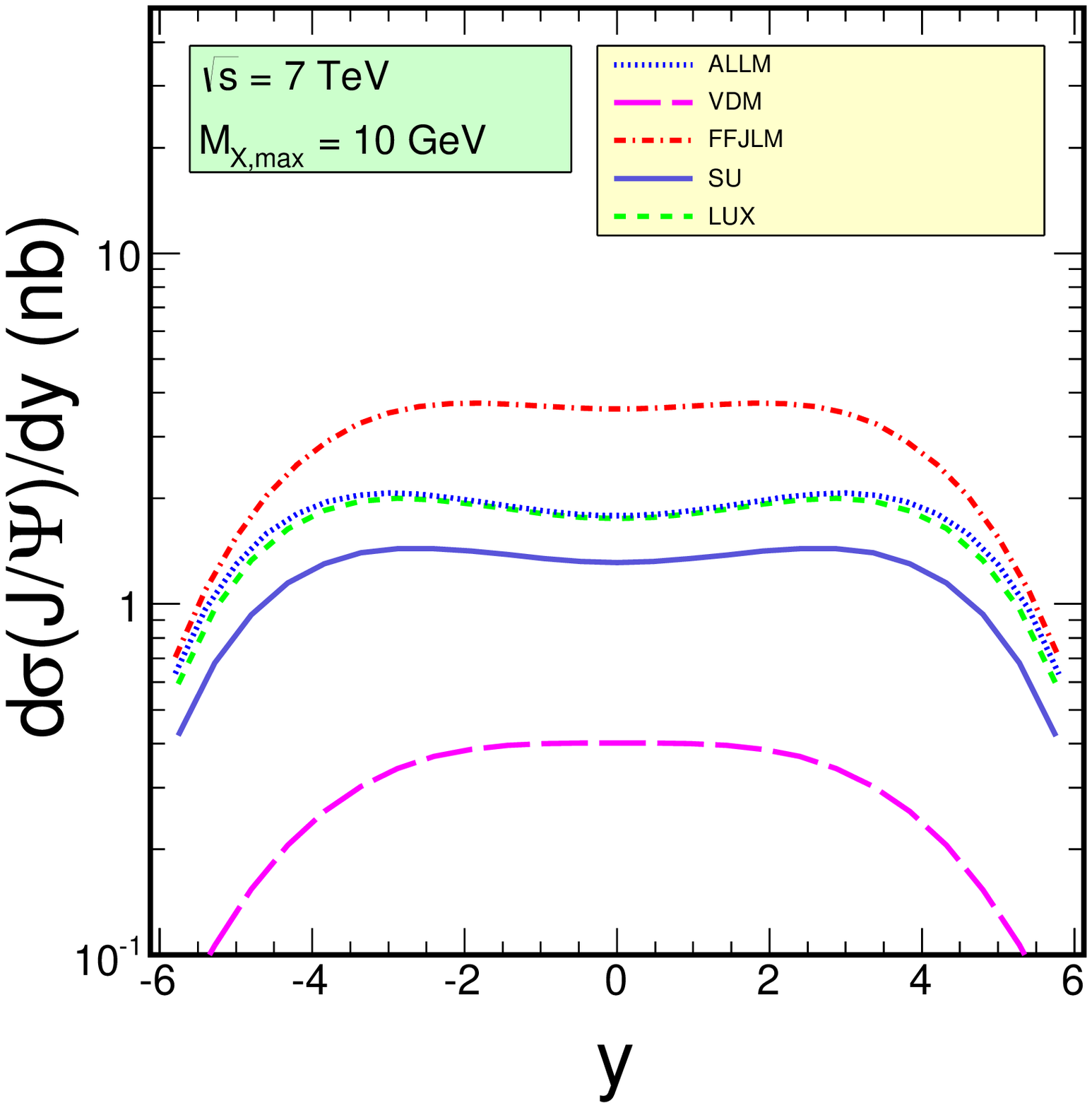}
\includegraphics[height=5.5cm]{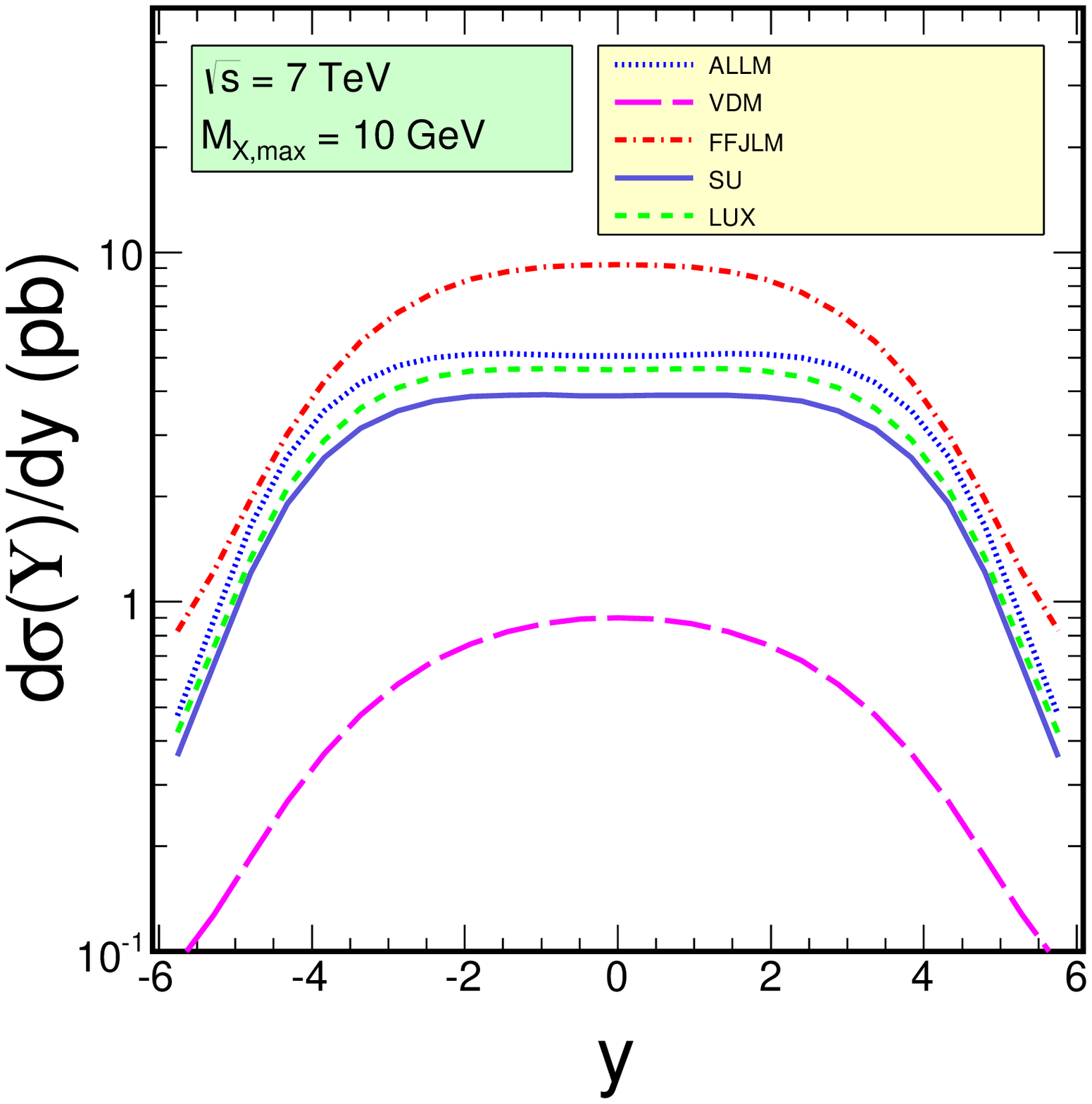}
\caption{Rapidity distribution for pp cm-energy $\sqrt{s} = 7 \, \rm{TeV}$
for the production of $\phi, J/\psi$ and $\Upsilon$ mesons for different parametrizations
of the proton structure function $F_2$.}
\label{fig:dsigdy_7TeV}
\end{figure}

\begin{figure}[!htb] 
	\includegraphics[width = 0.4\textwidth]{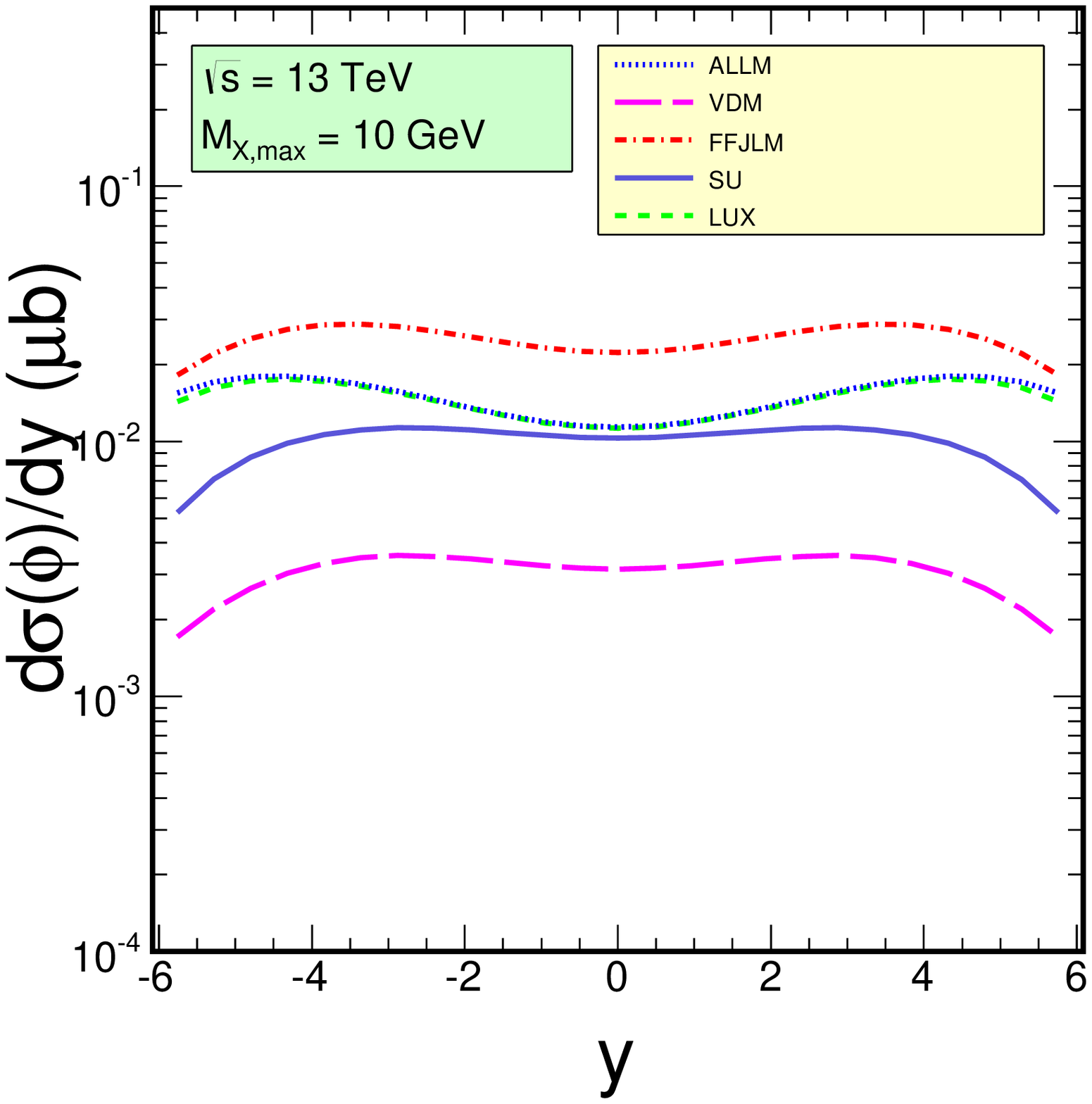}
	\includegraphics[width = 0.4\textwidth]{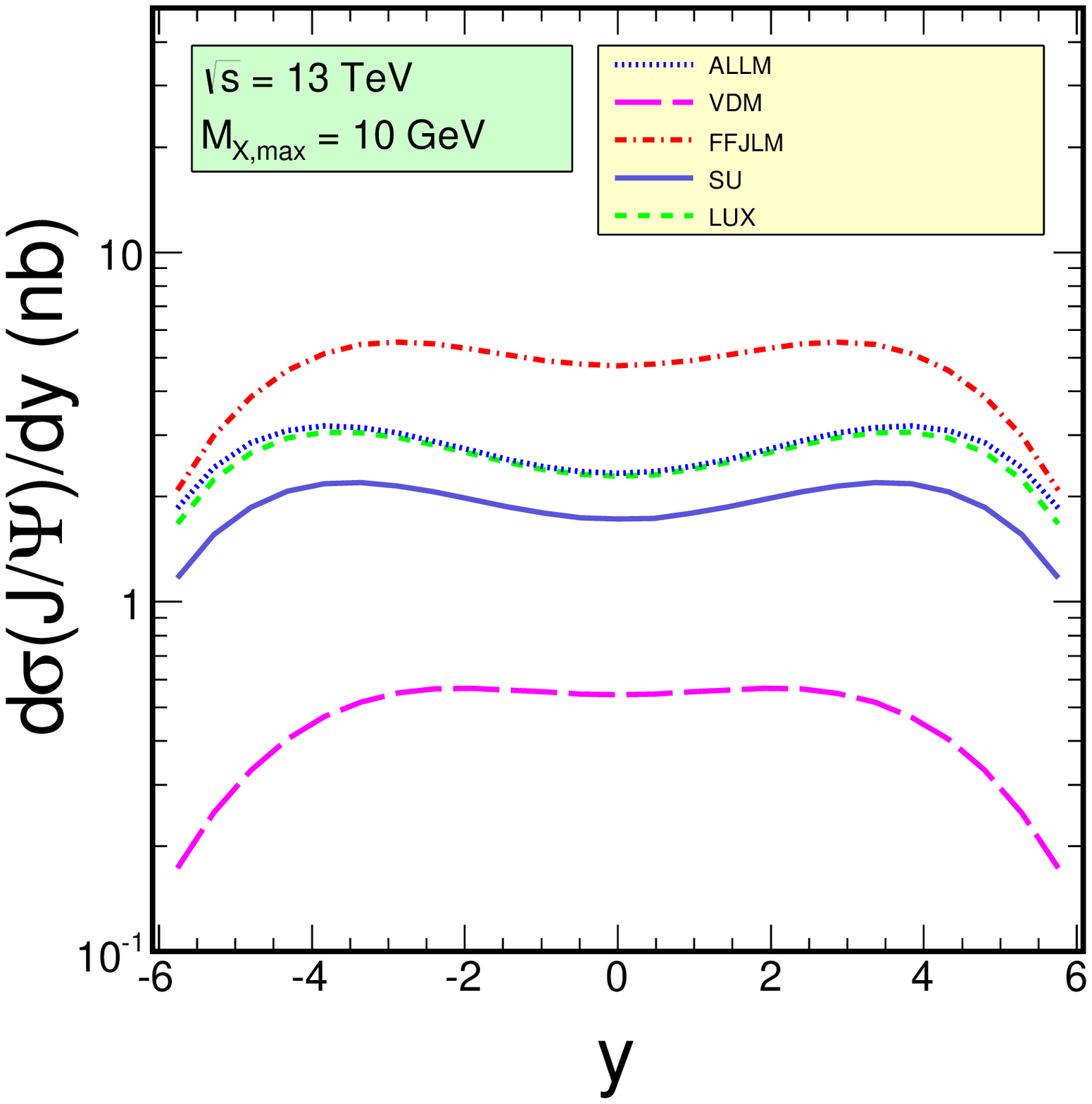}
	\includegraphics[width = 0.4\textwidth]{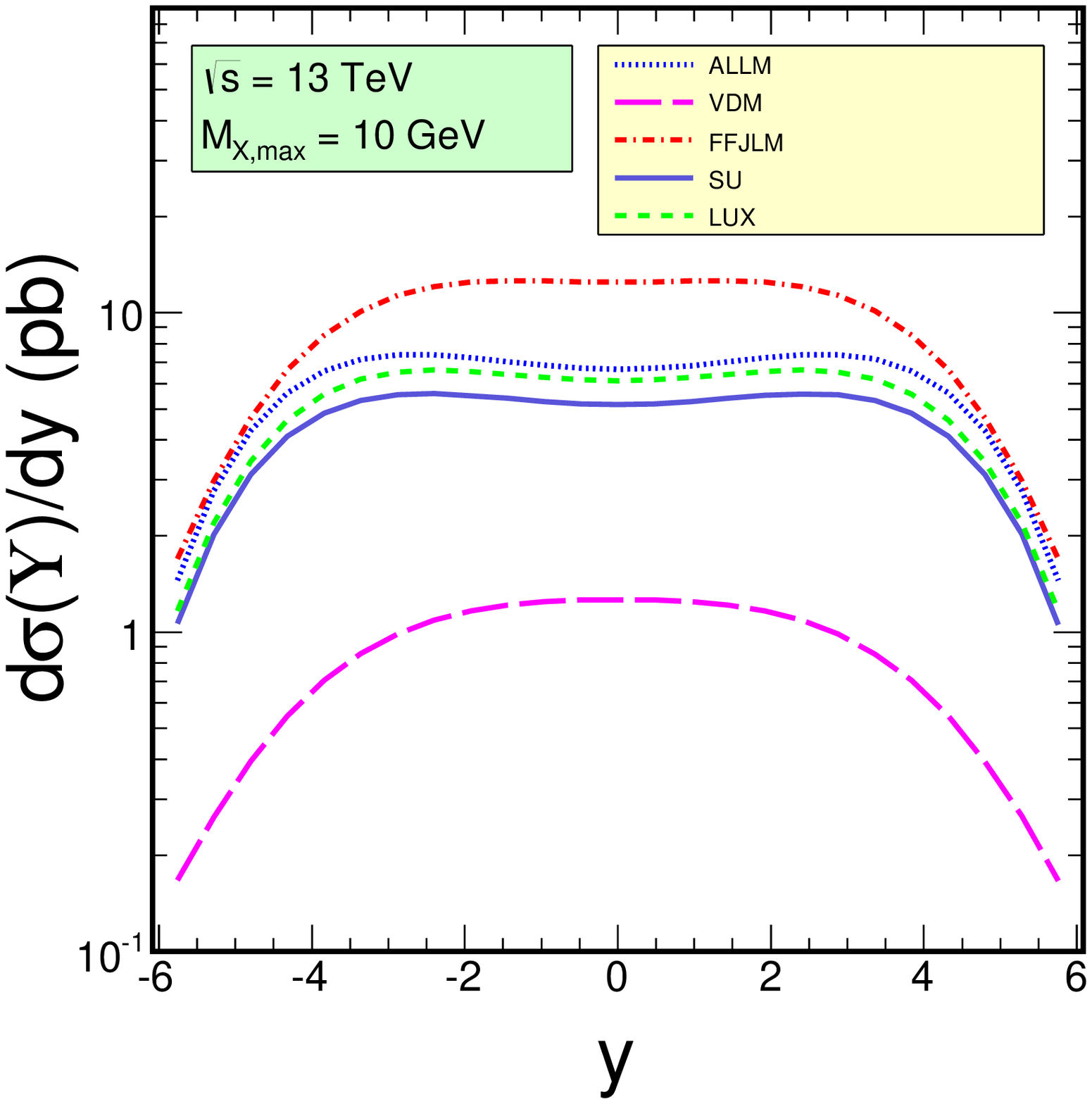}
	\caption[*]{Rapidity distribution for pp cm-energy $\sqrt{s} = 13 \, \rm{TeV}$
		for the production of $\phi, J/\psi$ and $\Upsilon$ mesons for different parametrizations
		of the proton structure function $F_2$.}
	\label{fig:dsigdy_13TeV}
\end{figure}

\begin{figure}[!htb] 
\includegraphics[width=.4\textwidth]{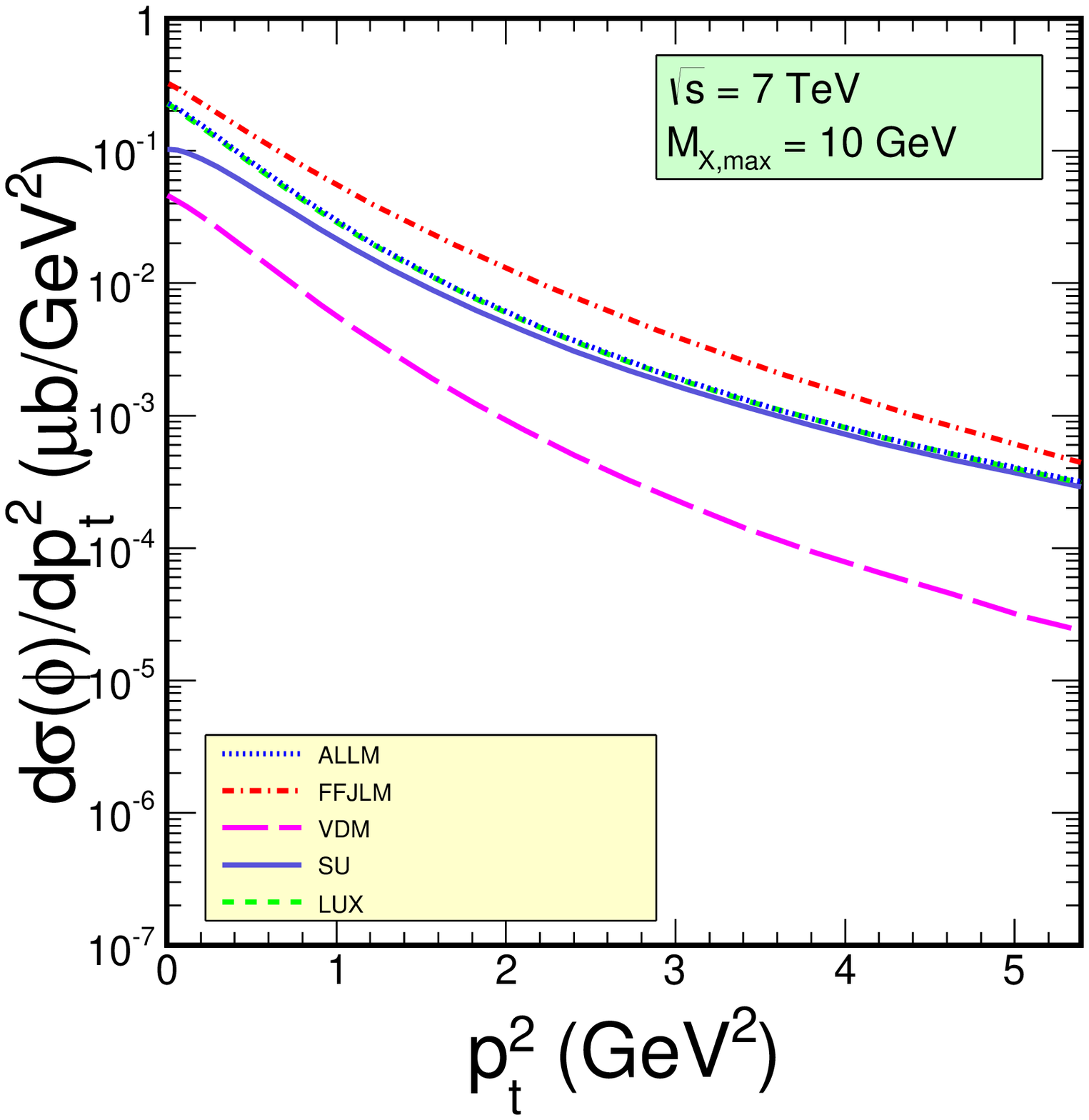}
\includegraphics[width=.4\textwidth]{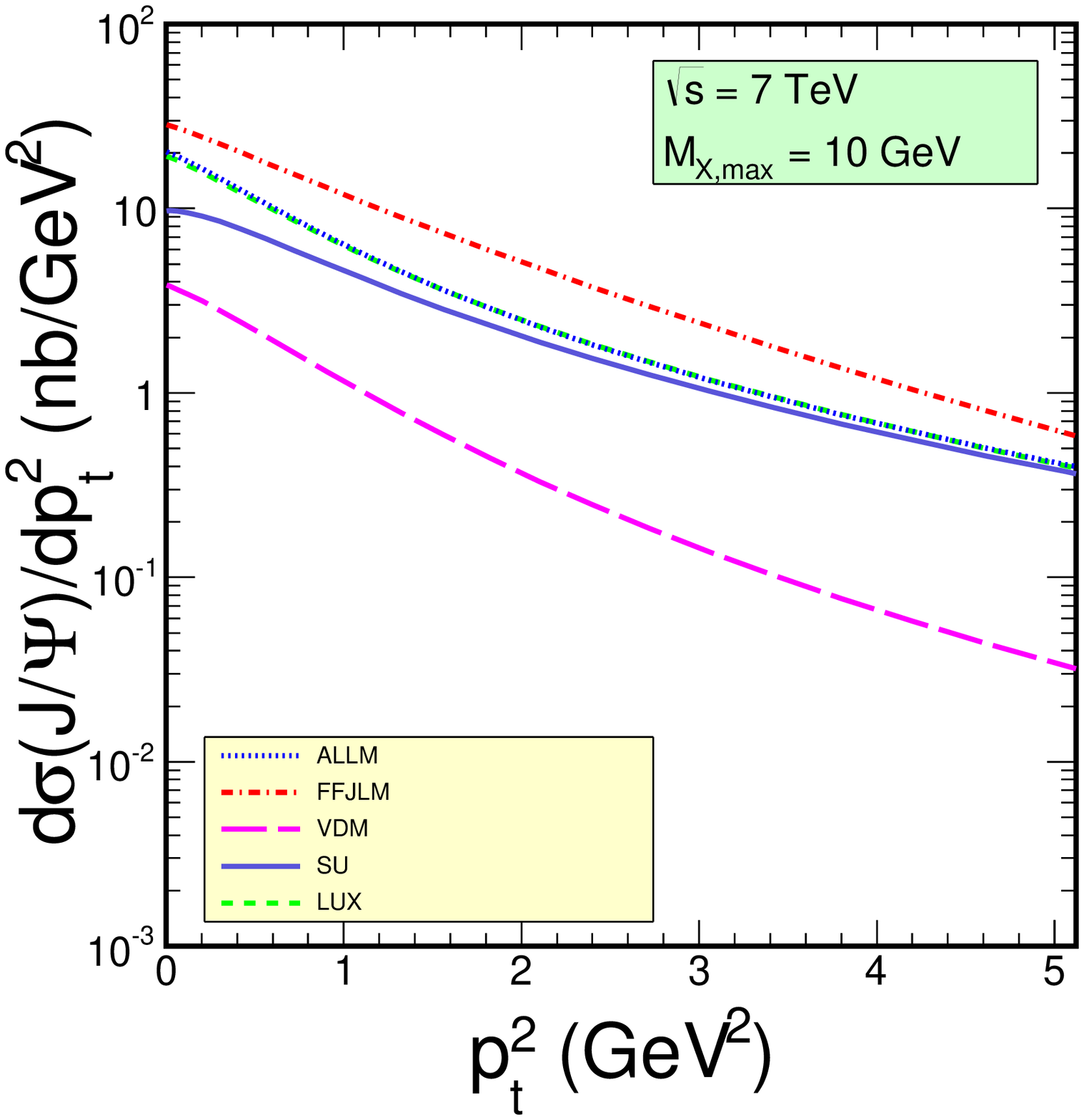}
\includegraphics[width=.4\textwidth]{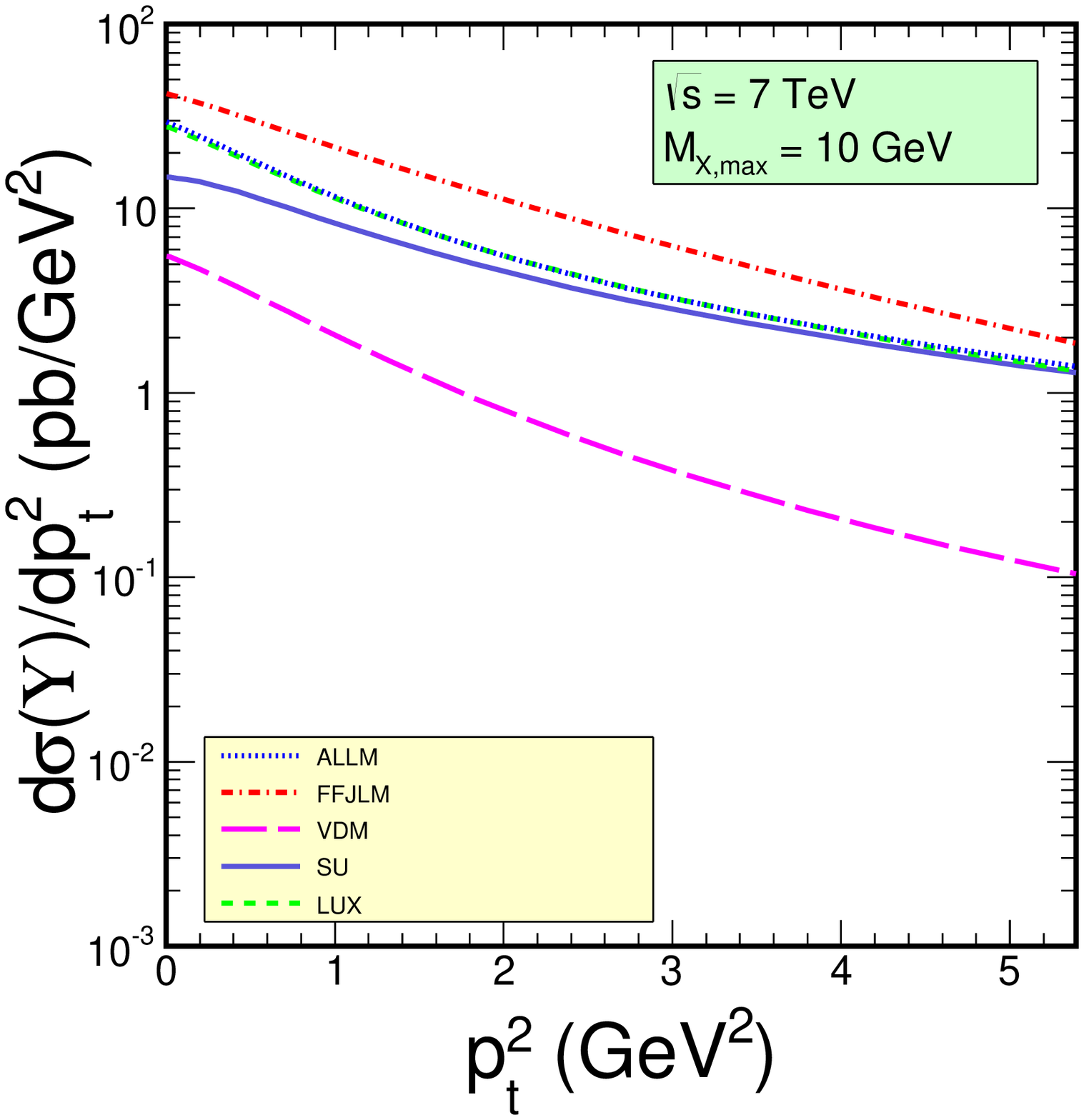}
\caption[*]{Transverse momentum  distribution of vector meson for
 pp cm-energy $\sqrt{s} = 7 \, \rm{TeV}$ for the production of $\phi, J/\psi$ and 
 $\Upsilon$ mesons for different parametrizations 
of the proton structure function $F_2$.}
\label{fig:dsigdpt2_7TeV}
\end{figure}

\begin{figure}[!htb] 
\includegraphics[height=.4\textwidth]{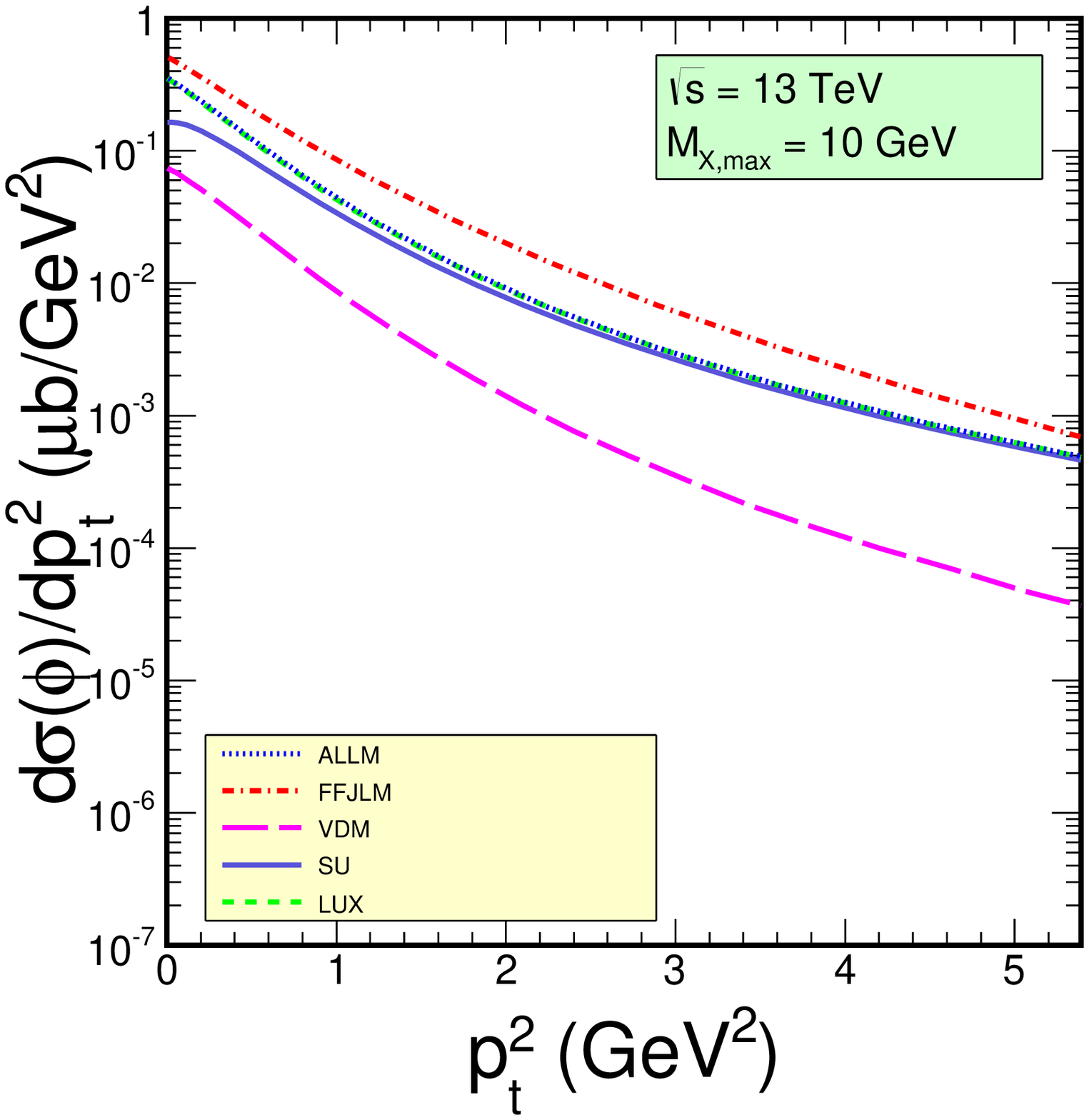}
\includegraphics[height=.4\textwidth]{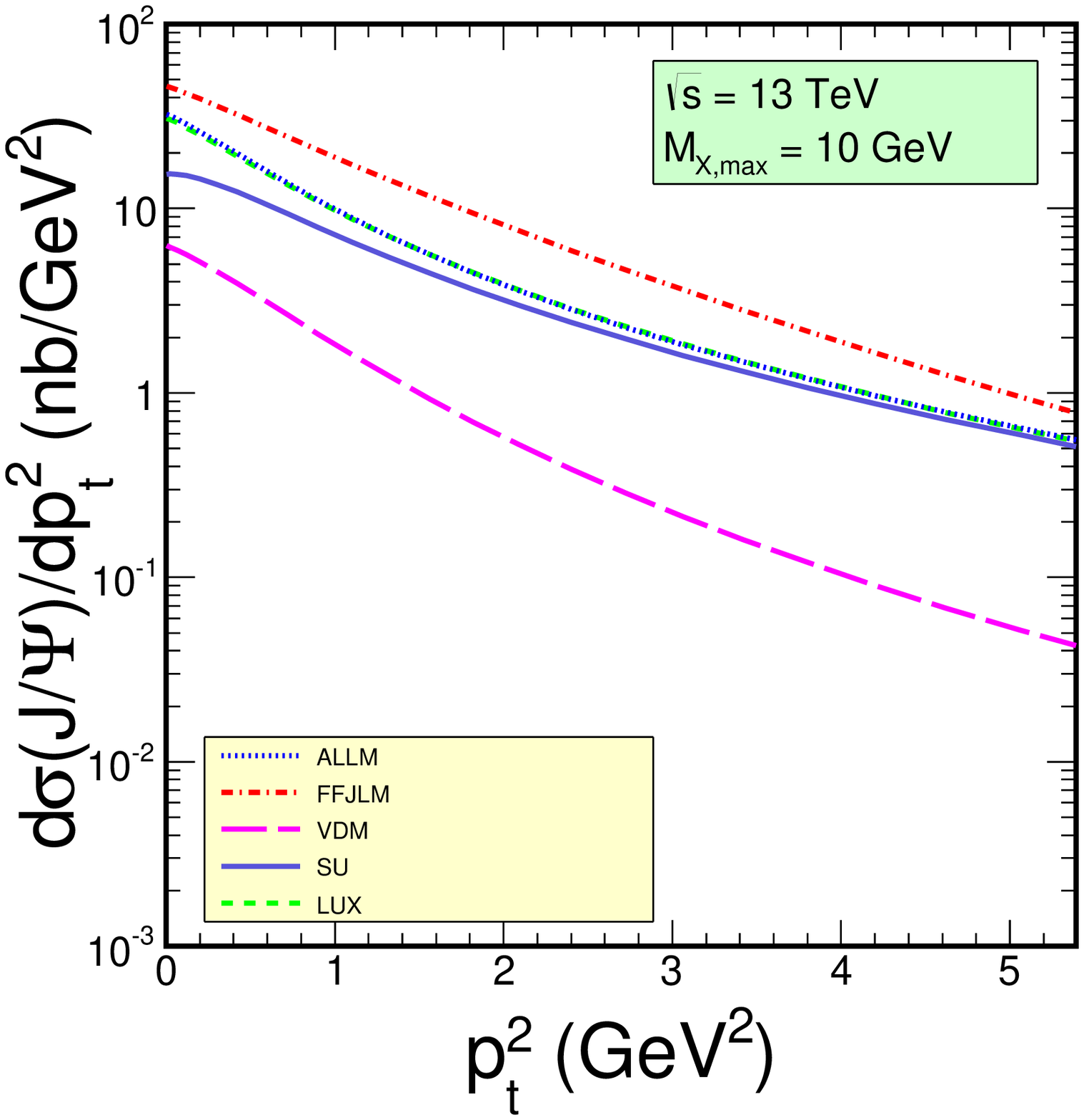}
\includegraphics[height=.4\textwidth]{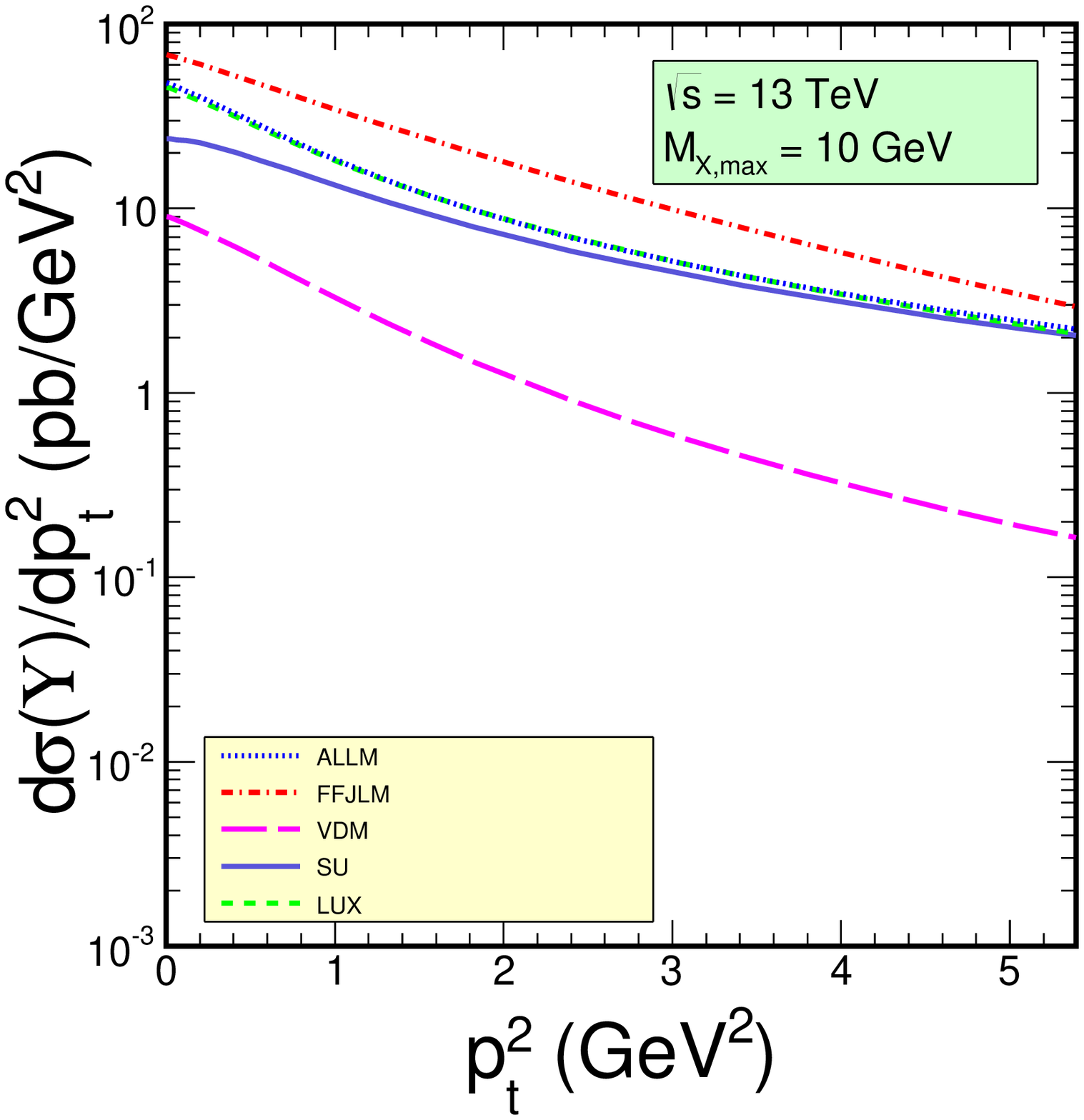}
\caption[*]{Transverse momentum distribution of vector meson for
 pp cm-energy $\sqrt{s} = 13 \, \rm{TeV}$ for the production of $\phi, J/\psi$ and 
 $\Upsilon$ mesons for different parametrizations 
of the proton structure function $F_2$.}
\label{fig:dsigdpt2_13TeV}
\end{figure}

We now wish to present results for rapidity and transverse-momentum distributions of mesons.
In Figs. \ref{fig:dsigdy_7TeV} and \ref{fig:dsigdy_13TeV} we show the rapidity dependent
cross section
\begin{eqnarray}
 {d \sigma (V) \over dy } \equiv \int_{M_{\rm thr}}^{M_{X, \rm max}} dM_X {d \sigma(pp \to VX;s) \over dy dM_X},
\end{eqnarray}
for $V = \phi, J/\psi , \Upsilon$ and $M_{X, \rm max} = 10 \, \rm{GeV}$, using different parametrizations 
of the proton structure function $F_2$. We observe a good agreement of the results obtained with 
different parametrizations. The rapidity distributions become narrower the heavier the produced meson.
In Figs. \ref{fig:dsigdpt2_7TeV} and \ref{fig:dsigdpt2_13TeV} we show the distributions in
transverse momentum squared of the meson. Again we integrate up to $M_{X, \rm max} = 10 \, \rm{GeV}$:
\begin{eqnarray}
 {d \sigma (V) \over dp_t^2 } \equiv 
 \int_{M_{\rm thr}}^{M_{X, \rm max}} dM_X {d \sigma(pp \to VX;s) \over dp_t^2 dM_X} \ \ .
\end{eqnarray}
Also the transverse momentum distributions show a good agreement for the
different parametrizations of the proton structure function.
We see that up to $p_t^2 \lsim 2 \, \rm{GeV}^2$ the $p_t^2$-distributions have an
approximate exponential behaviour $\propto \exp[-B_{\rm inel} p_t^2]$.

\begin{figure}[!htb] 
\includegraphics[height=6.75cm]{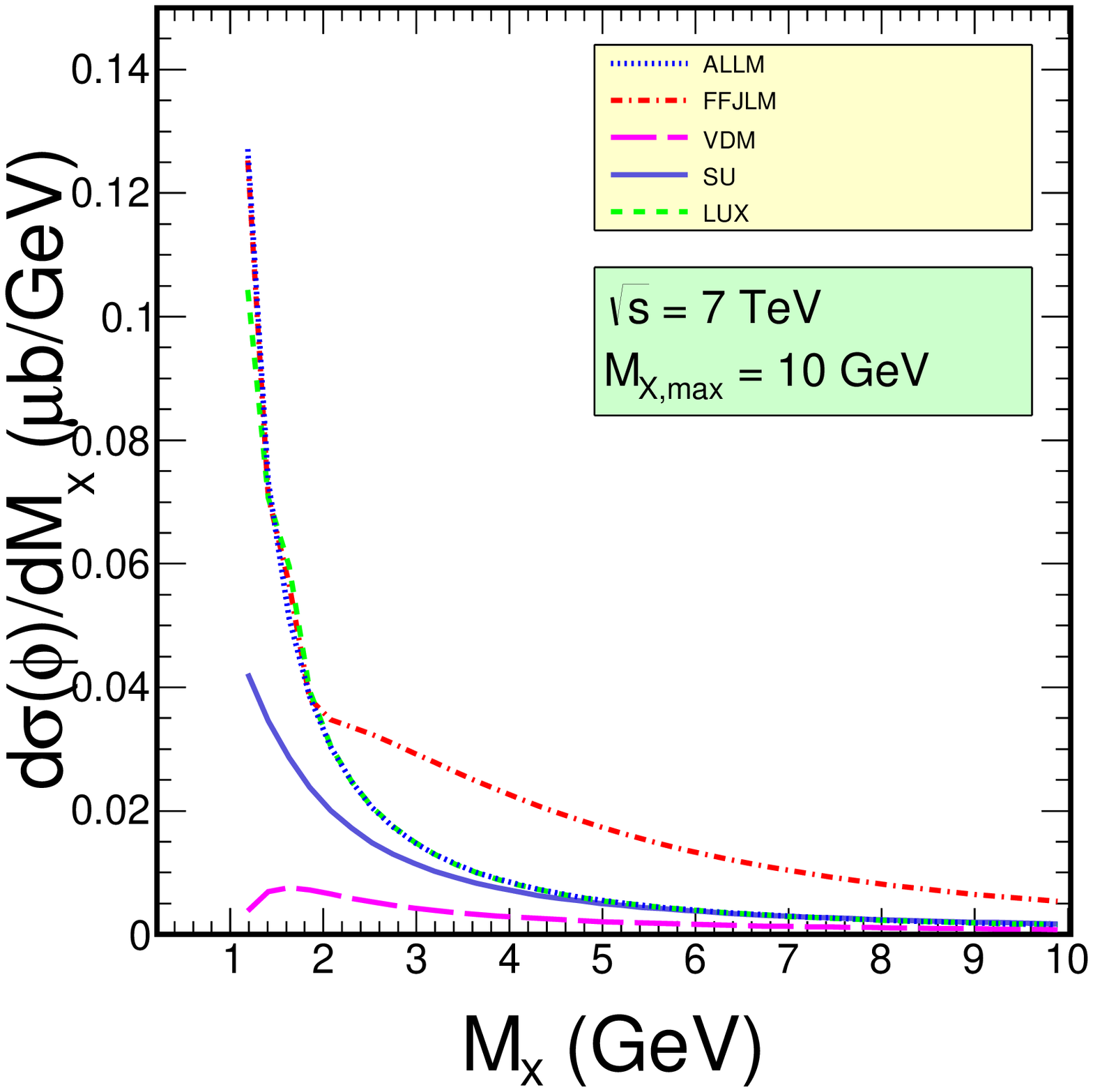}
\includegraphics[height=6.75cm]{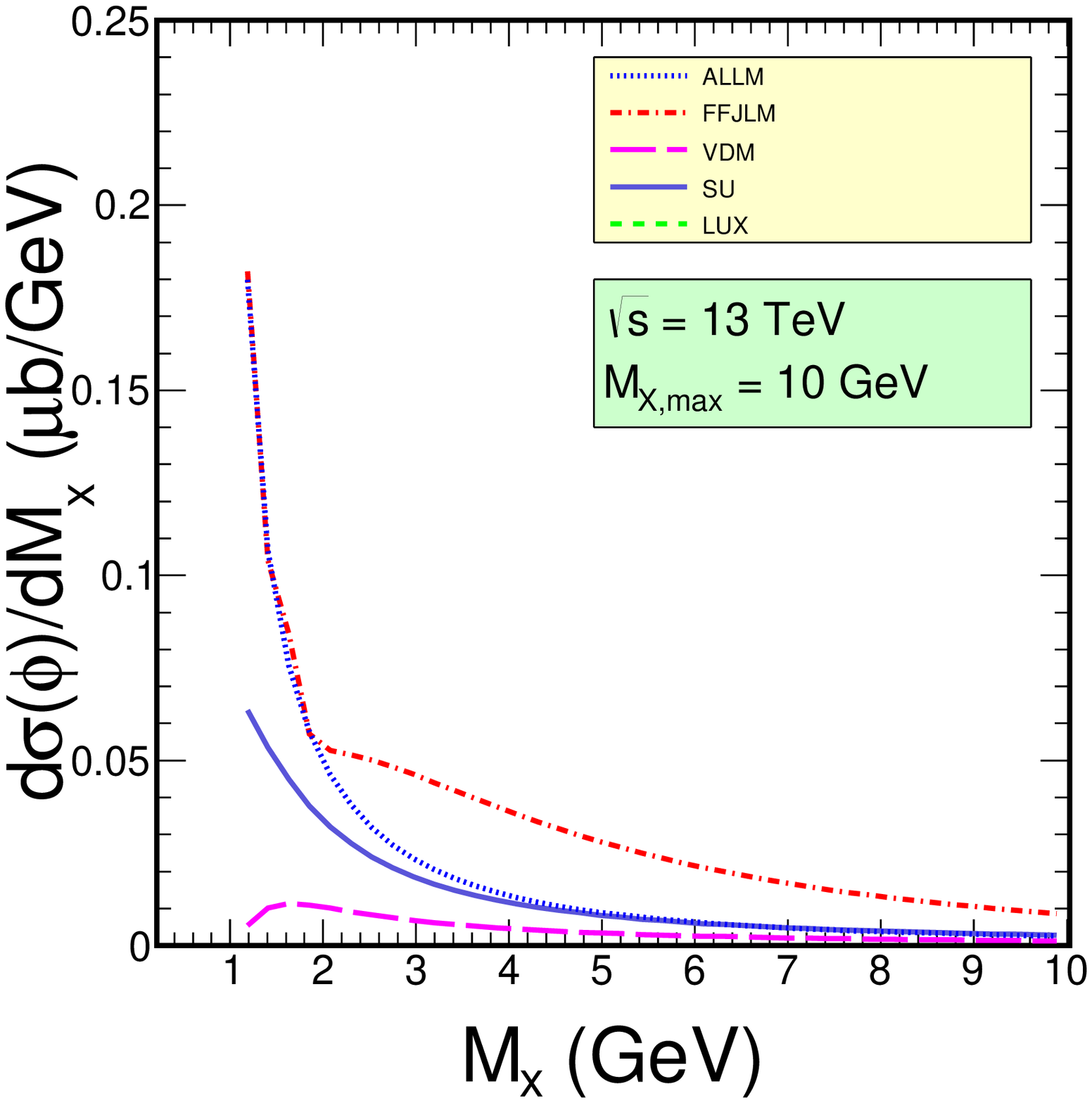}
\includegraphics[height=6.75cm]{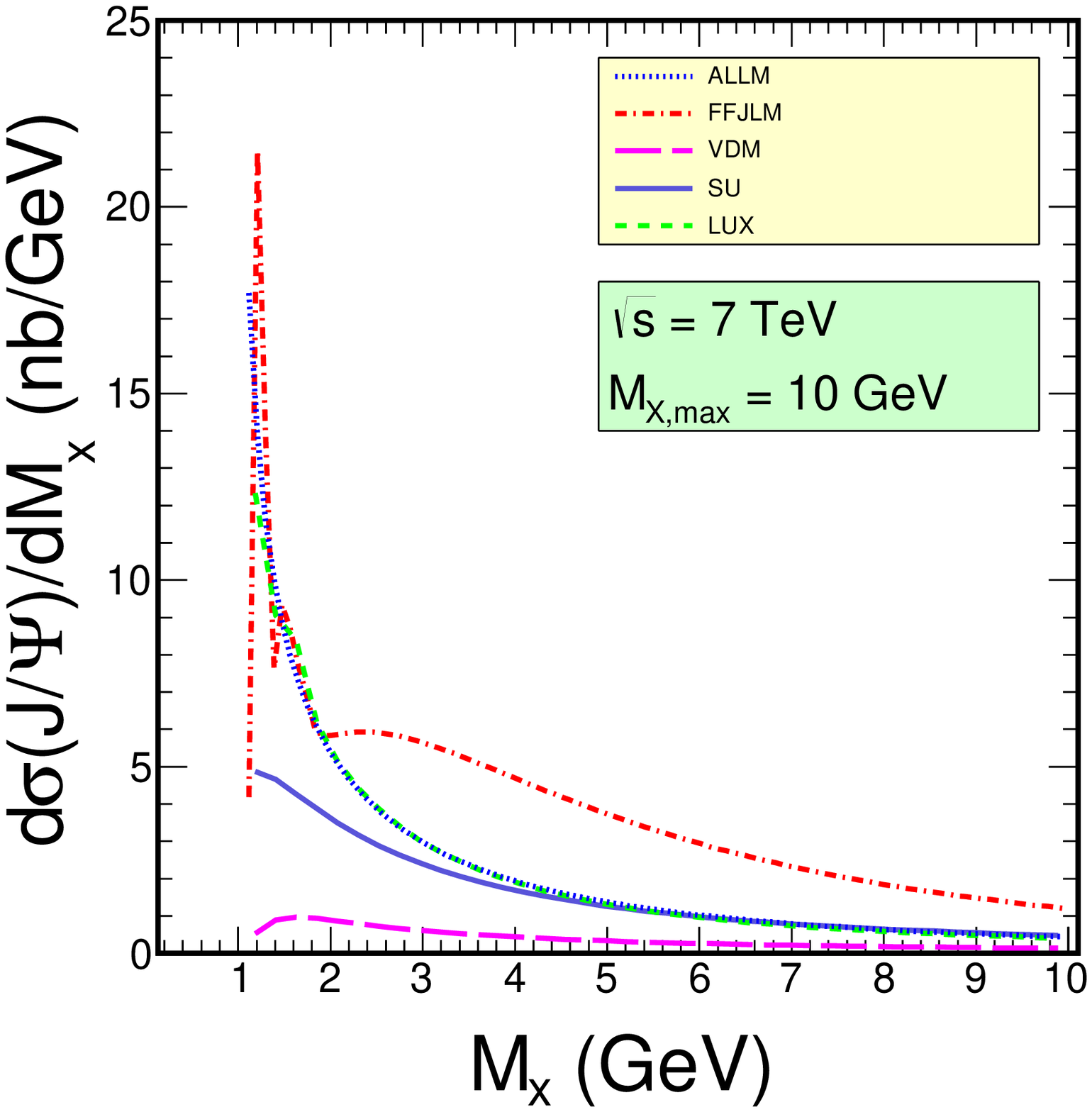}
\includegraphics[height=6.75cm]{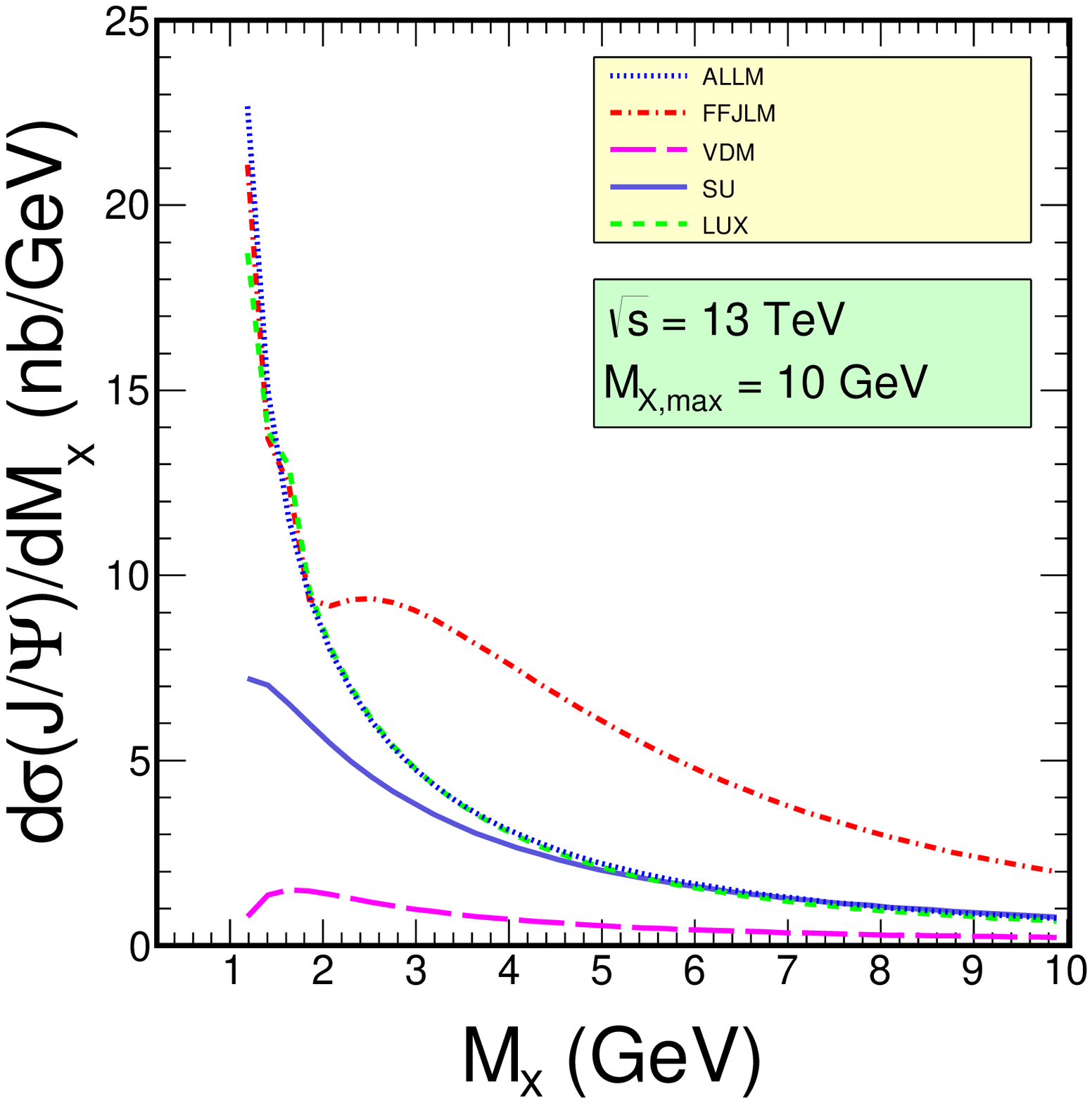}
\includegraphics[height=6.75cm]{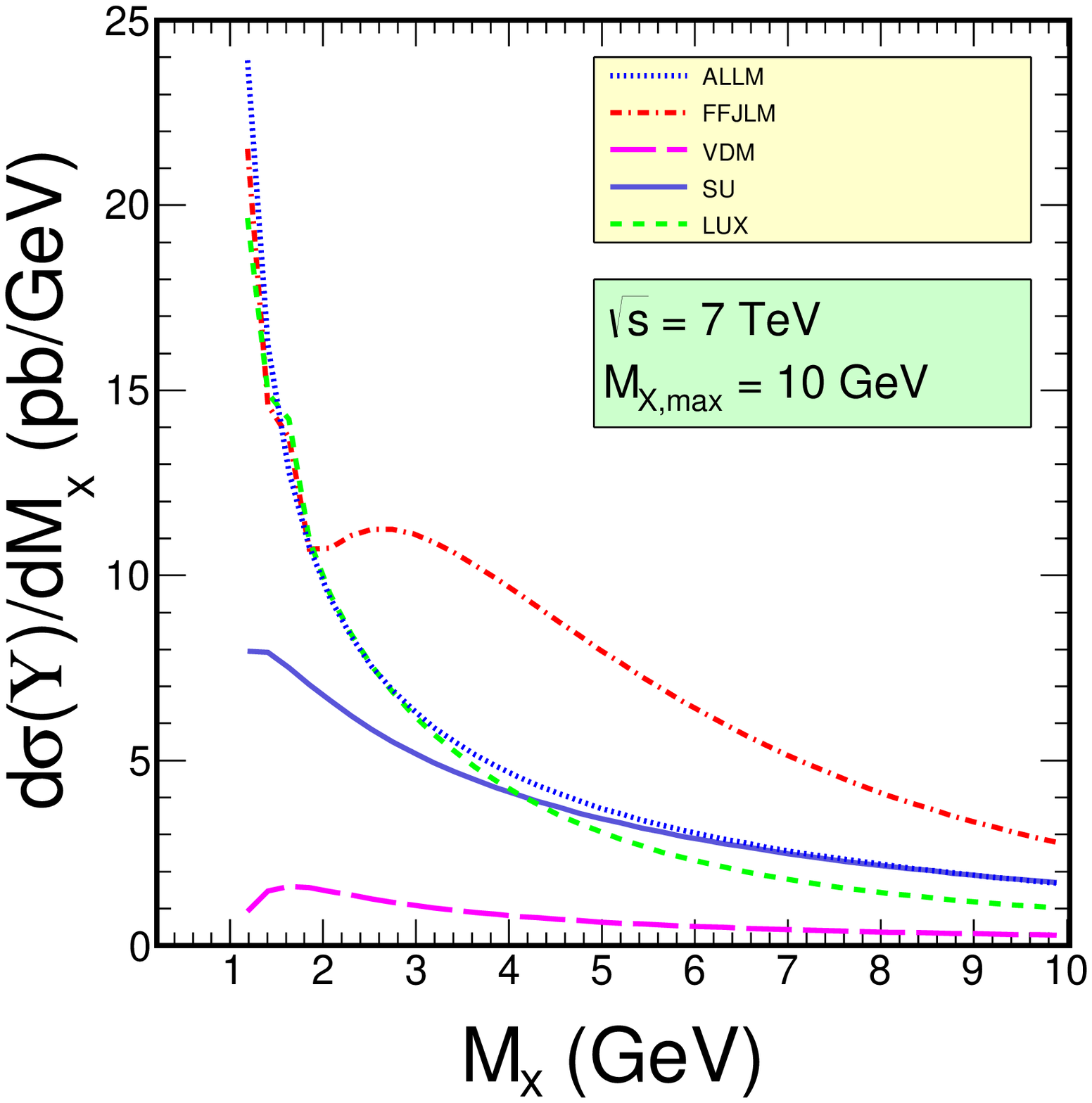}
\includegraphics[height=6.75cm]{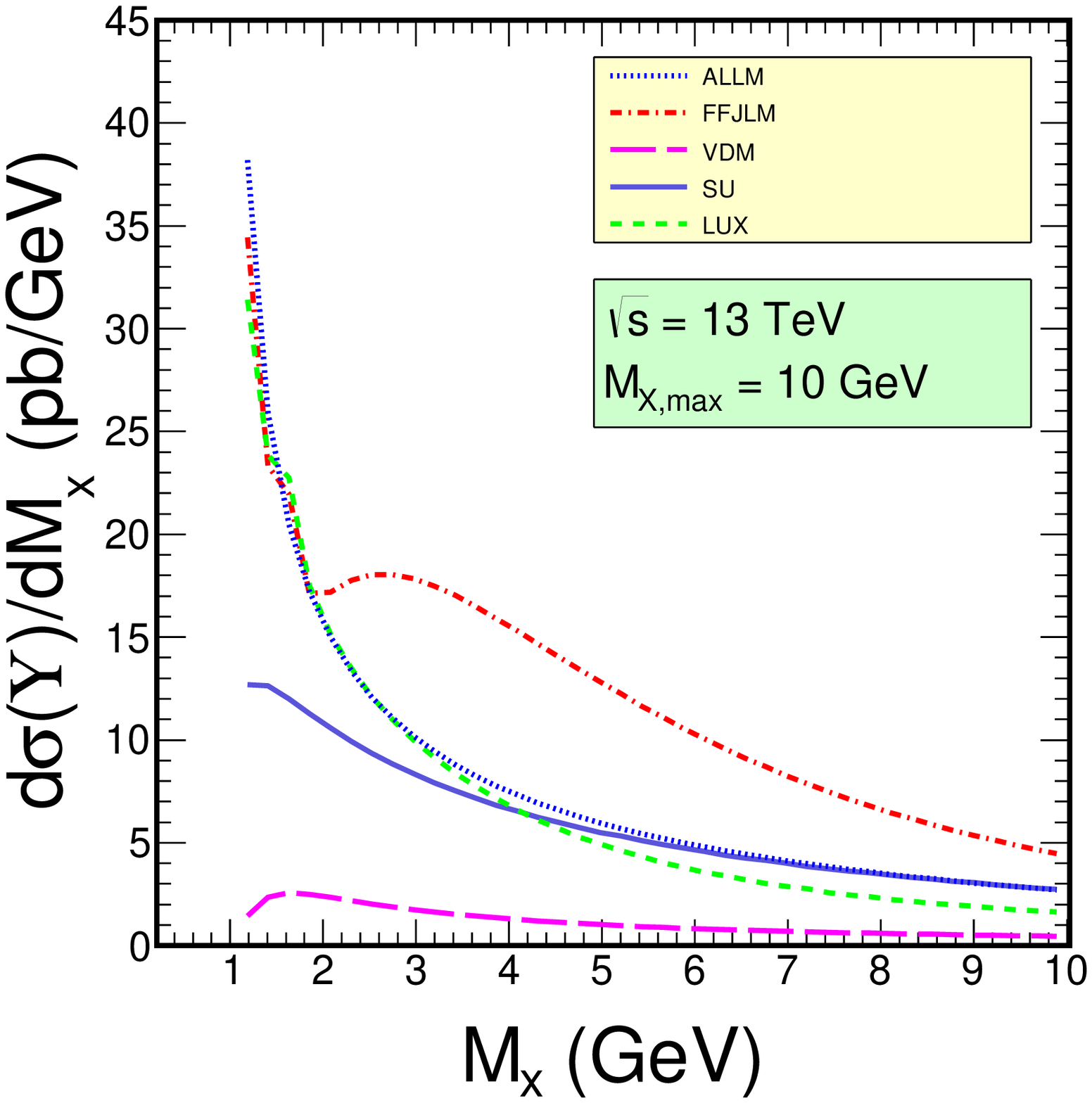}
\caption[*]{Distribution of the invariant mass $M_X$ of the excited system 
for the pp cm-energy $\sqrt{s} = 7 \, \rm {TeV}$ (left panels) and
 $\sqrt{s} = 13 \, \rm {TeV}$ (right panels).}
\label{fig:dsig_dMX}
\end{figure}

In Fig. \ref{fig:dsig_dMX} we show the distribution in the invariant mass $M_X$ of the excited 
system. Here we see, that the Fiore-fit behaves very differently from the other 
parametrizations at $M_X > 2 \ \rm{GeV}$. Due to this unphysical behaviour, it cannot be
used for large $M_X$. On the other hand, the Szczurek-Uleshchenko parametrization appears
to underestimate the cross section in the resonance region of small $M_X$. Here the 
LUX-type fit, ALLM and the Fiore parametrizations agree quite well.

\begin{figure}[!htb] 
\includegraphics[height=6.75cm]{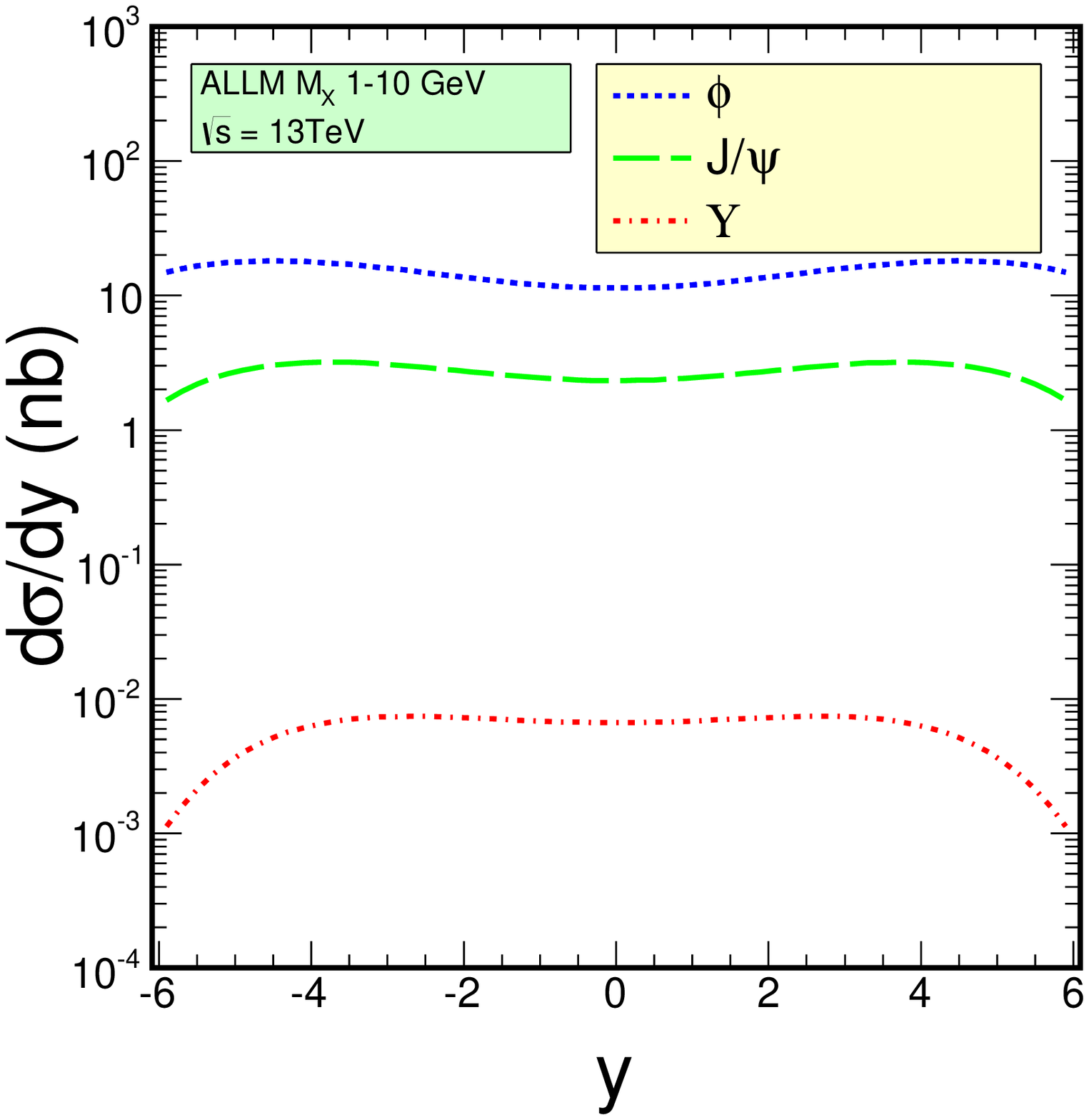}
\includegraphics[height=6.75cm]{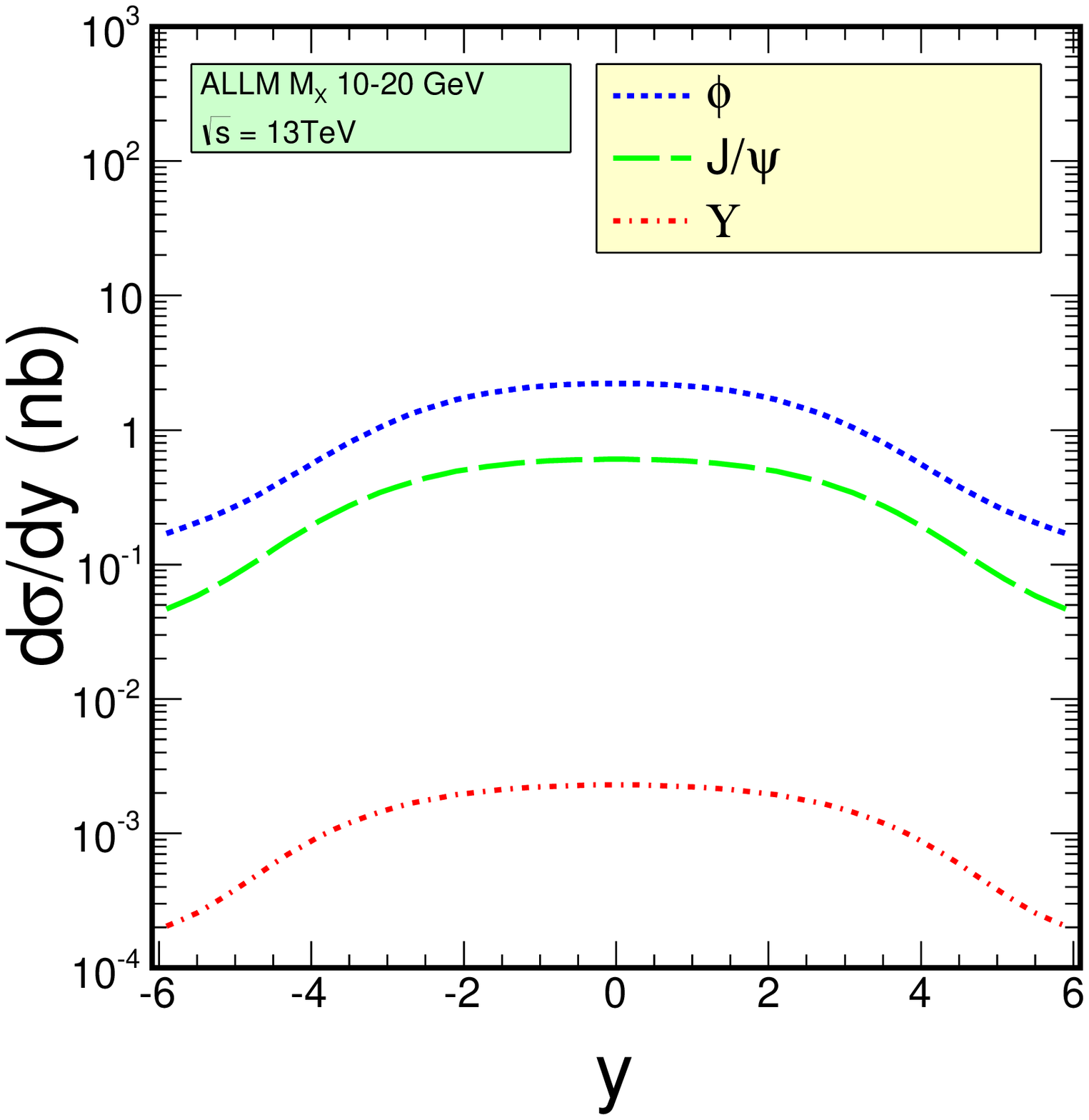}
\includegraphics[height=6.75cm]{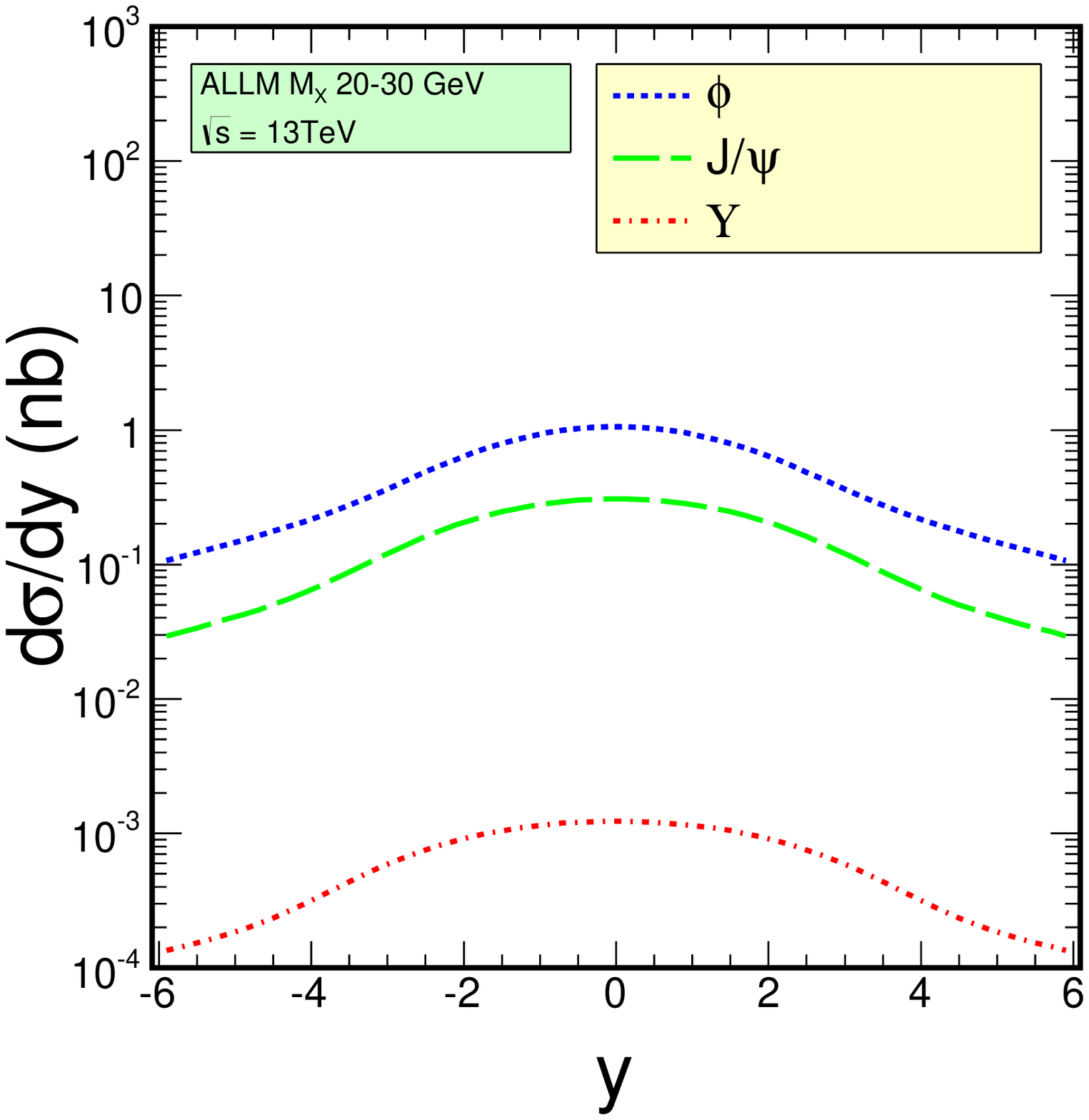}
\includegraphics[height=6.75cm]{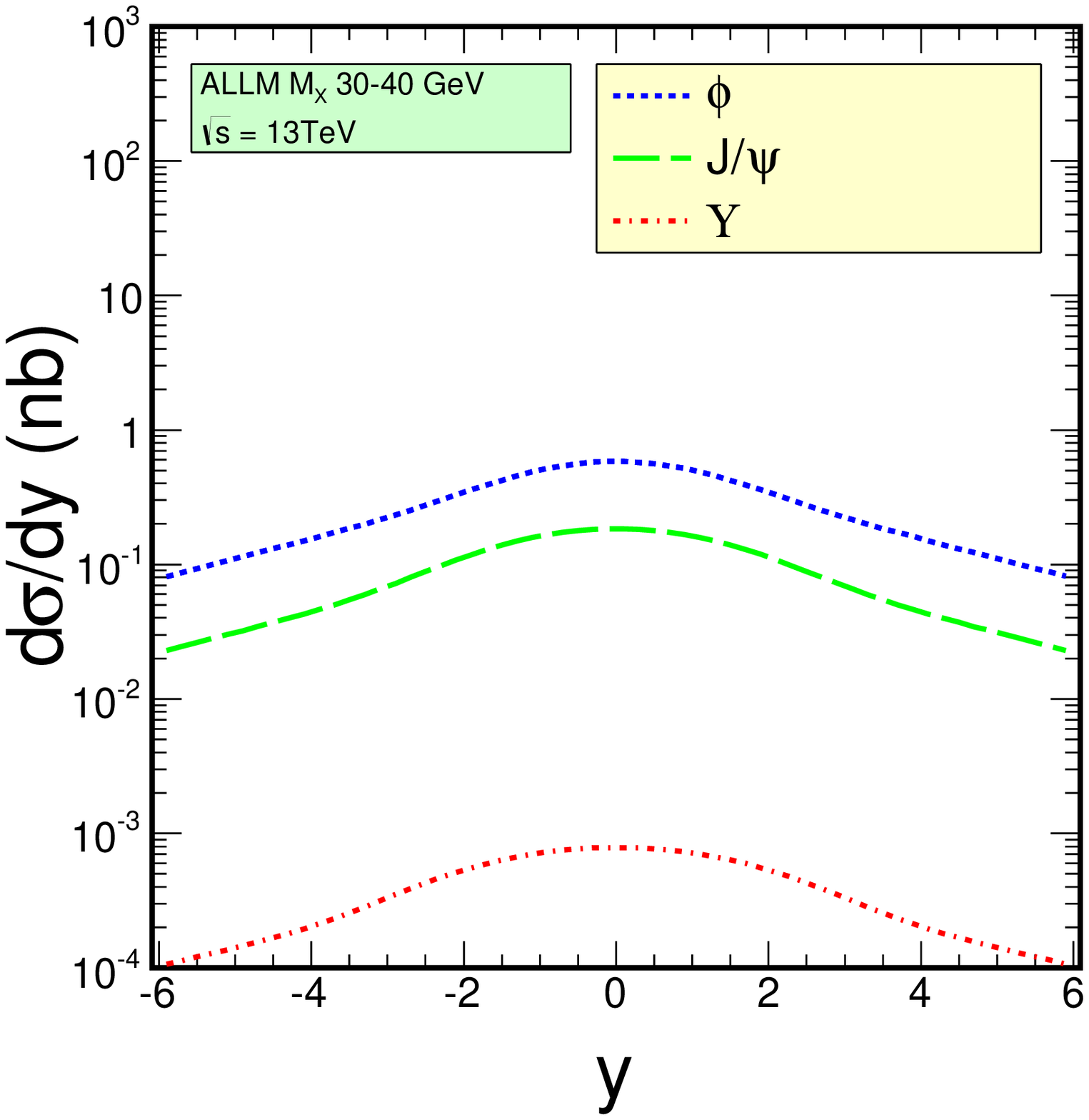}
\includegraphics[height=6.75cm]{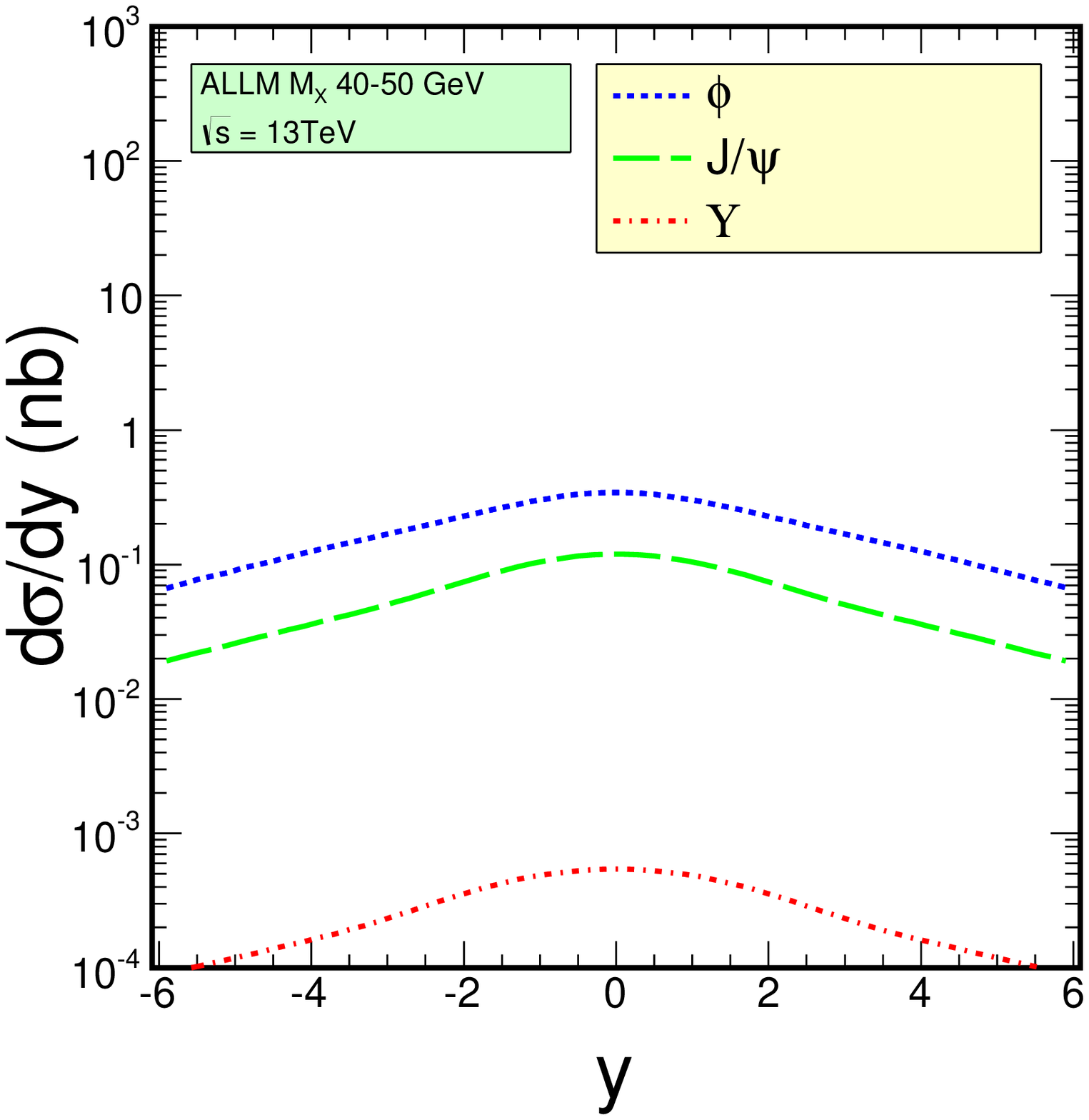}
\caption[*]{Rapidity distribution of vector mesons for different bins of the
invariant mass of the excited system for the ALLM parametrisation.}
\label{fig:y_MX_bins_ALLM}
\end{figure}

Let us have a closer look at the correlation of the $M_X$-dependence with rapidity 
of the vector meson. In Fig. \ref{fig:y_MX_bins_ALLM}
we show the rapidity distribution of mesons for ALLM structure function of proton and
different bins of $M_X \in [ M_{X,\rm min}, M_{X\rm, max}]$.
We observe that rapidity distributions for bins with $M_X > 10 \, \rm{GeV}$ are peaked 
at midrapidity and are somewhat narrower than the contribution of $M_X \leq 10 \, {\rm GeV}$.

\begin{figure}[!htb] 
\includegraphics[height=6.75cm]{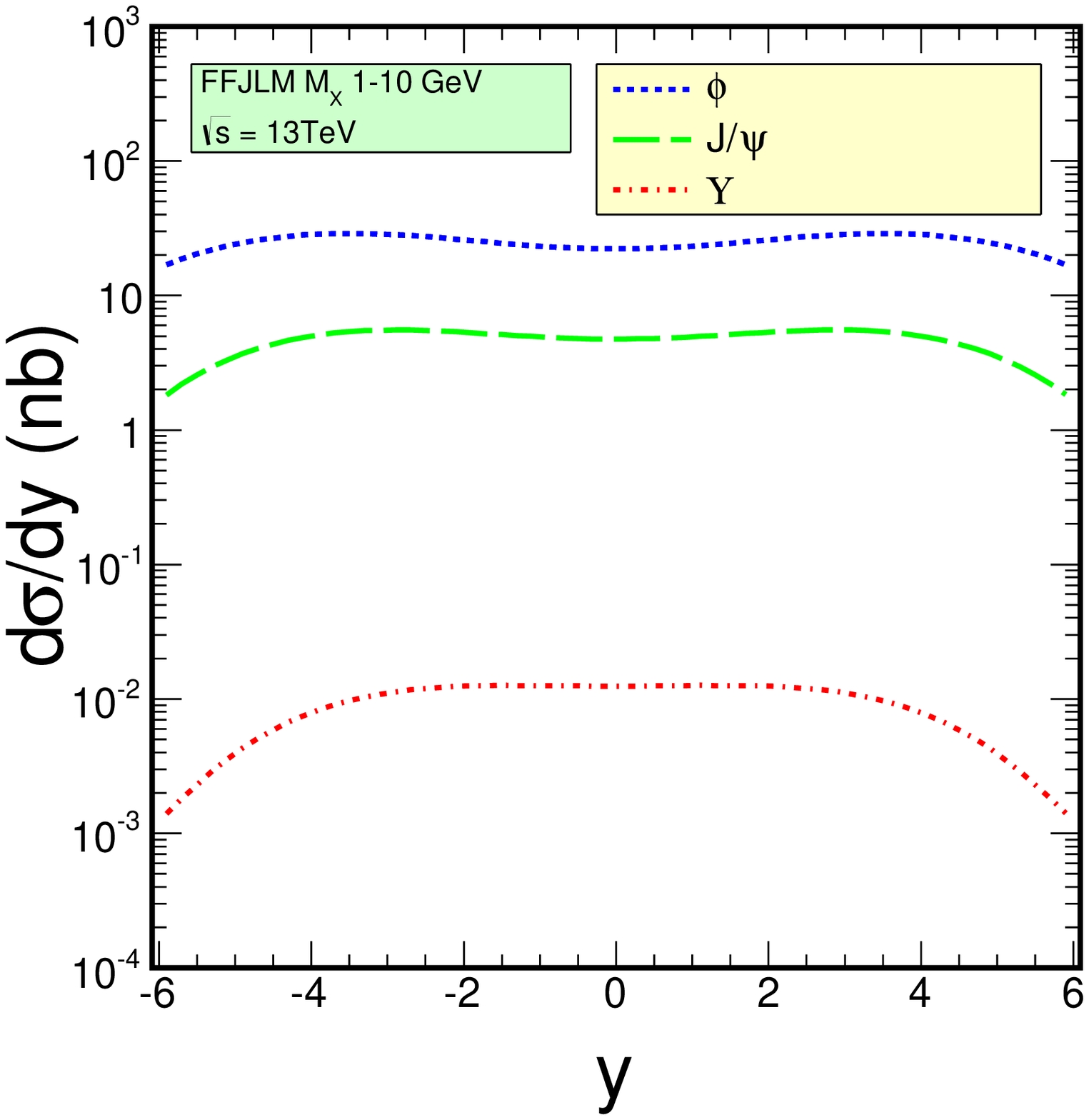}
\includegraphics[height=6.75cm]{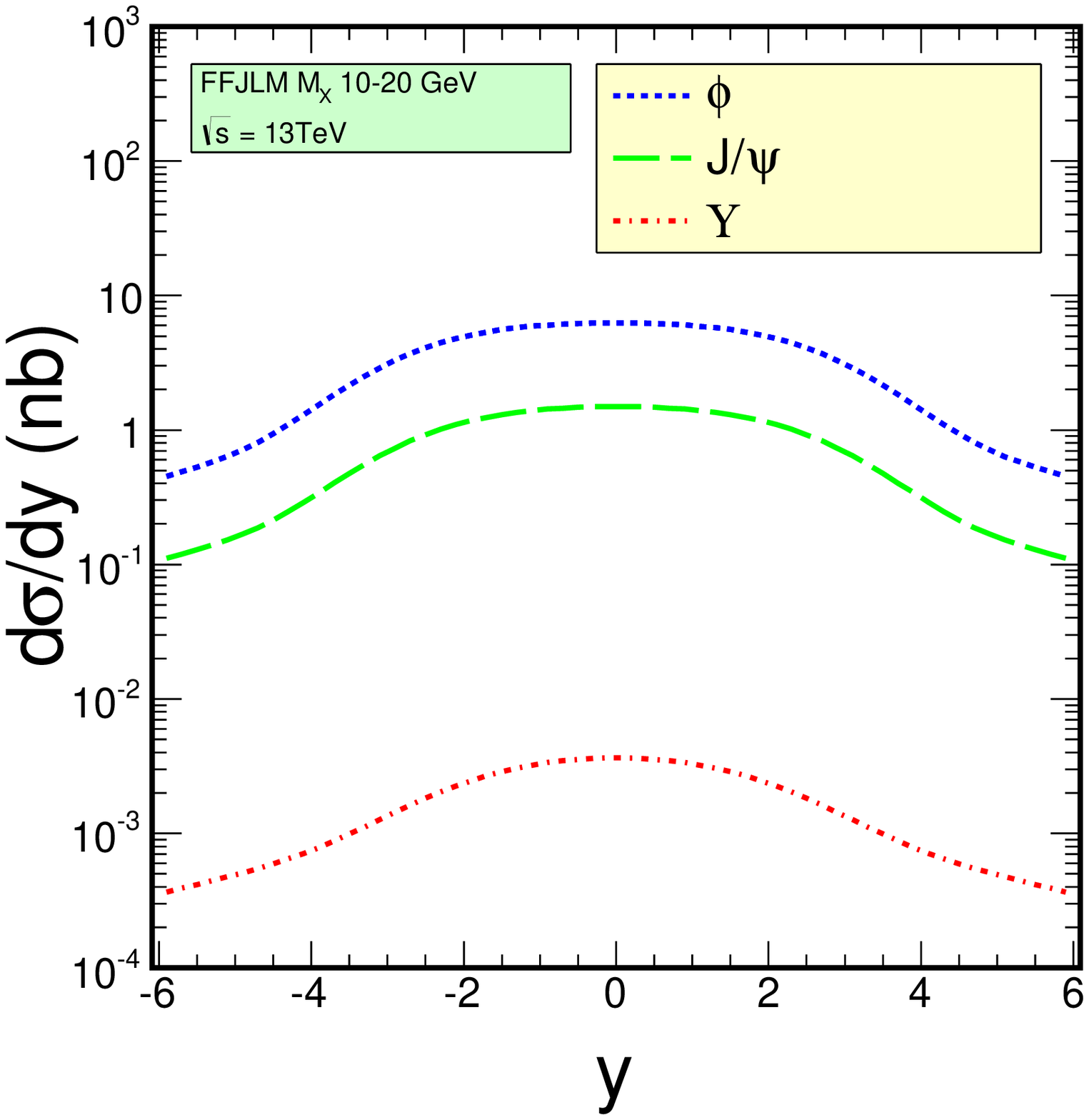}
\includegraphics[height=6.75cm]{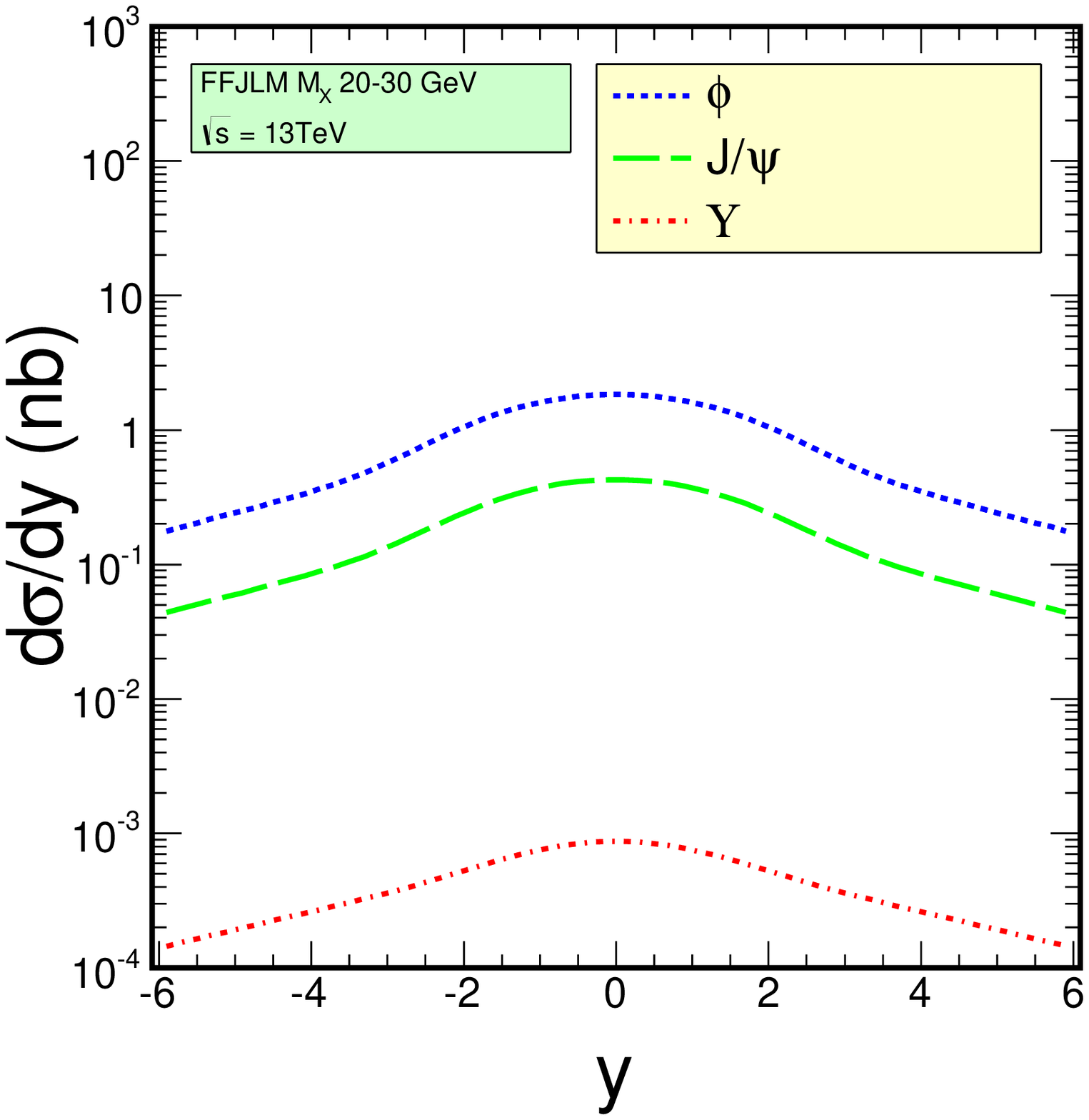}
\includegraphics[height=6.75cm]{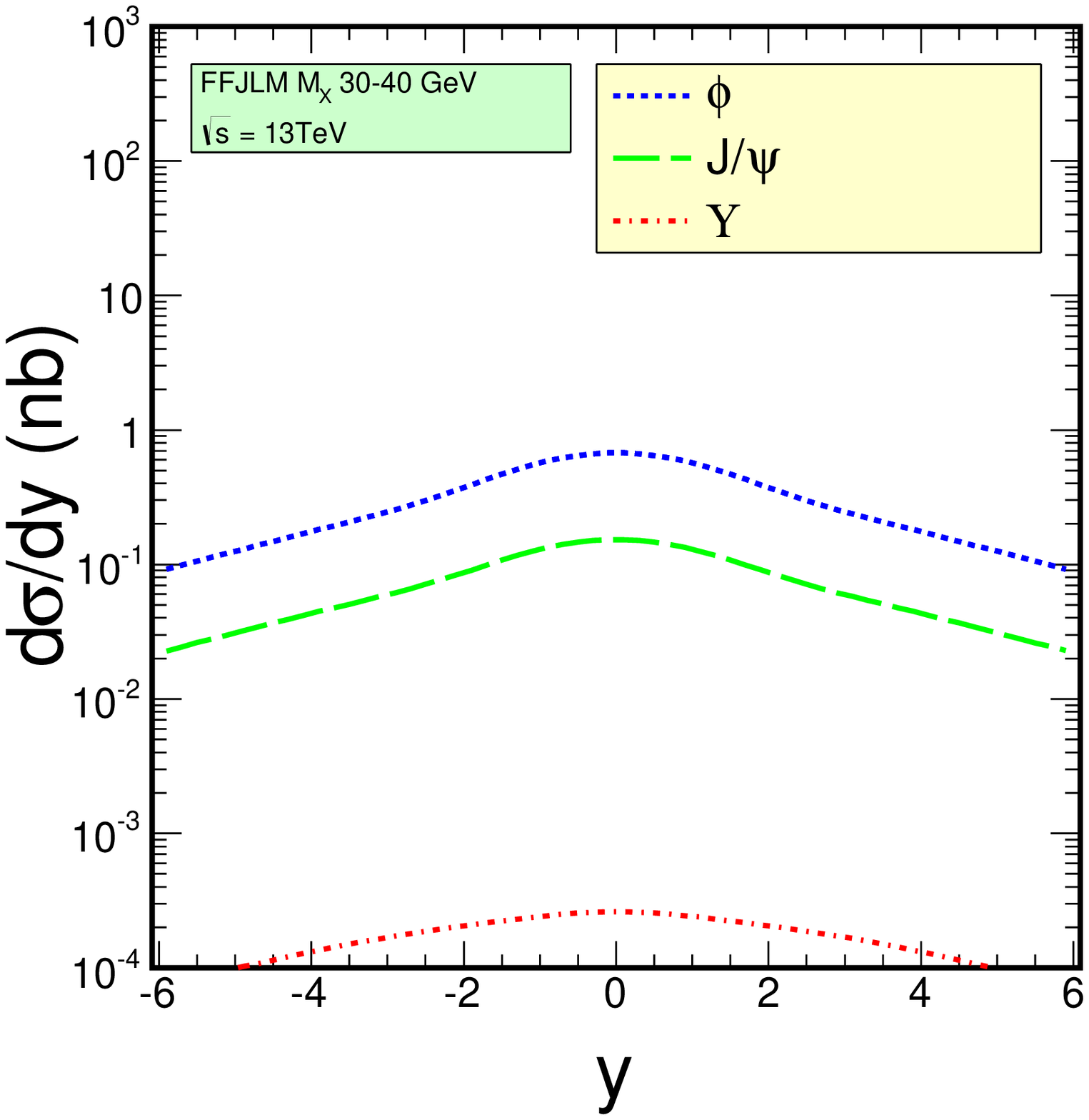}
\includegraphics[height=6.75cm]{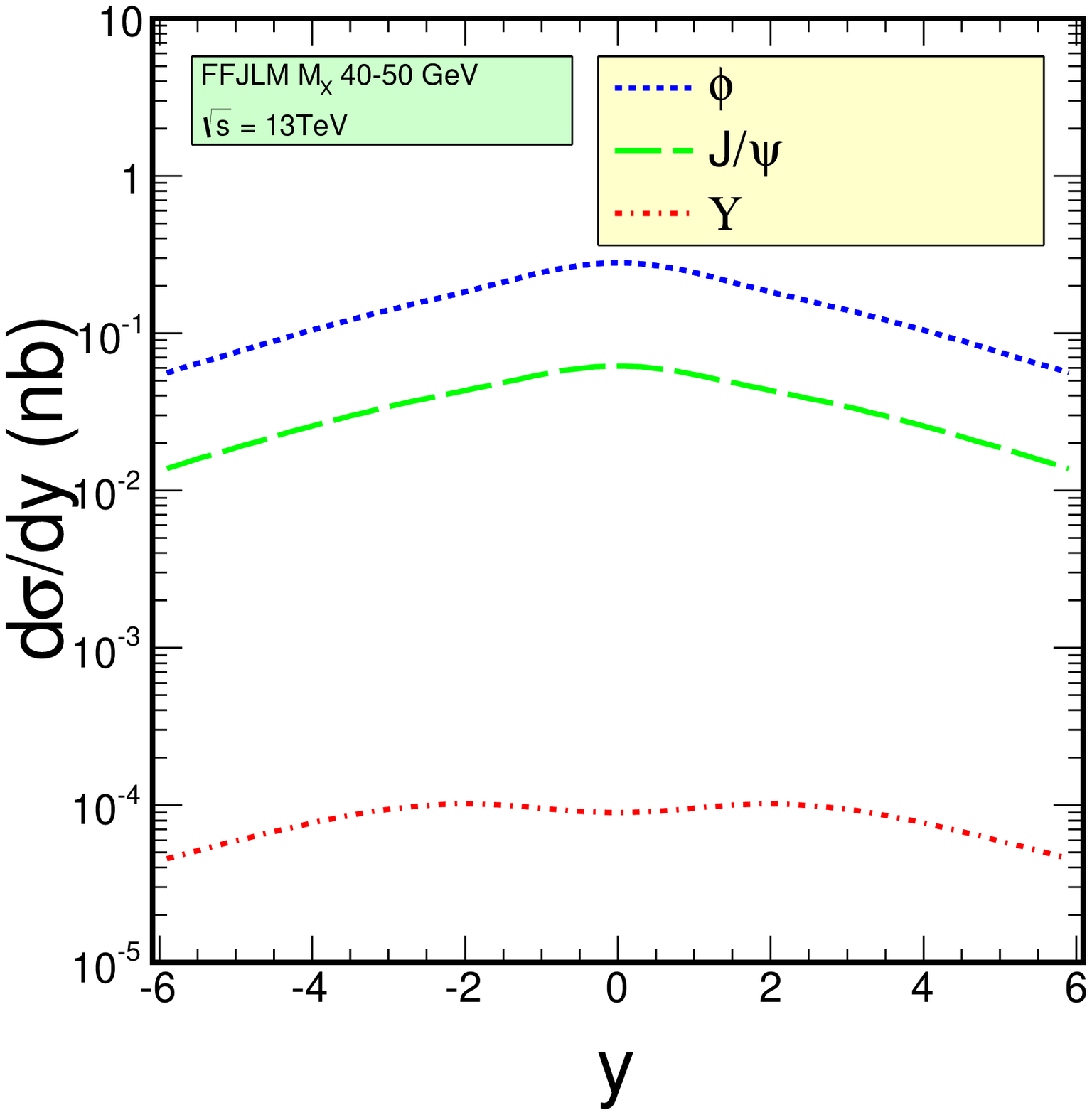}
\caption[*]{Rapidity distribution of vector mesons for different bins of the
	invariant mass of the excited system for the FFJLM parametrization.}
\label{fig:y_MX_bins_Fiore}
\end{figure}

In Fig. \ref{fig:y_MX_bins_Fiore} we show the rapidity distribution of mesons for different bins of $M_X$
for FFJLM structure function of proton, similary as Fig. \ref{fig:y_MX_bins_ALLM}.

\begin{figure}[!htb] 
\includegraphics[height=6.75cm]{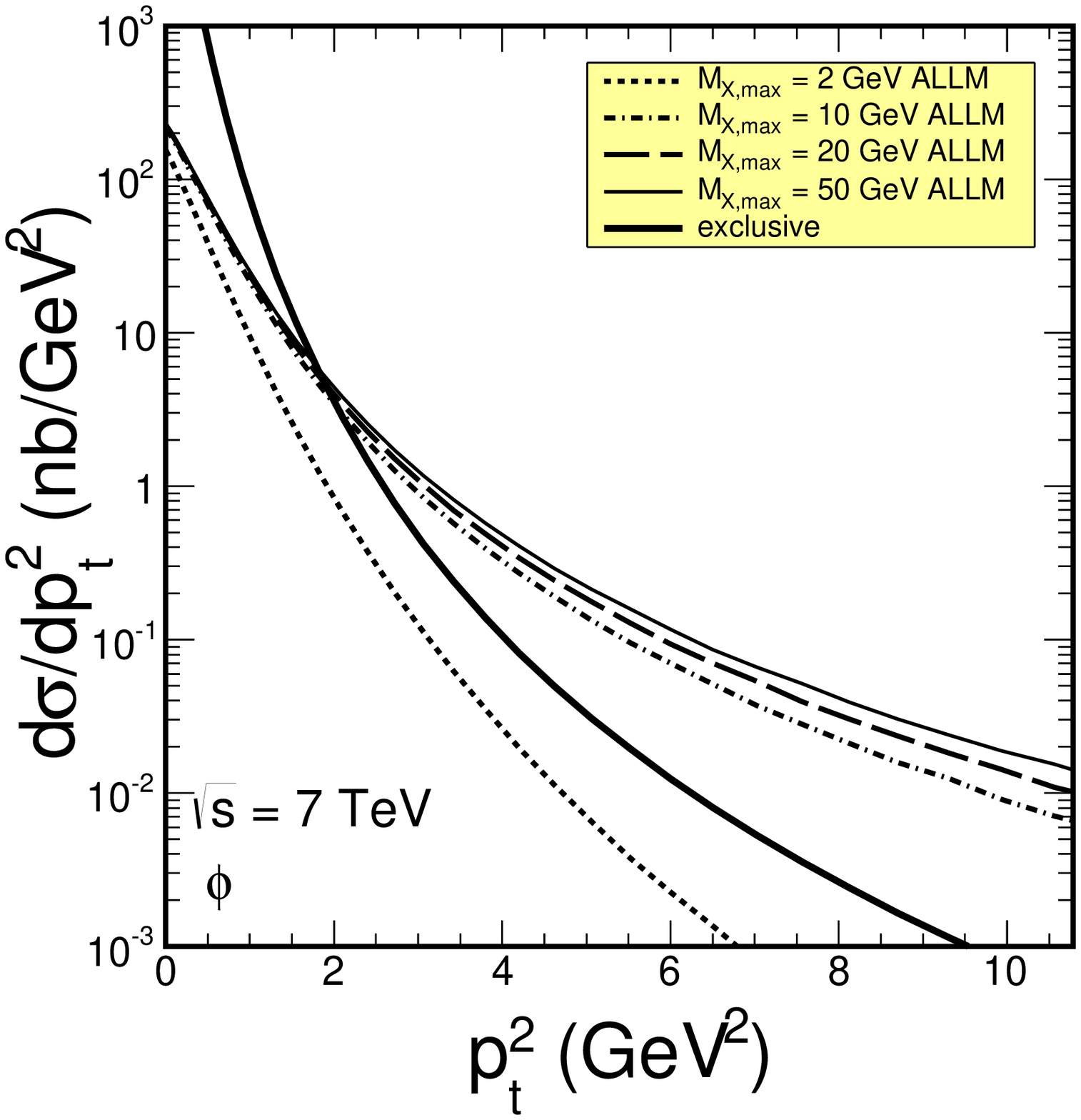}
\includegraphics[height=6.75cm]{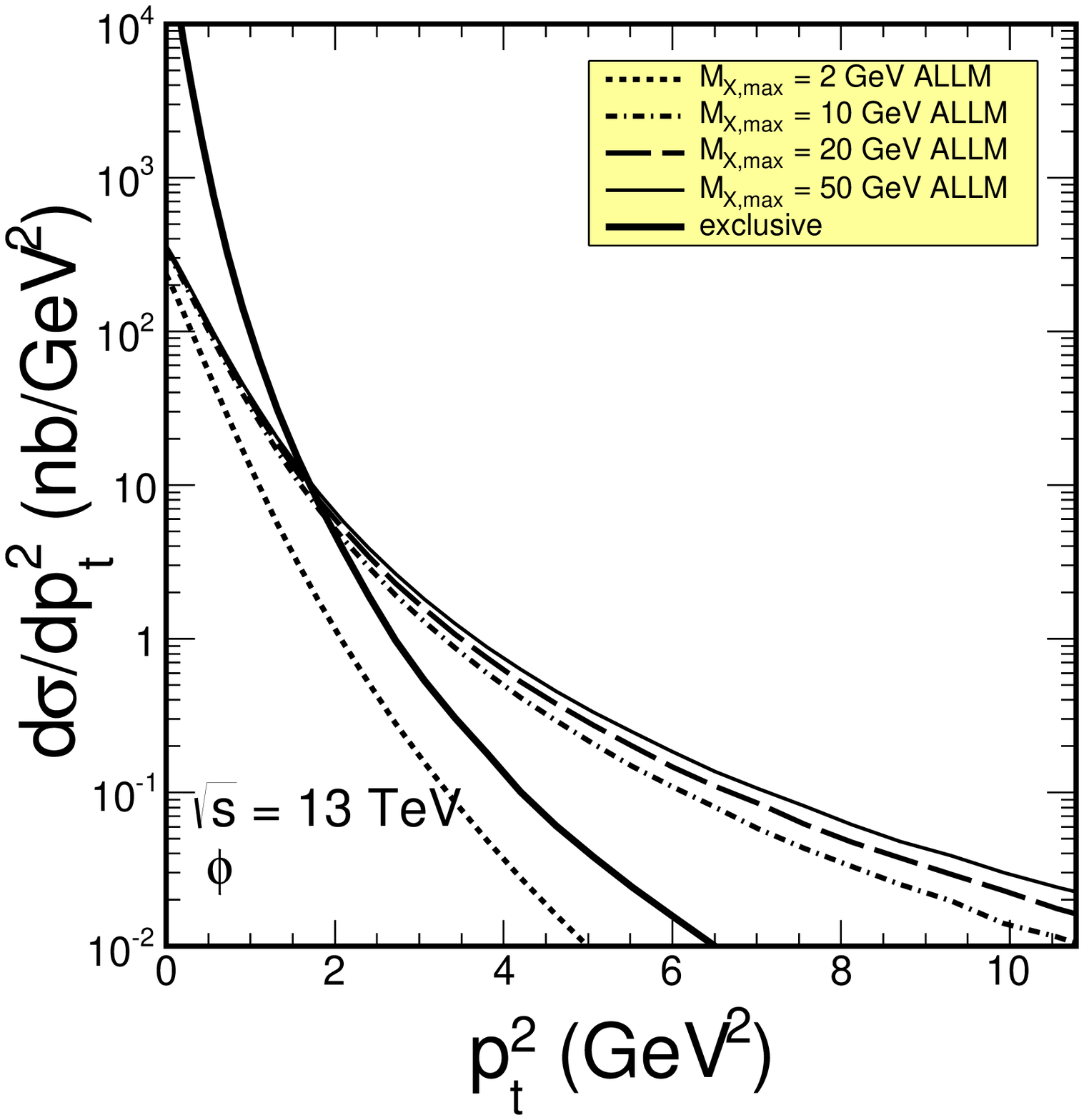}
\includegraphics[height=6.75cm]{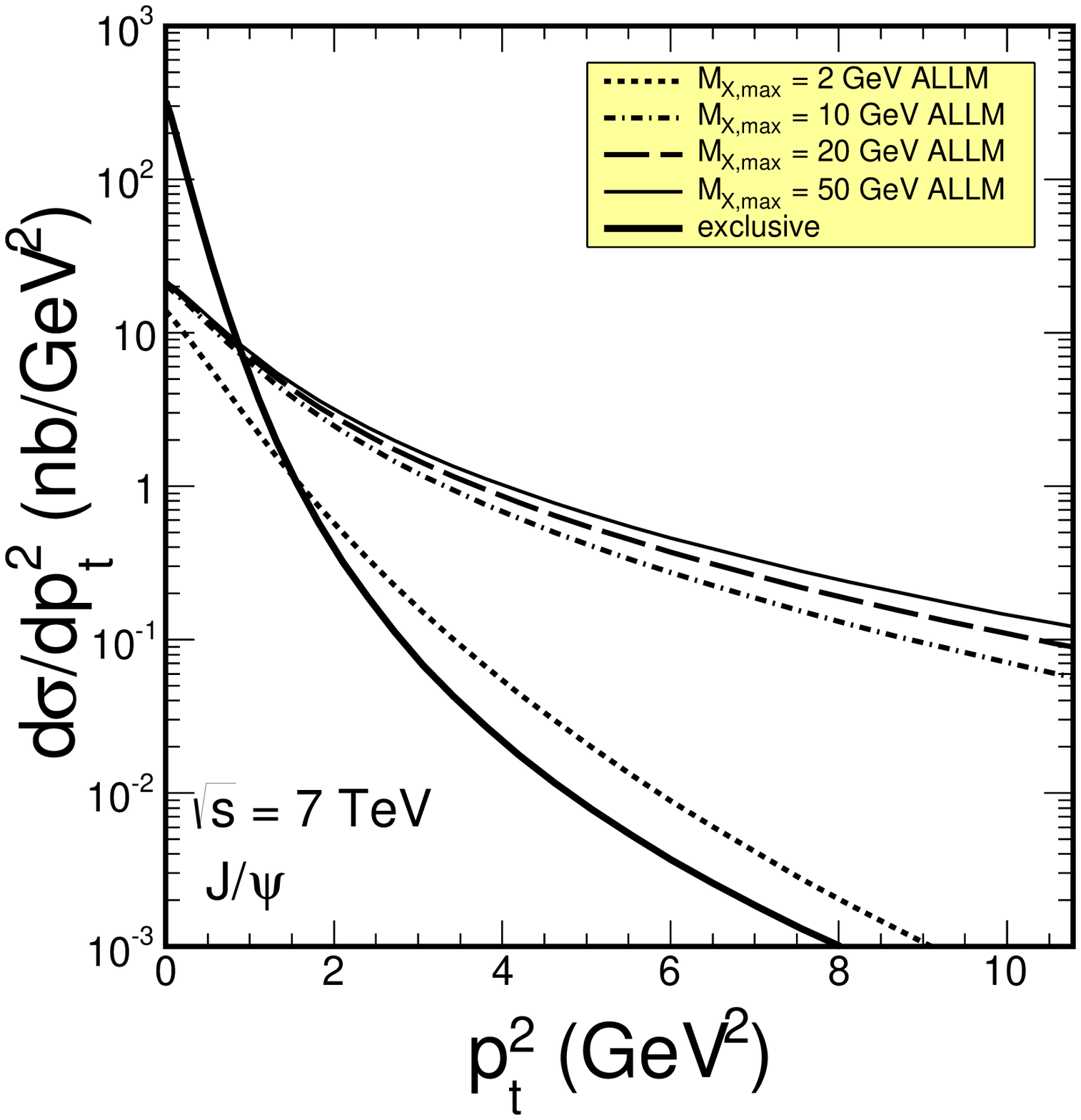}
\includegraphics[height=6.75cm]{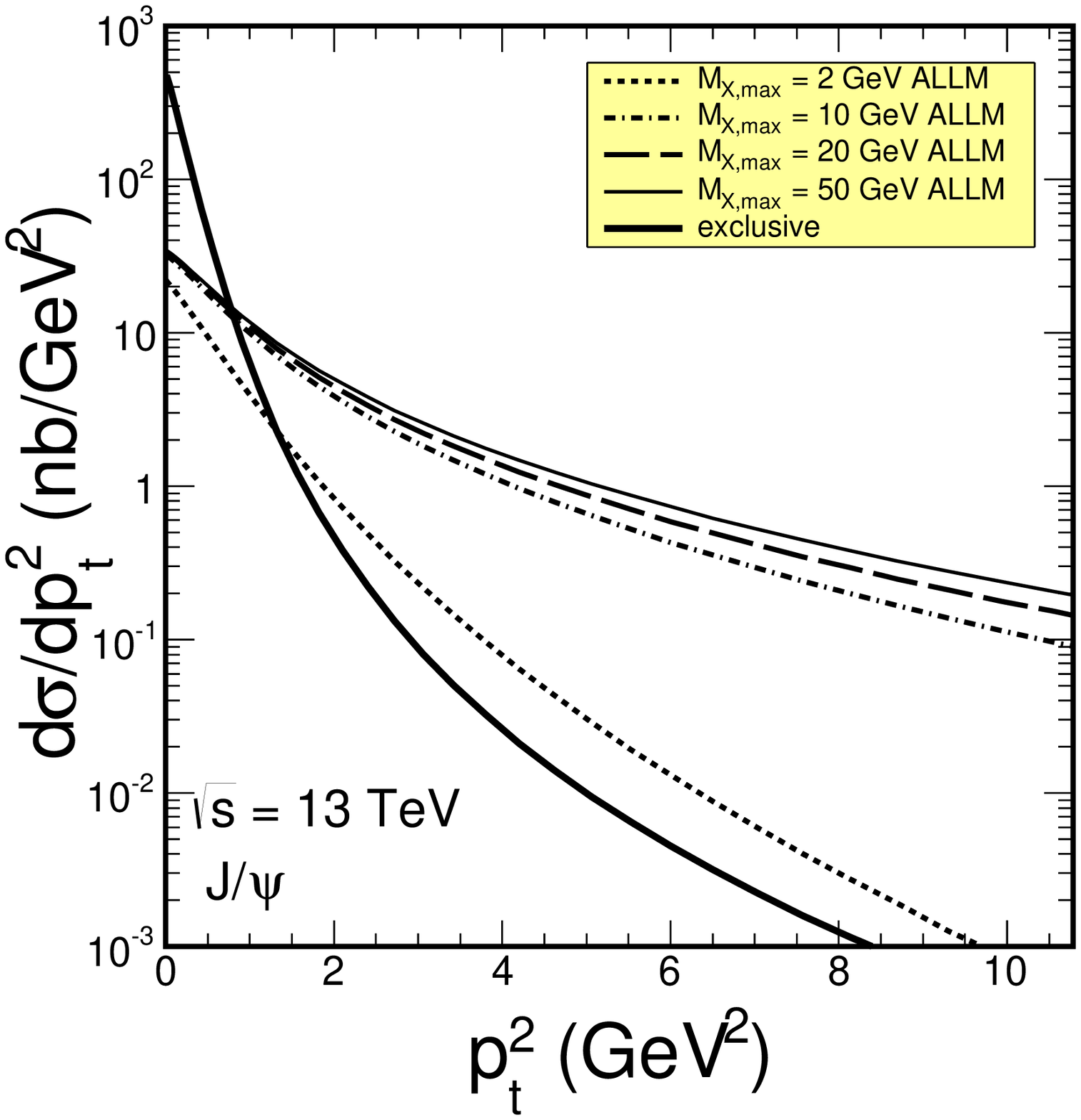}
\includegraphics[height=6.75cm]{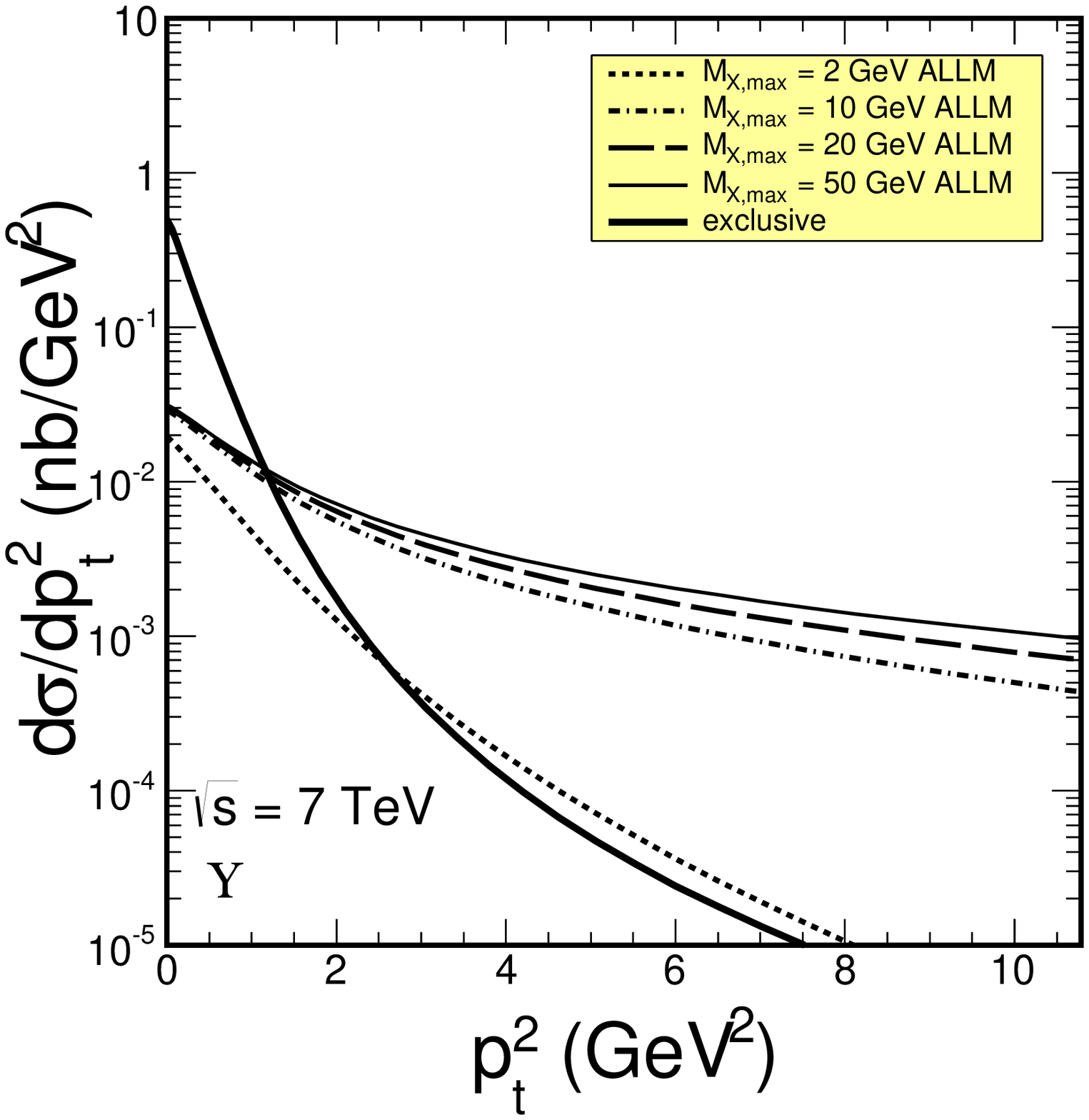}
\includegraphics[height=6.75cm]{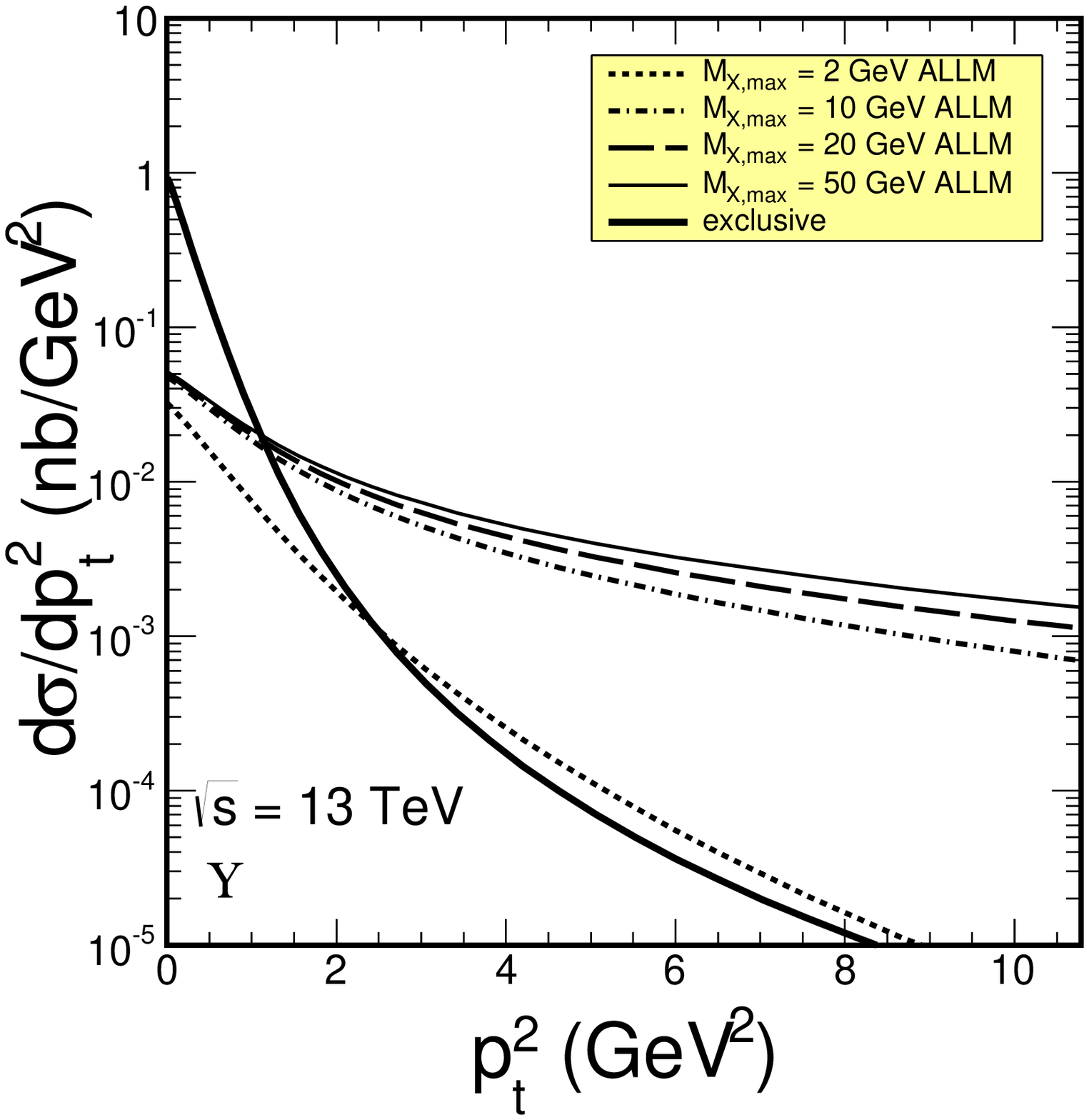}
\caption[*]{
$p_t^2$-distribution of vector mesons for different upper limits on the
invariant mass of the excited system.
}
\label{fig:dsig_dpt2_MX}
\end{figure}

We return to the transverse momentum distributions in Fig. \ref{fig:dsig_dpt2_MX},
where we plot the $p_t^2$-distribution with $M_X$ integrated up to different 
values of $M_{x, \rm max}$. We observe that the contribution from the resonance region
of $M_X < 2 \, \rm{GeV}$ has a much softer tail at large $p_t^2$ than
when the large-mass contribution is added. For comparison, we also show the 
$p_t^2$-distribution of the exclusive $pp \to ppV$ contribution \cite{CSS2015}.
The theoretical analysis for exclusive photoproduction in proton-proton callisions
was shown also in papers \cite{Motyka:2008ac,Jones:2013pga,Goncalves:2015pki}.
The shape of the $p_t^2$-distributions only weakly depends on energy.

Under the conditions of LHC experiments, like \cite{Aaij:2013jxj,Aaij:2015kea},
the dissociative diffractive production is a background to the fully exclusive production
with intact protons. 

\begin{figure}[!htb] 
\includegraphics[height=6.75cm]{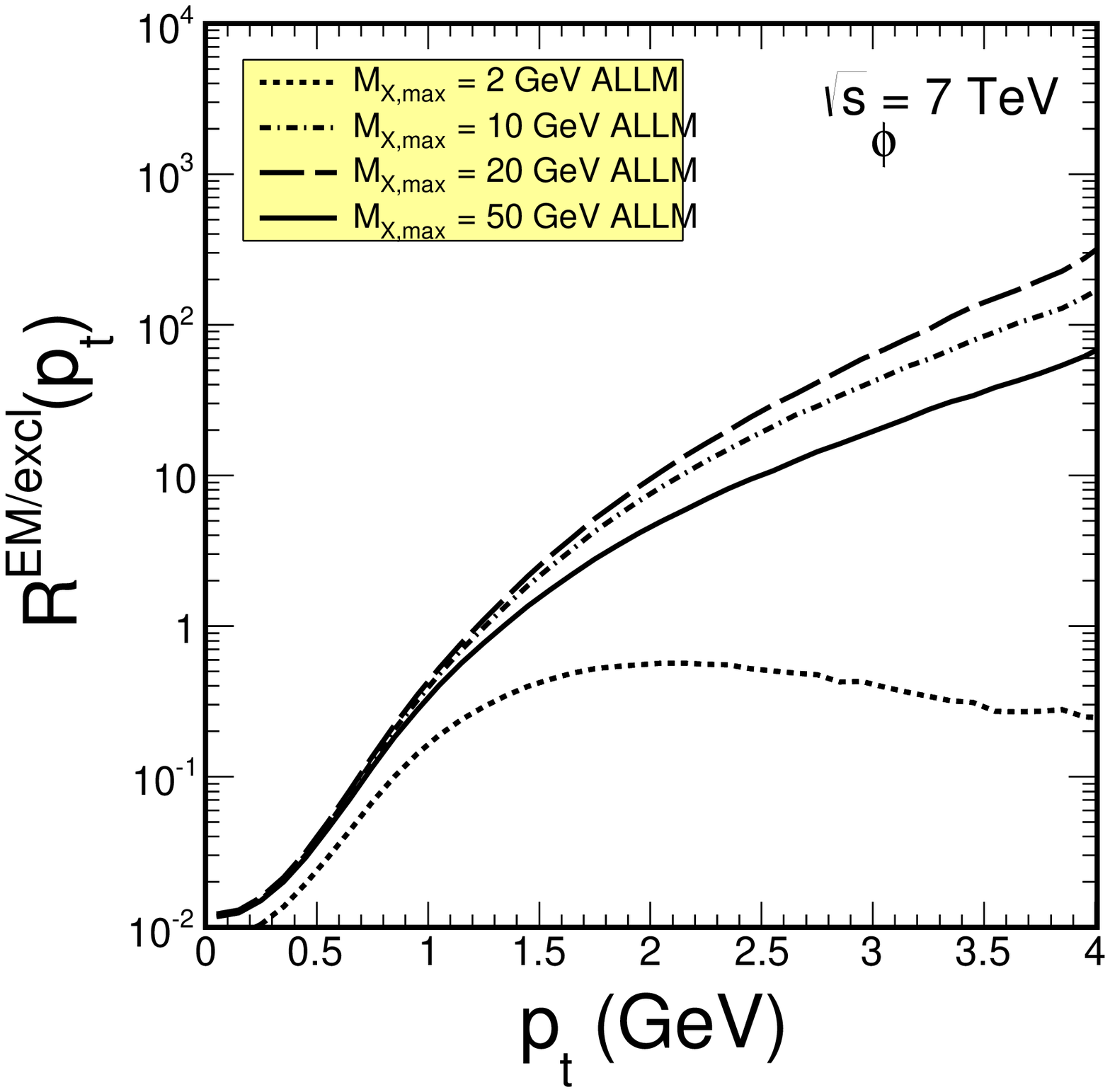}
\includegraphics[height=6.75cm]{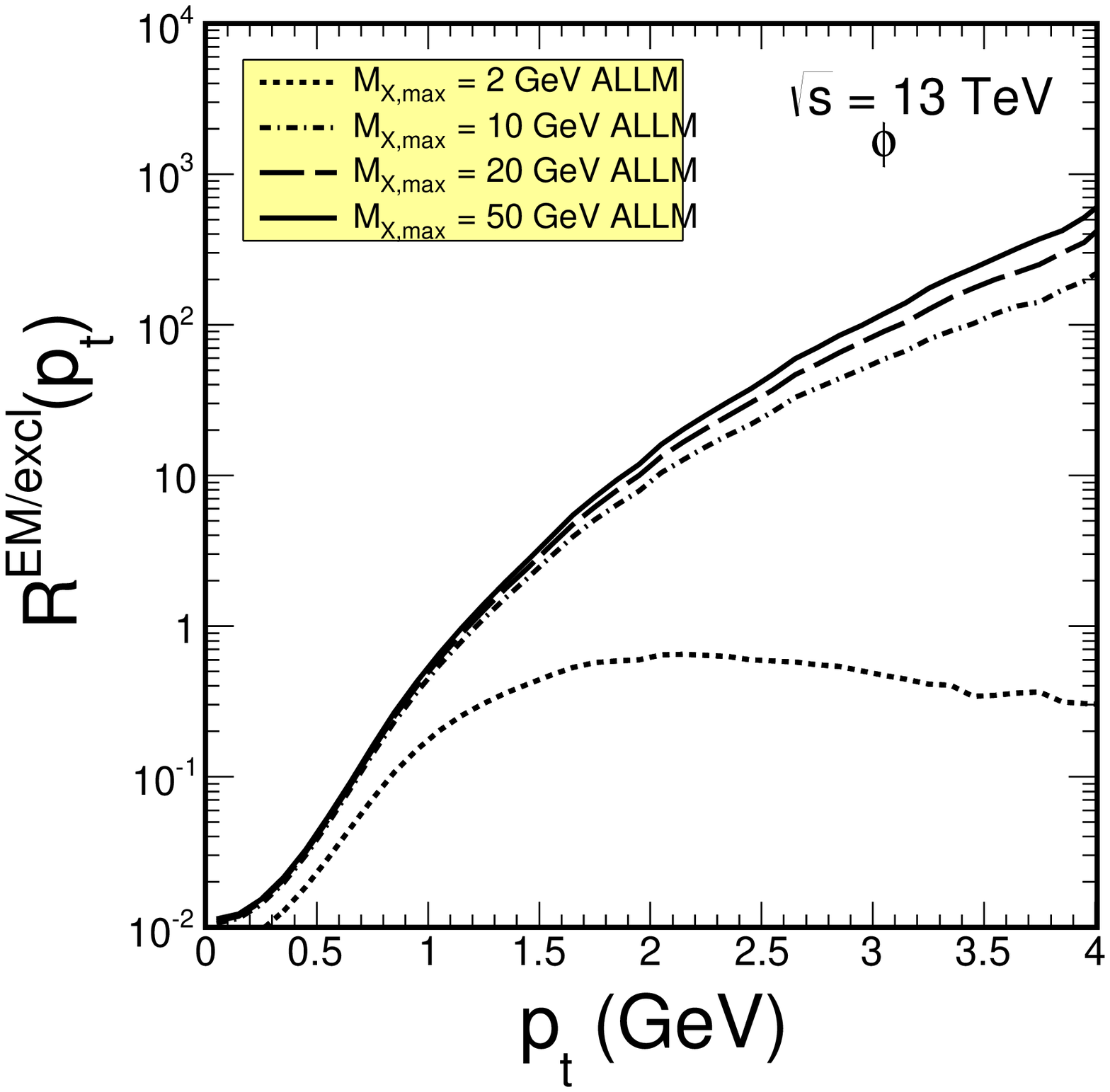}
\includegraphics[height=6.75cm]{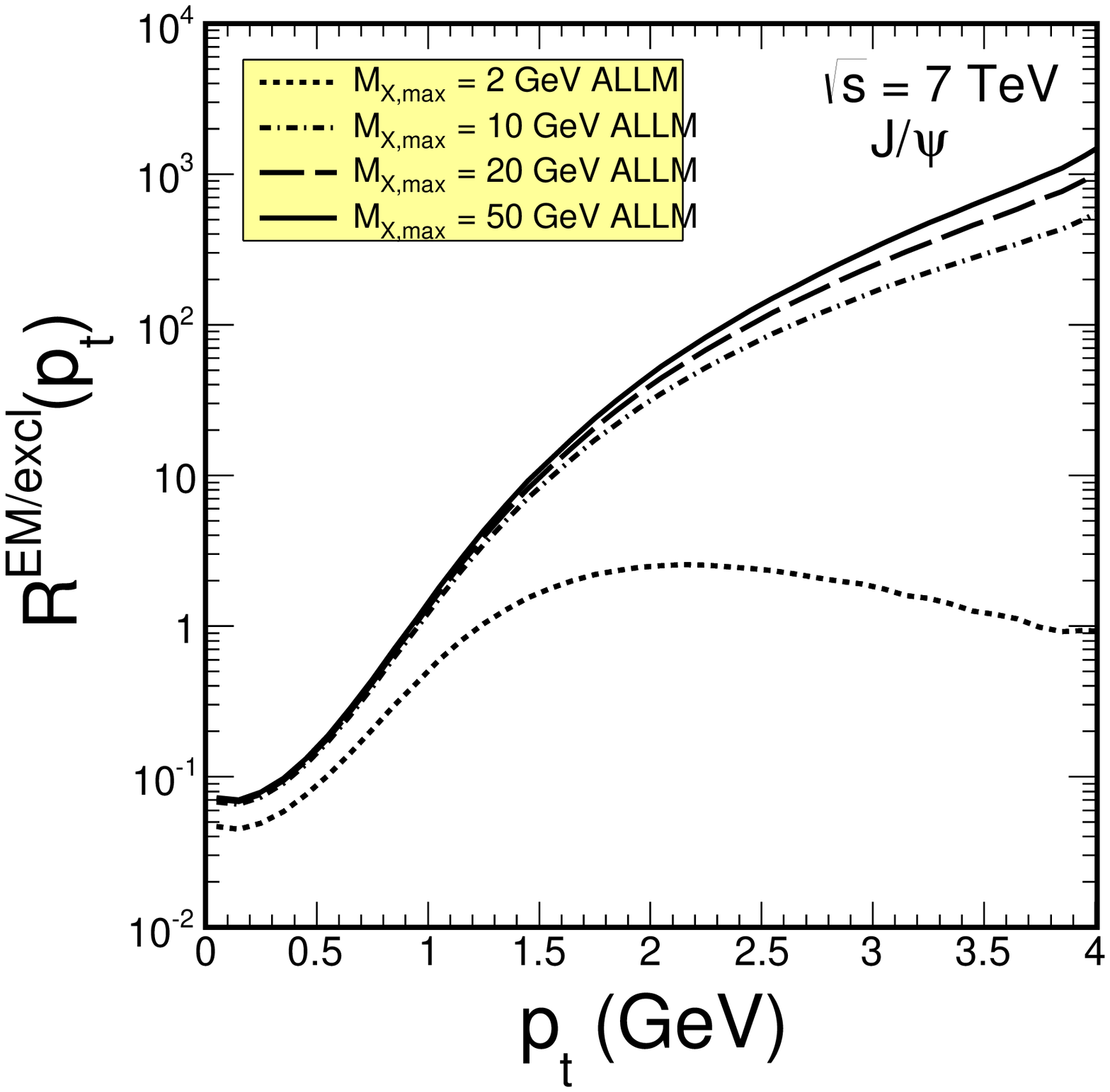}
\includegraphics[height=6.75cm]{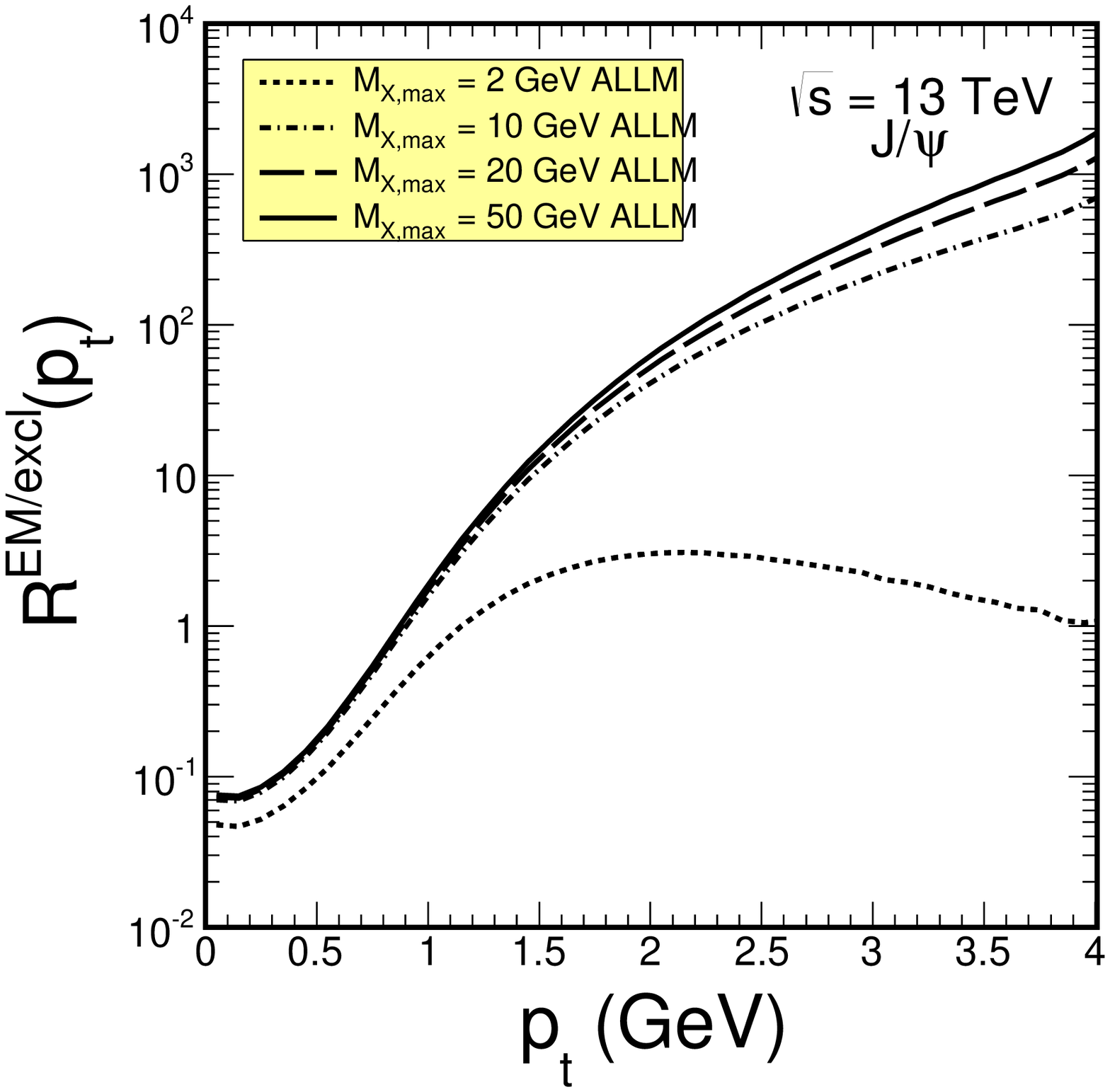}
\includegraphics[height=6.75cm]{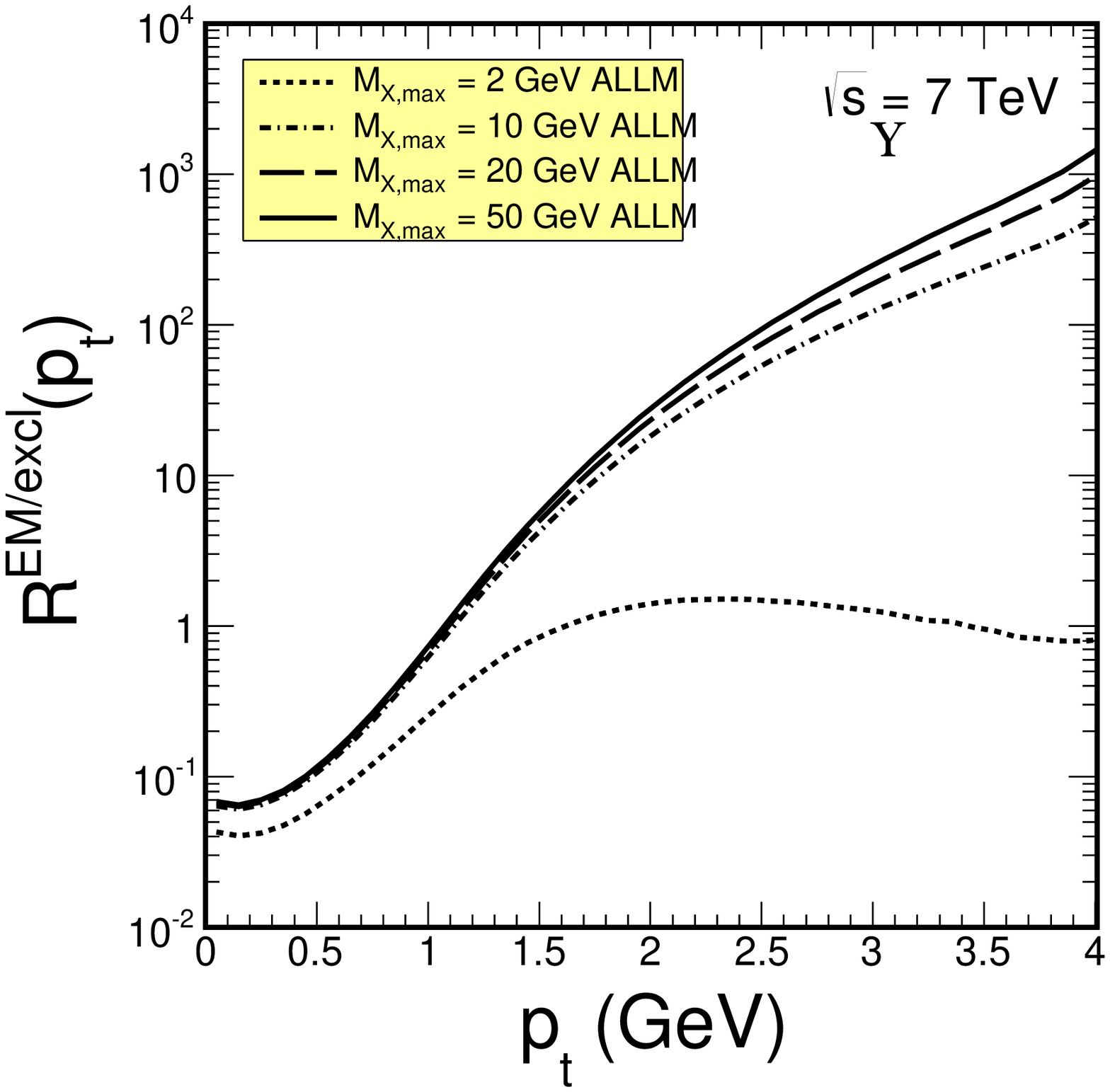}
\includegraphics[height=6.75cm]{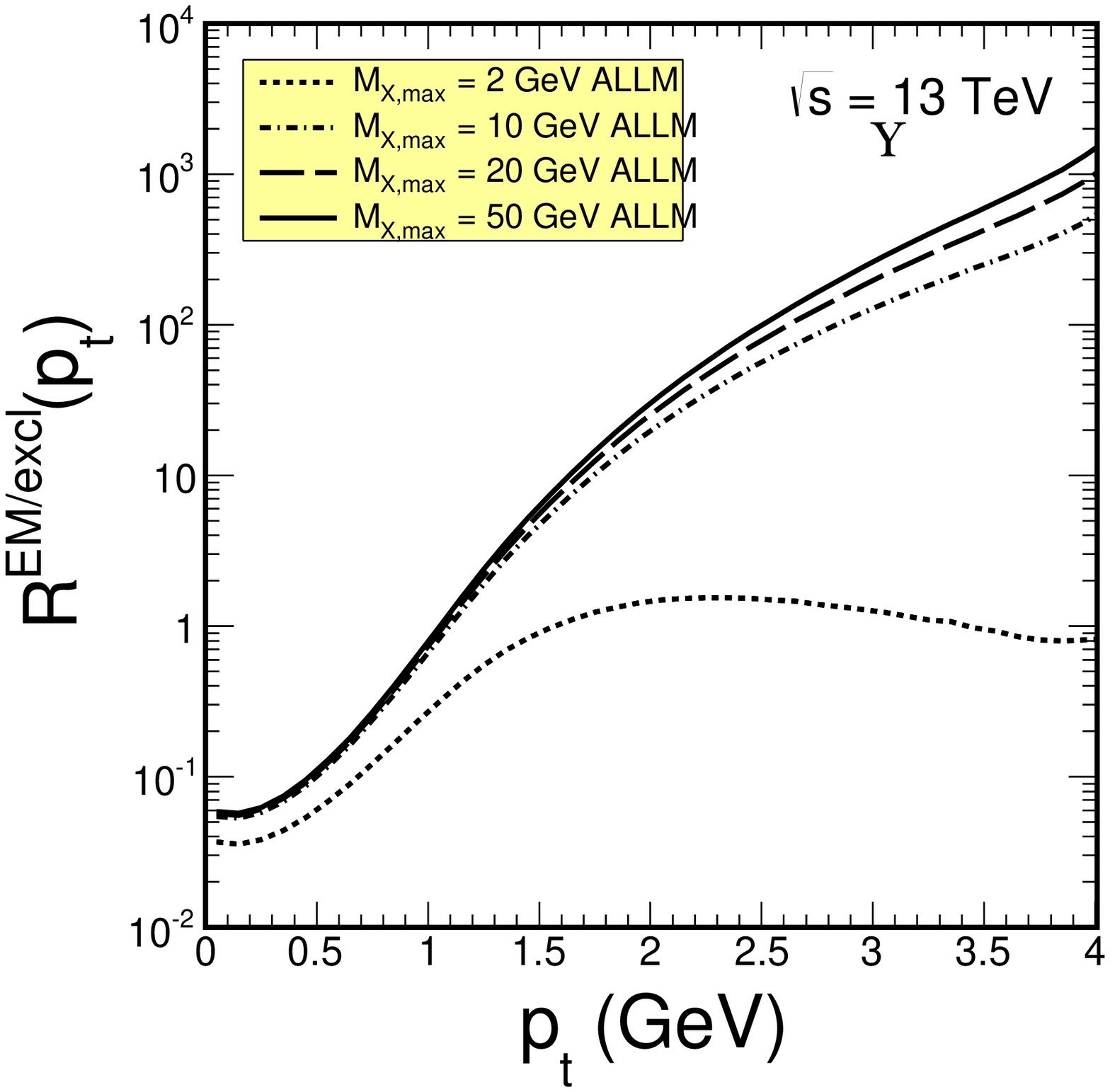}
\caption[*]{Ratio of inelastic diffractive to exclusive vector meson production
as a function of transverse momentum for different upper limits on the excited mass $M_X$.}
\label{Ratio_pt}
\end{figure}

\begin{figure}[!htb] 
\includegraphics[height=6.75cm]{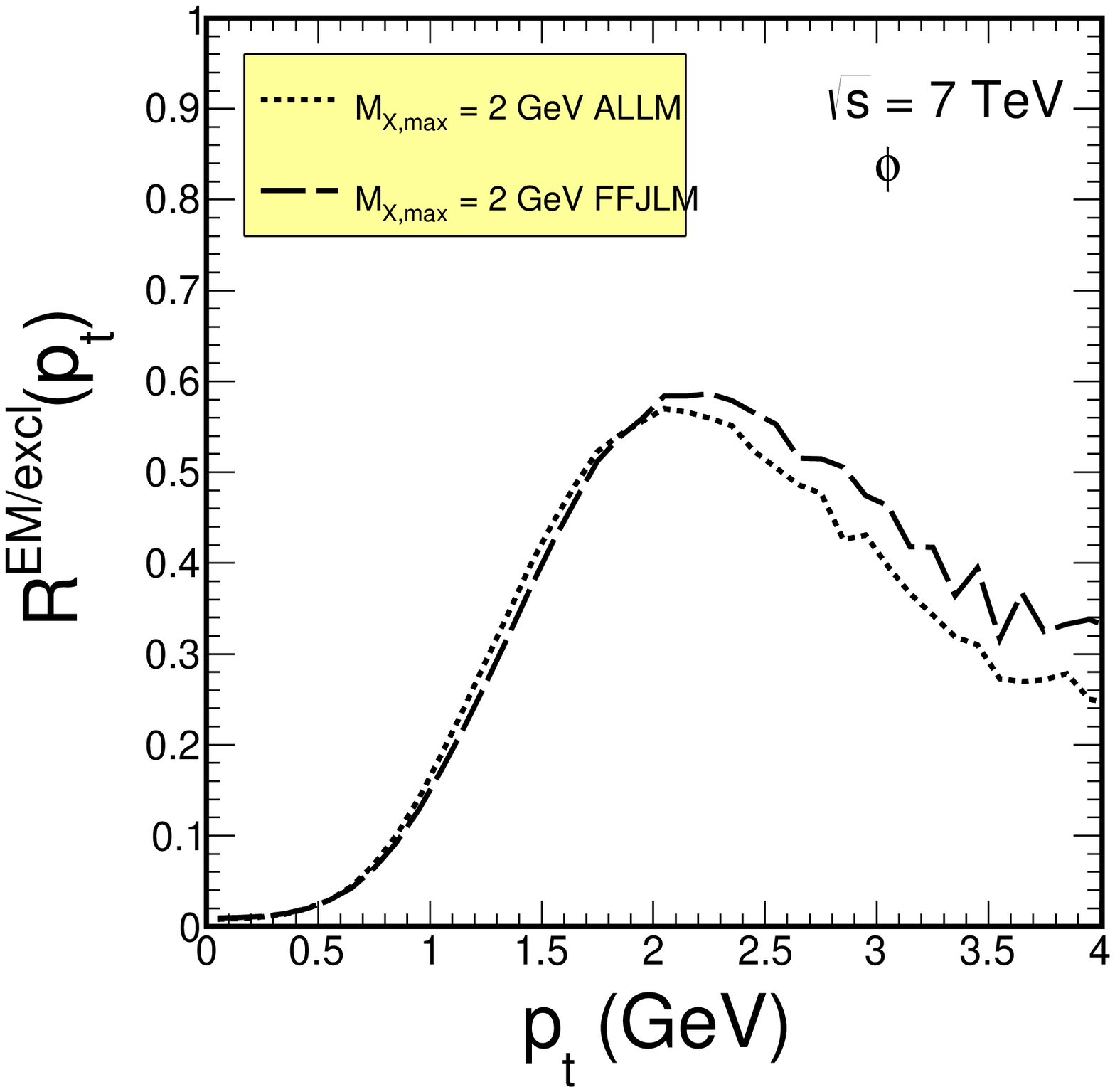}
\includegraphics[height=6.75cm]{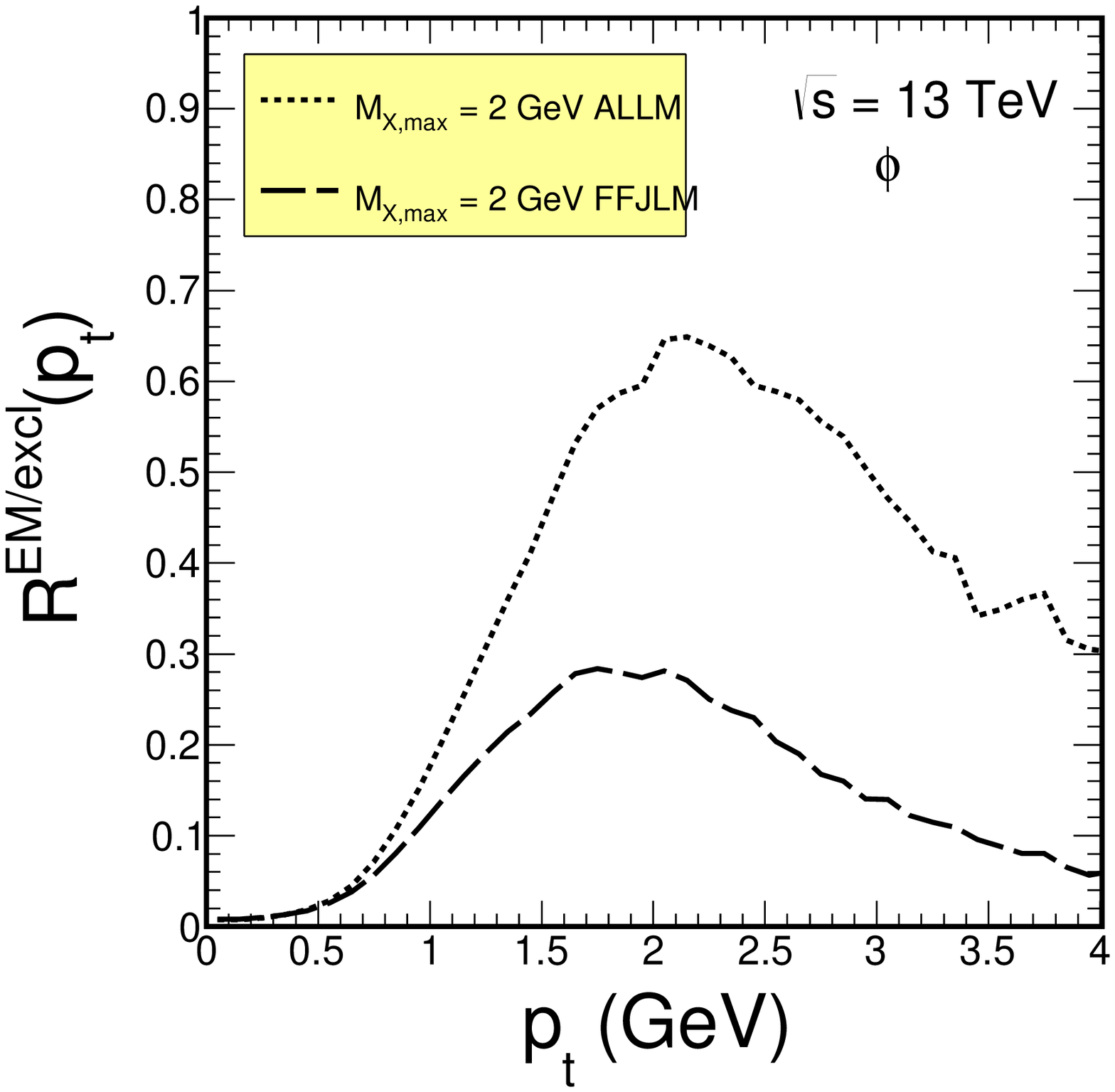}
\includegraphics[height=6.75cm]{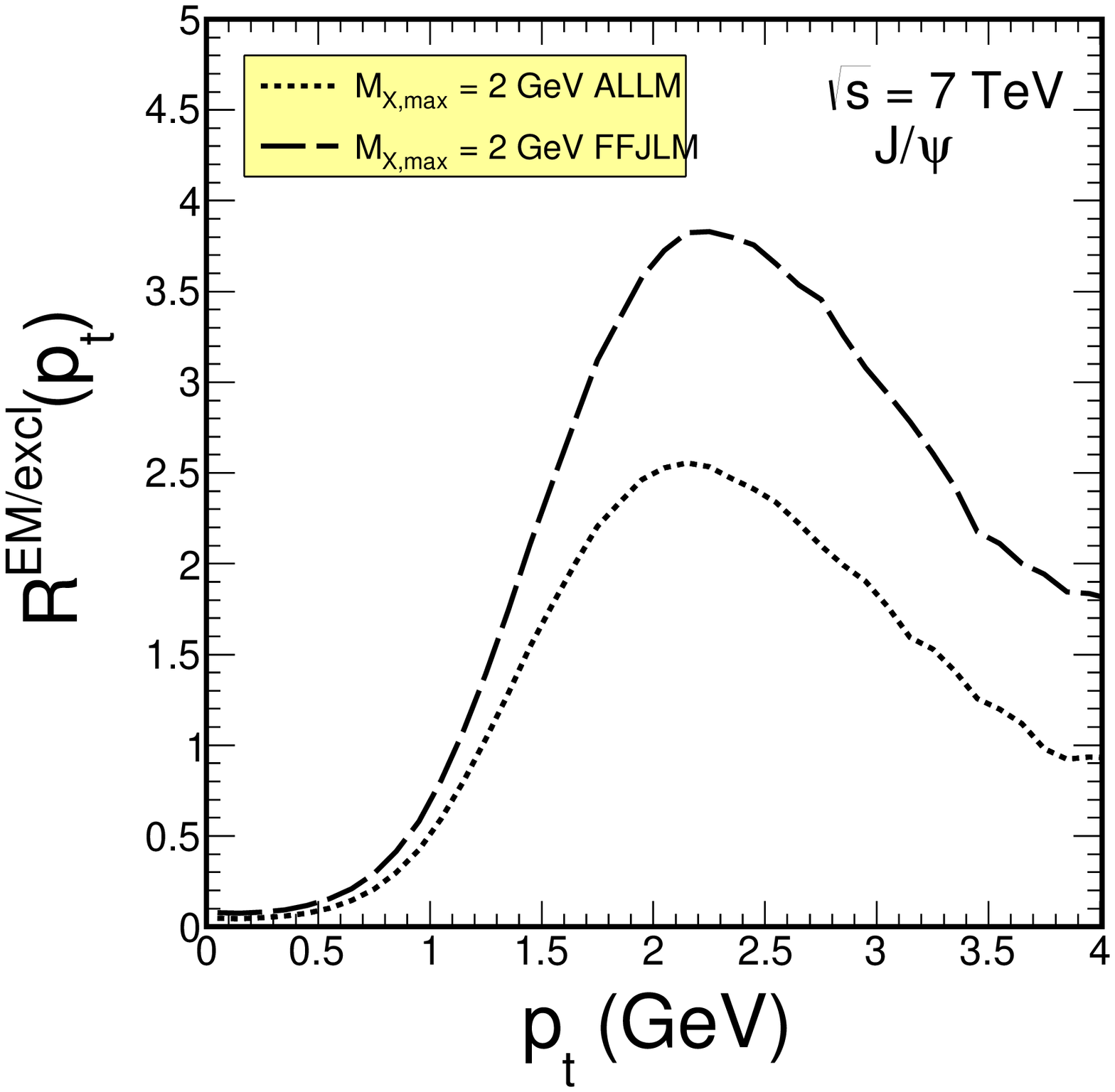}
\includegraphics[height=6.75cm]{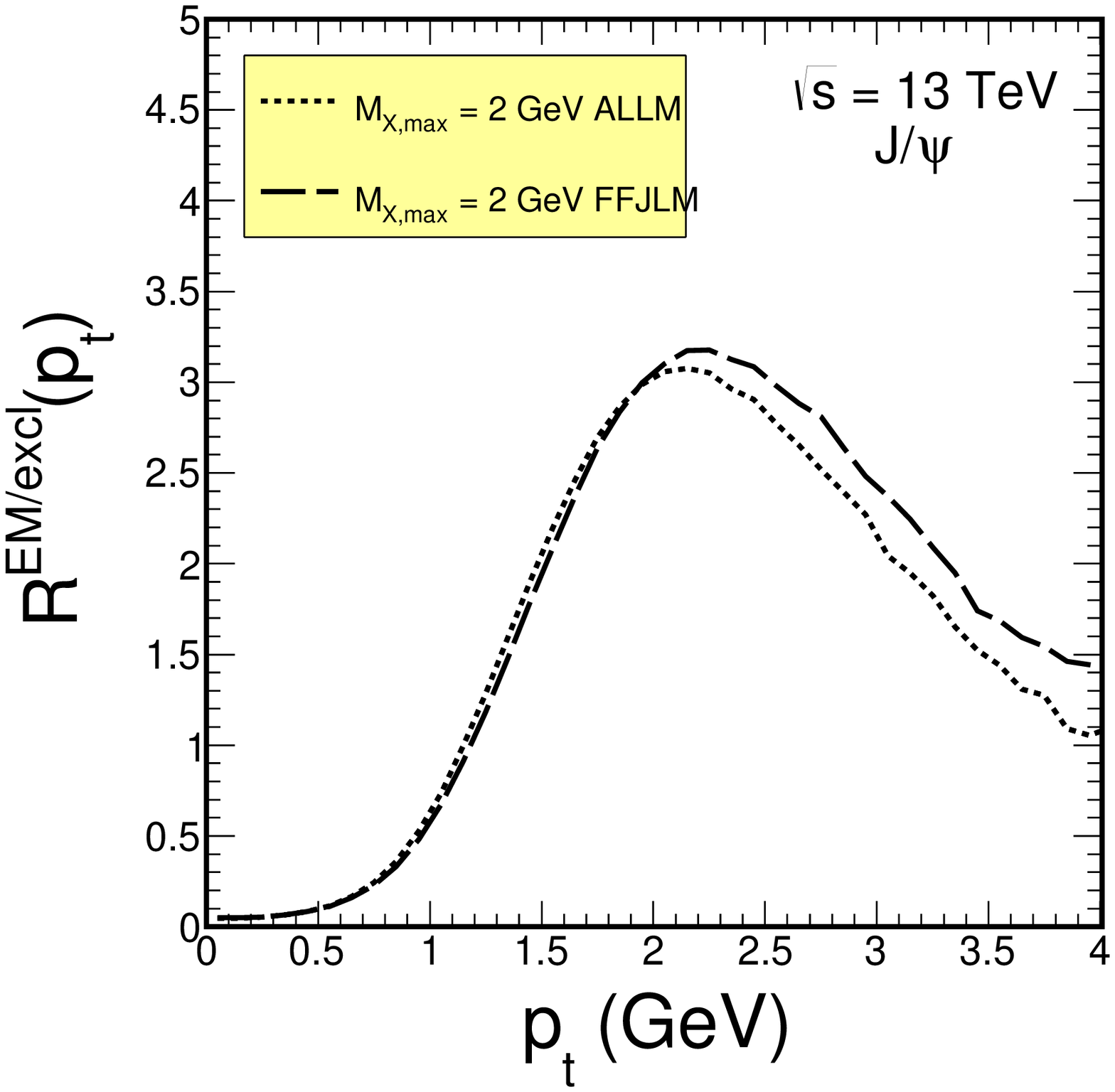}
\includegraphics[height=6.75cm]{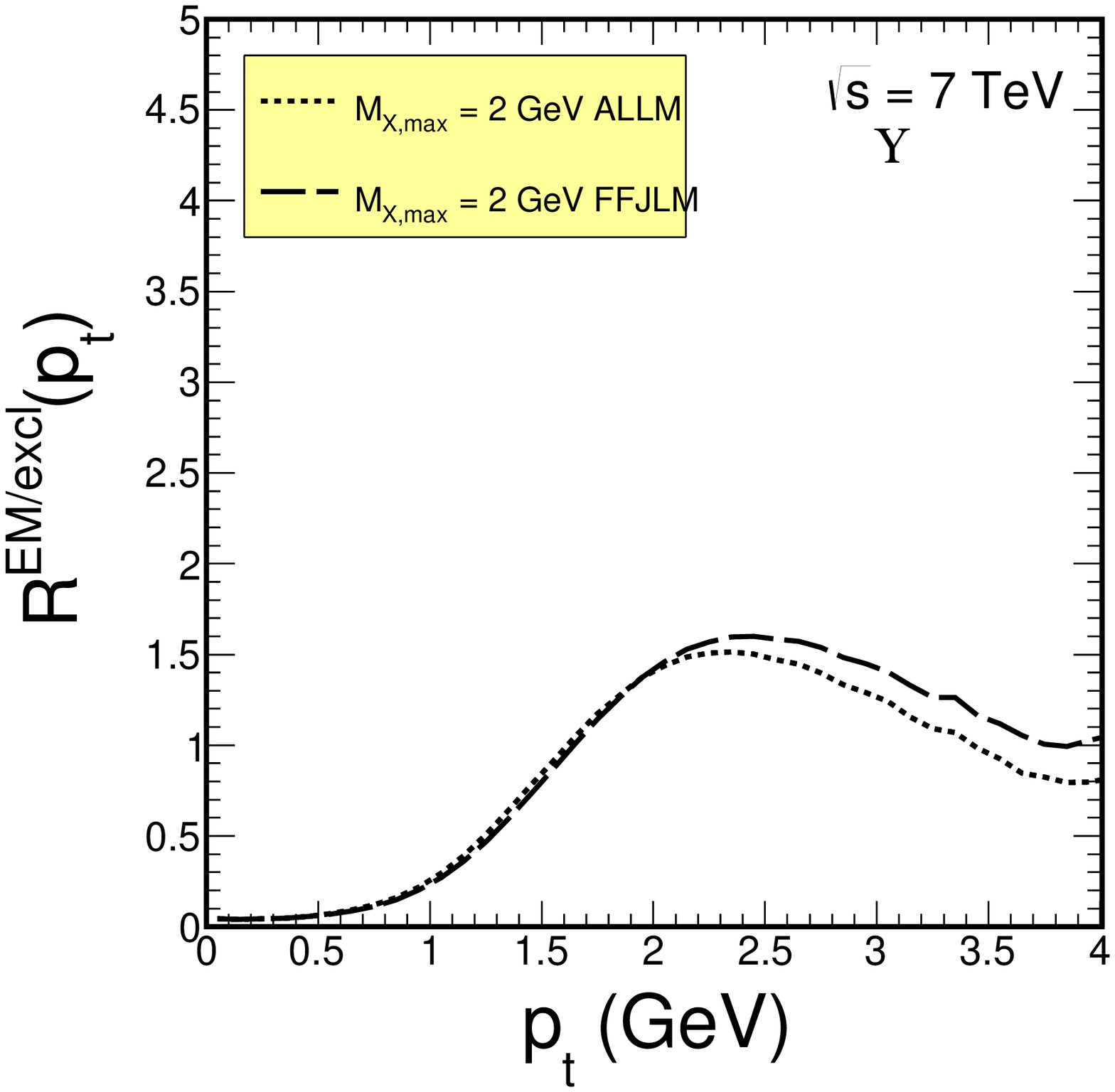}
\includegraphics[height=6.75cm]{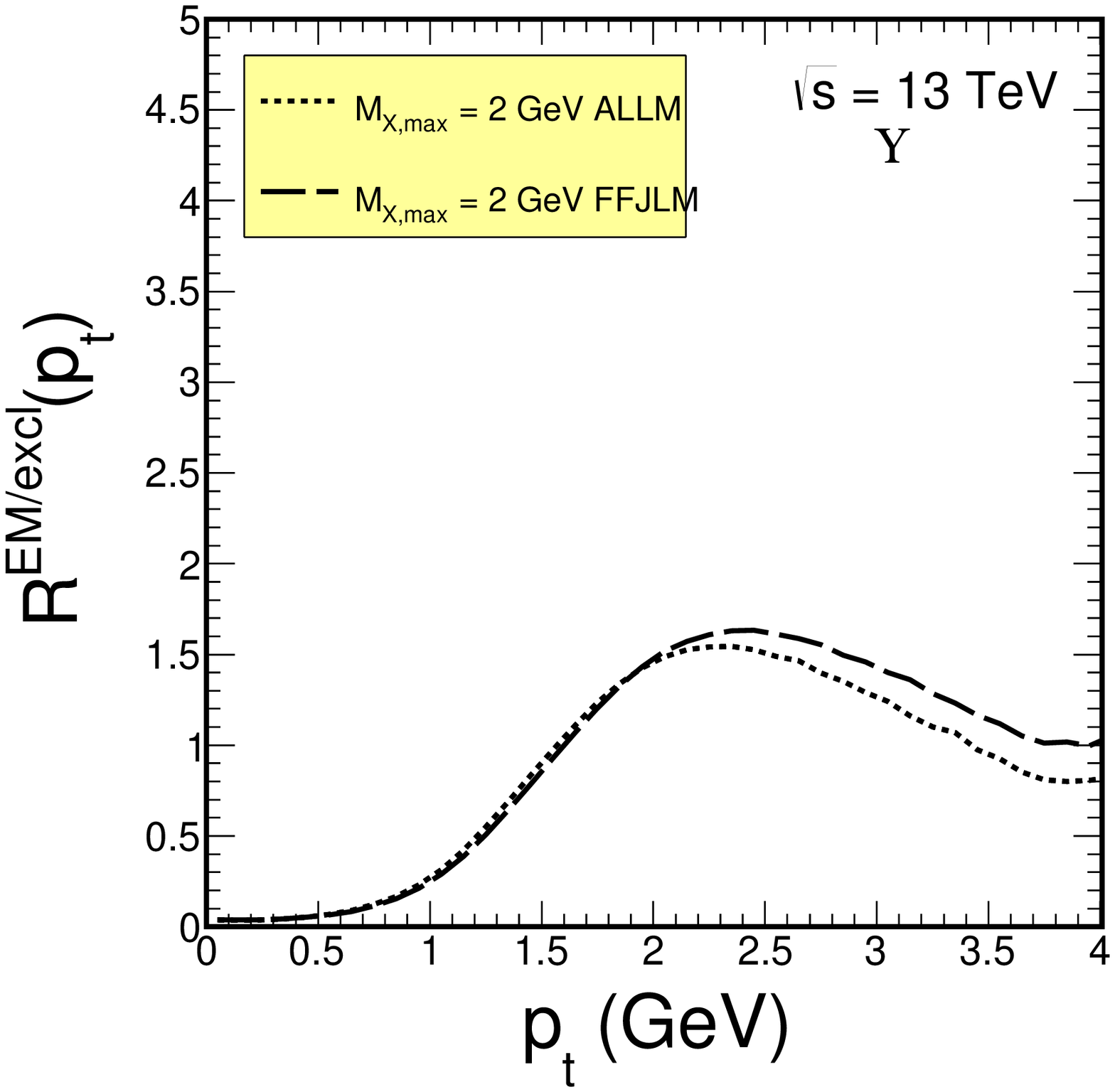}
\caption[*]{Ratio of inelastic diffractive to exclusive vector meson production
as a function of transverse momentum for low proton excited masses, integrated up to 
$M_X= 2 \, \rm{GeV}$.
}
\label{Ratio_pt_low_MX}
\end{figure}

In order to highlight the difference of the inelastic constribution and the exclusive one,
we introduce the ratios:
\begin{eqnarray}
 R^{\rm EM/excl.}(p_t,M_{X,\rm max}) &=& 
 \displaystyle {\int_{M_{\rm thr}}^{M_{X,\rm max}} dM_X 
 { d \sigma (pp \to pVX) \over dp_t dM_X}
 \over { d \sigma (pp \to pVp ) \over dp_t}}
 \ \ ,
  \nonumber \\
  R^{\rm EM/excl.}(y, M_{X,\rm max}) &=& 
 \displaystyle {\int_{M_{\rm thr}}^{M_{X,\rm max}} dM_X 
 { d \sigma (pp \to pVX) \over y dM_X}
 \over { d \sigma (pp \to pVp ) \over dy}}
 \ \ .
\end{eqnarray}
In Fig. \ref{Ratio_pt} we show the ratio $ R^{\rm EM/excl.}$
as a function of $p_t$ for different upper limits on $M_X$, we
see that as soon high mass states are included, the inelastic contribution
dominates at $p_t \gsim 1 \, \rm{GeV}$.
In Fig. \ref{Ratio_pt_low_MX} we have a closer look at $ R^{\rm EM/excl.}$
for the excitation of low-mass states $M_X < 2 \, \rm GeV$. Here we use 
both the ALLM and Fiore parametrizations. We see that for $\phi$ production the 
ratio is always smaller than one, while for heavier mesons, inelastic 
production will dominate at $p_t \gsim 1.5 \, \rm GeV$.

\begin{figure}[!htb] 
\includegraphics[height=6.75cm]{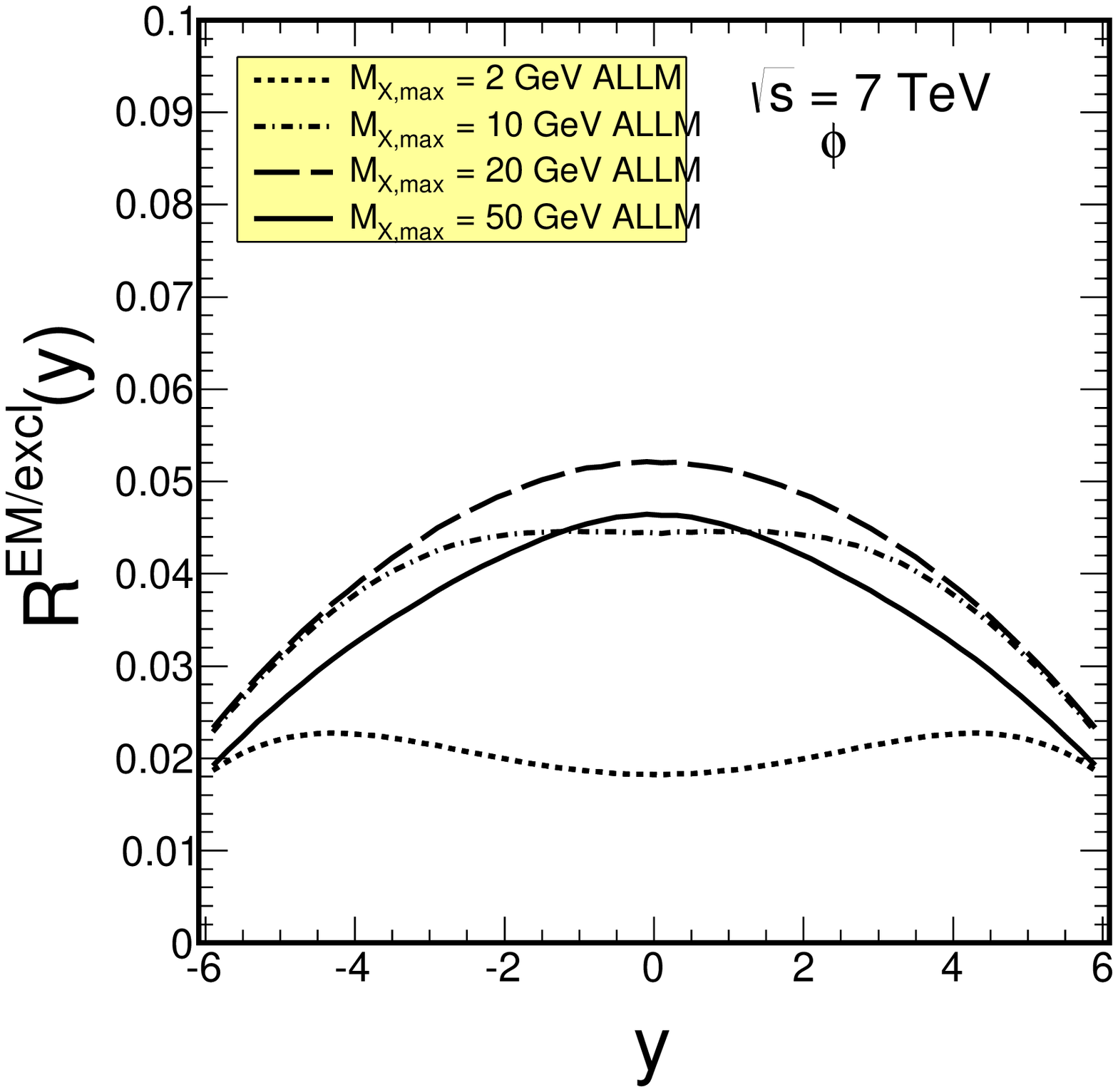}
\includegraphics[height=6.75cm]{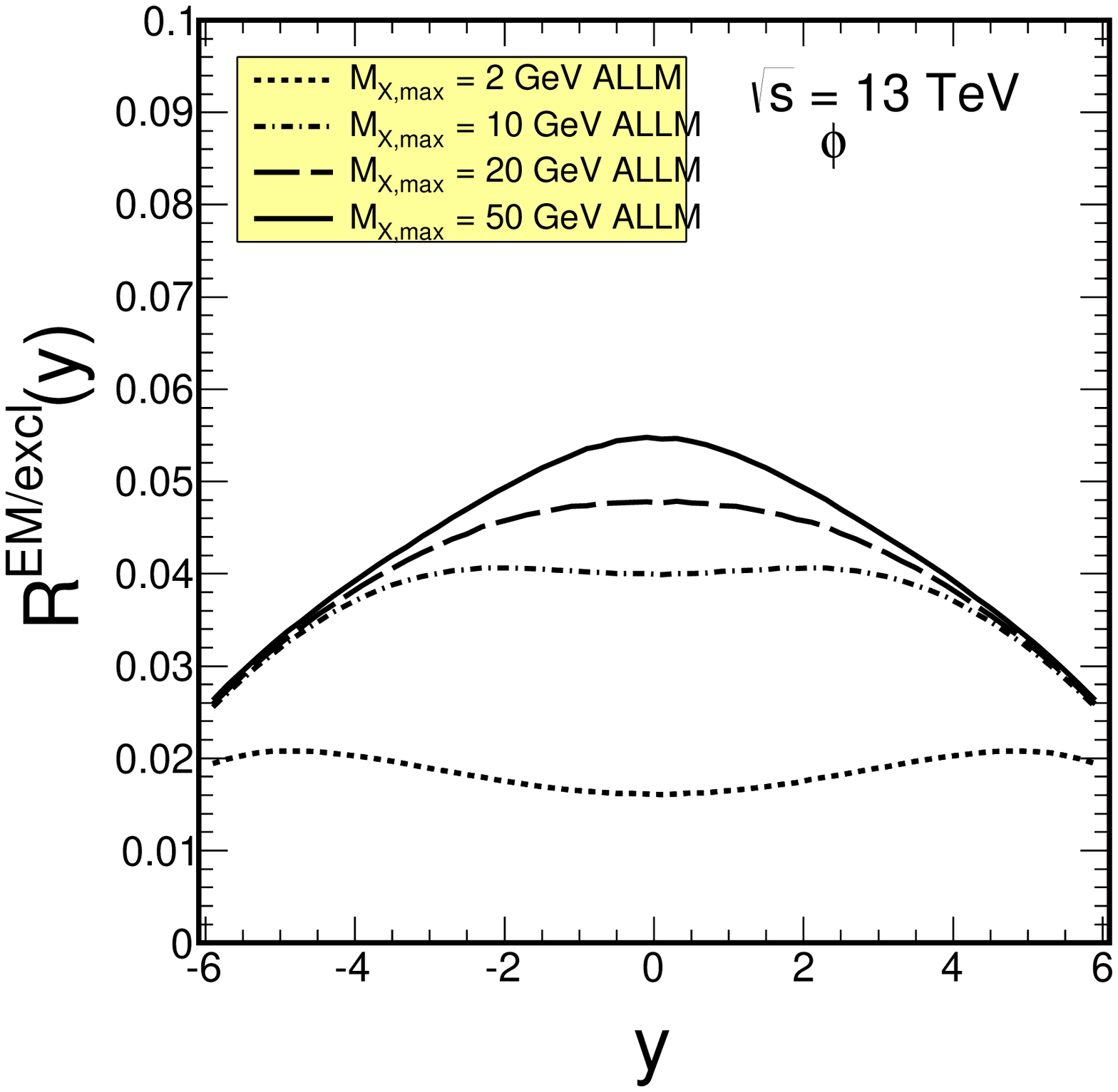}
\includegraphics[height=6.75cm]{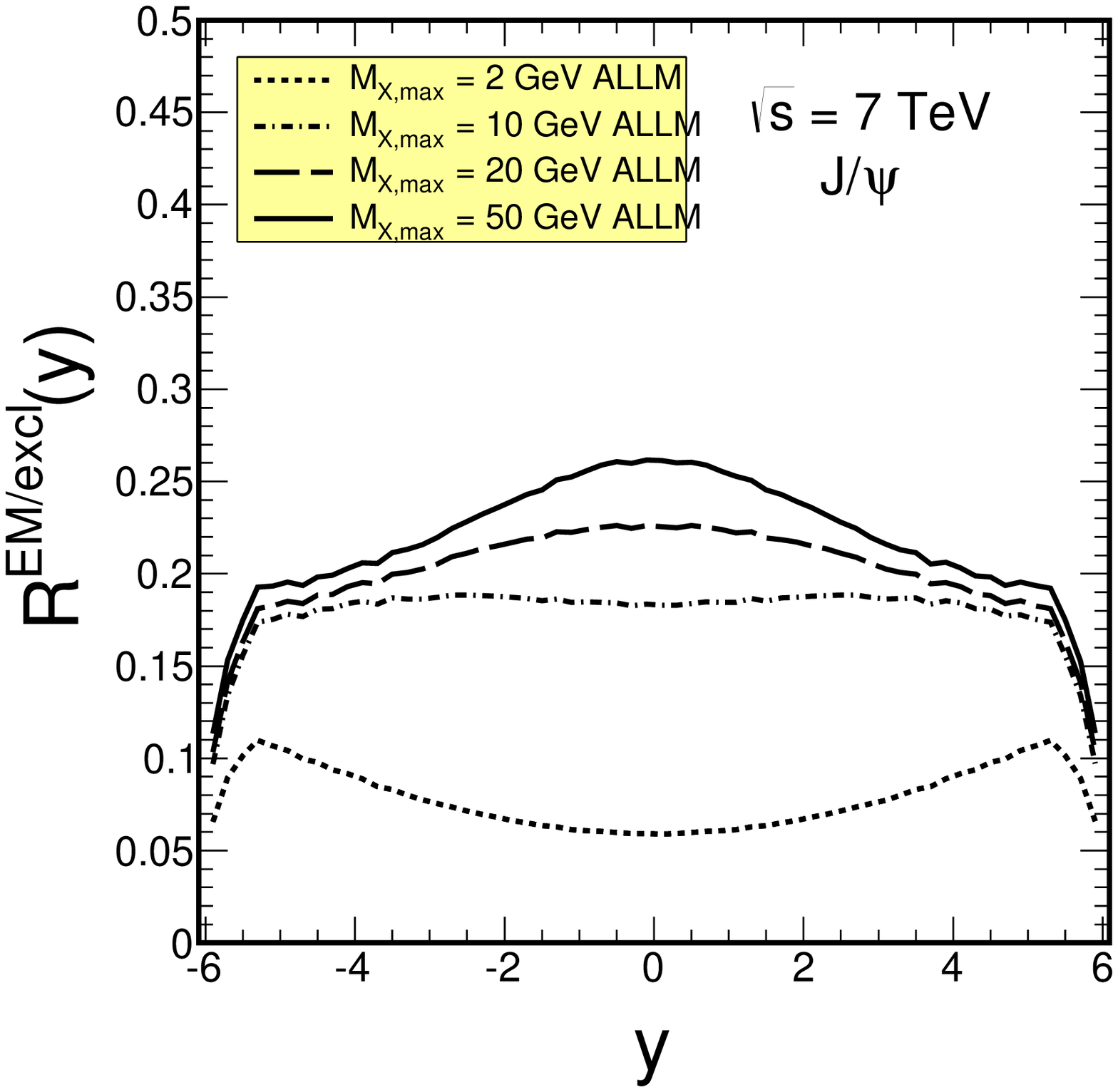}
\includegraphics[height=6.75cm]{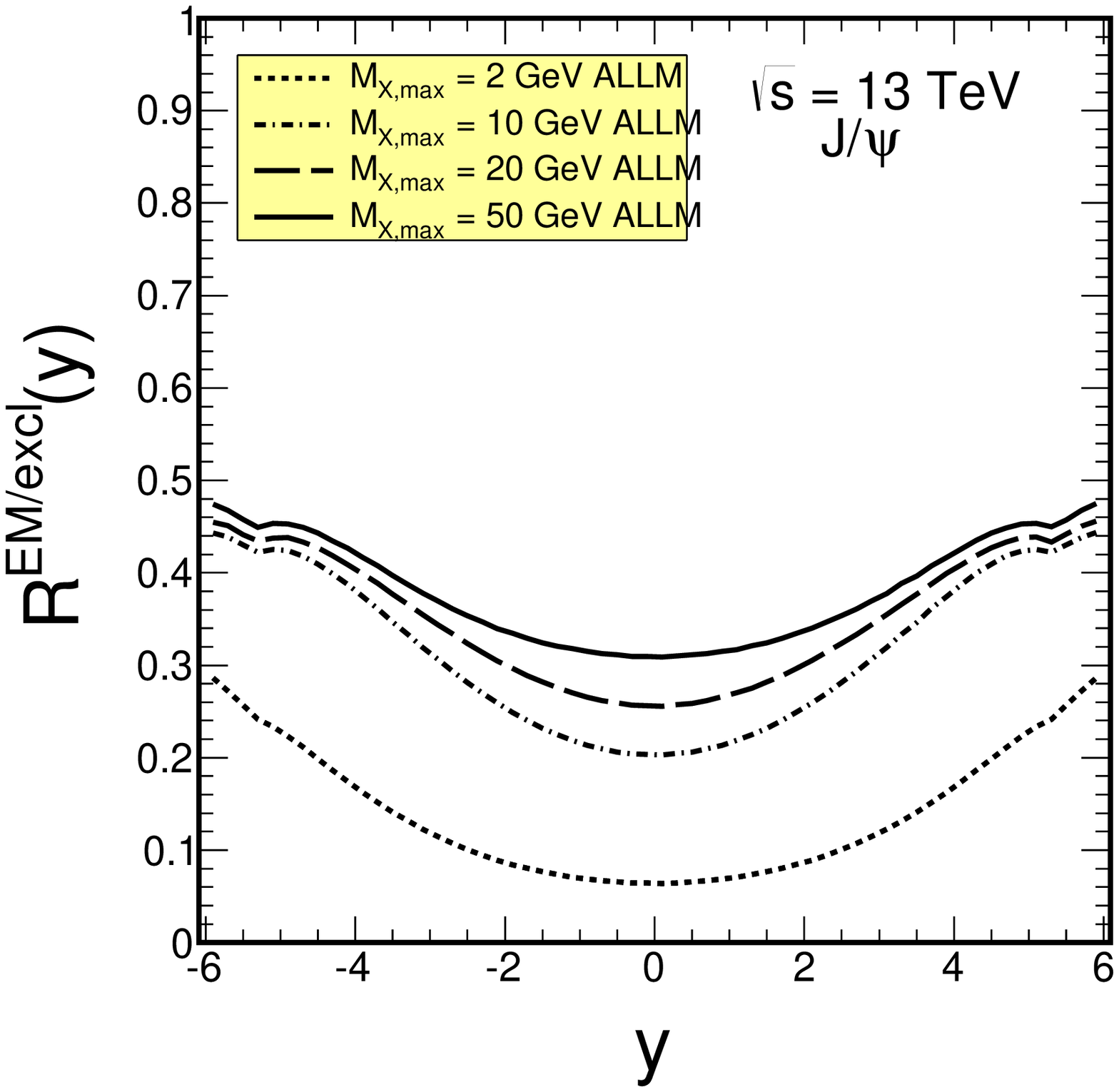}
\includegraphics[height=6.75cm]{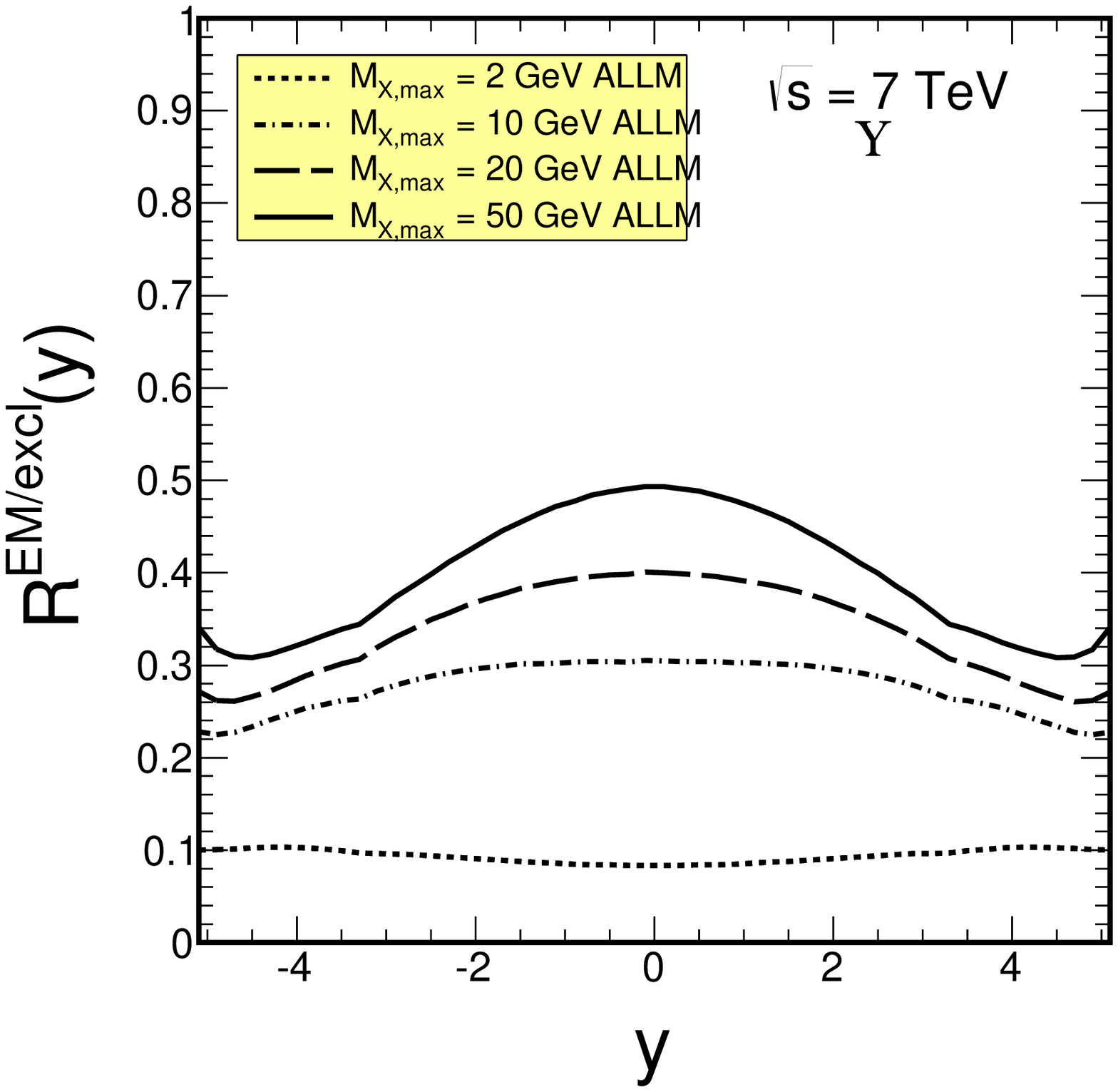}
\includegraphics[height=6.75cm]{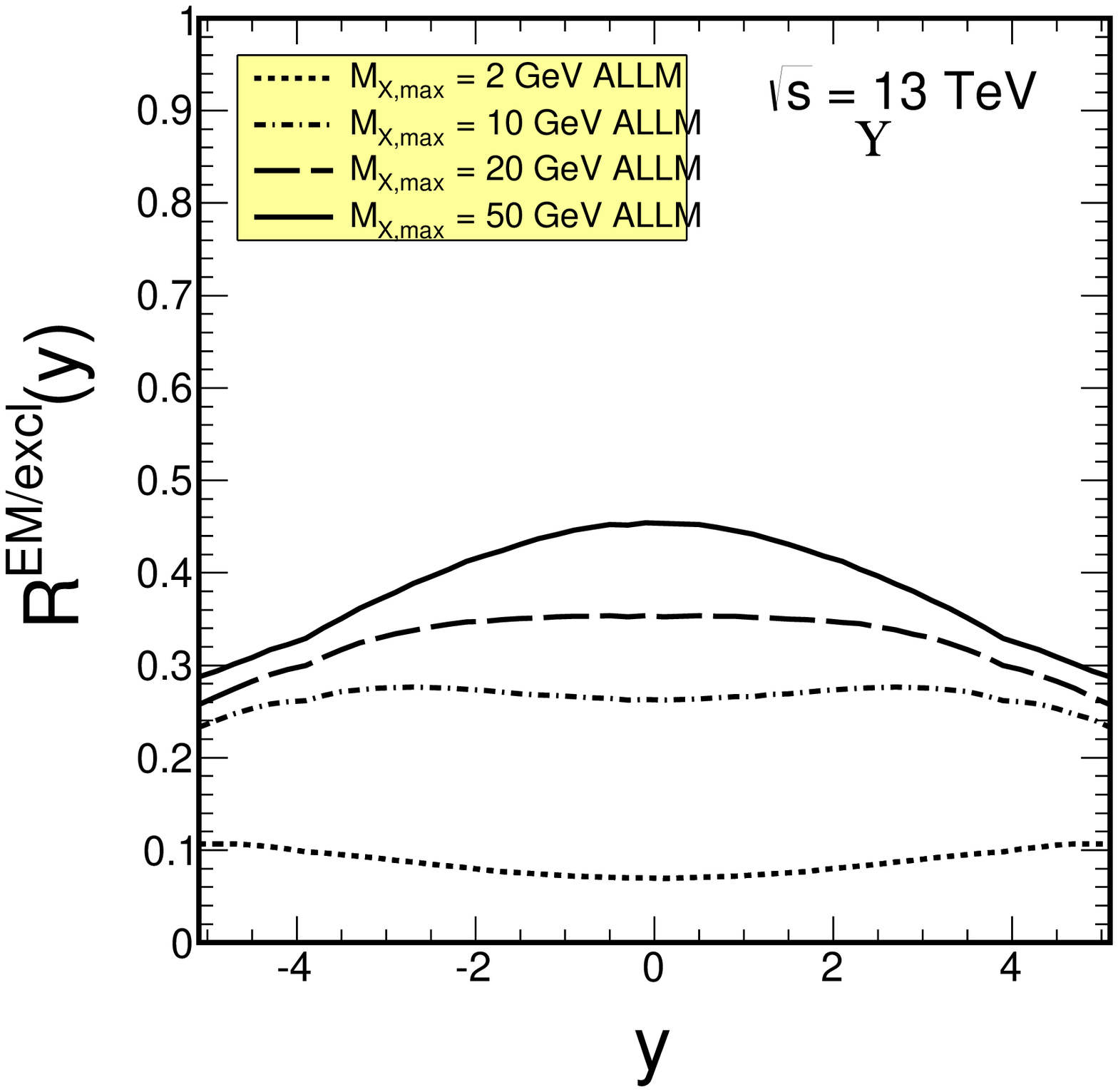}
\caption[*]{Ratio of inelastic diffractive to exclusive vector meson production
as a function of rapidity for different upper limits on the excited mass $M_X$.}
\label{Ratio_y}
\end{figure}

\begin{figure}[!htb] 
\includegraphics[height=6.75cm]{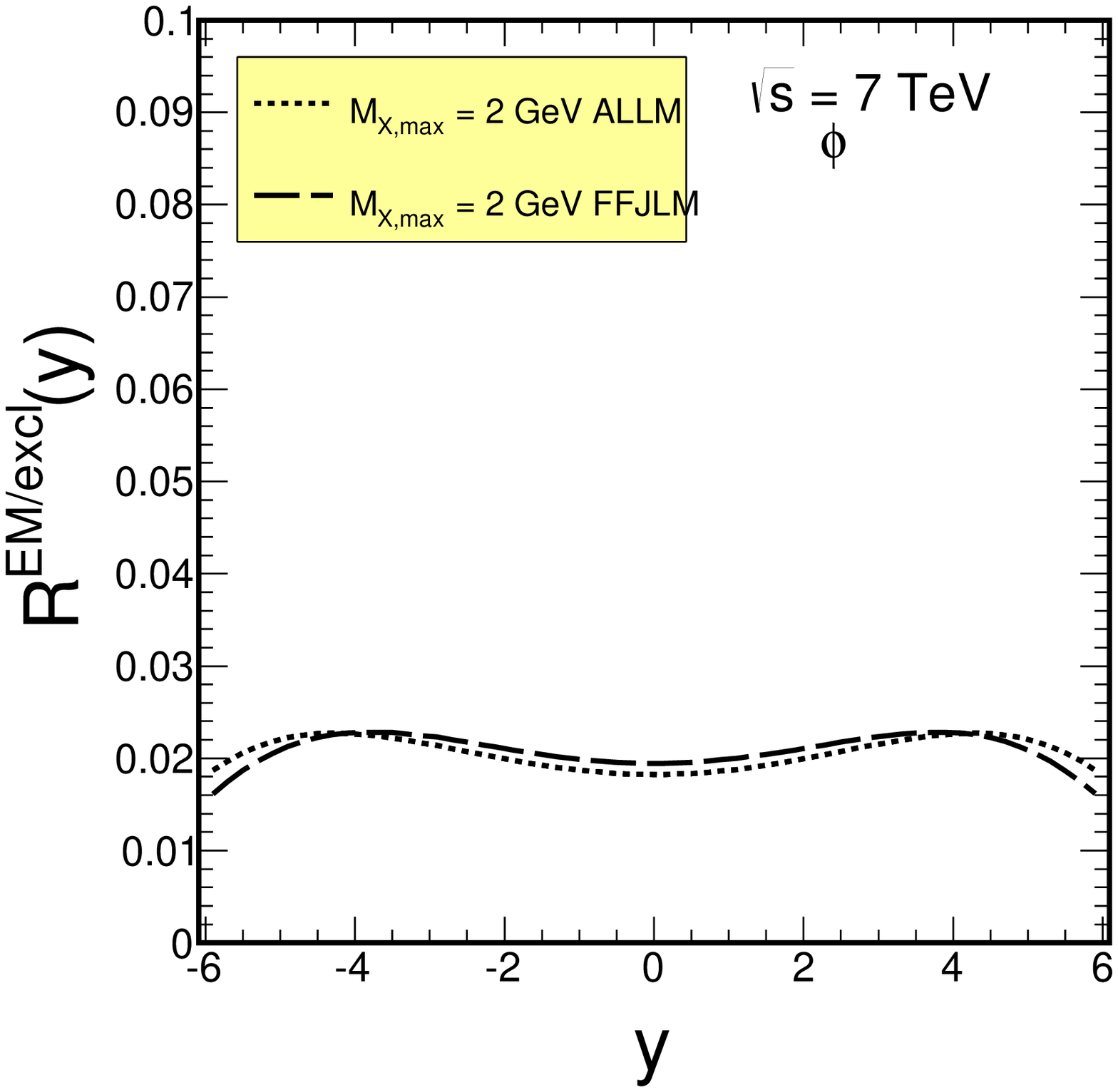}
\includegraphics[height=6.75cm]{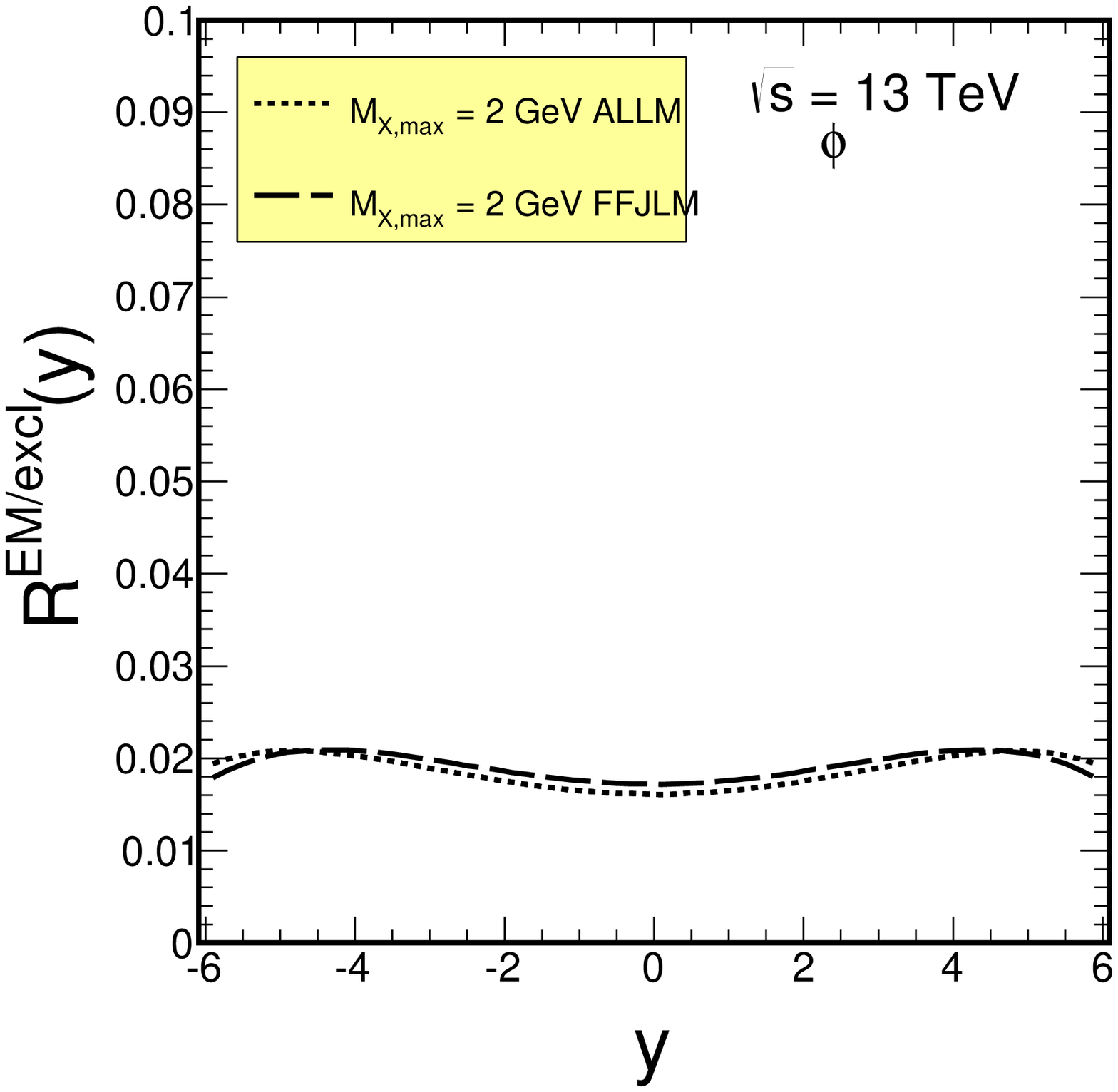}
\includegraphics[height=6.75cm]{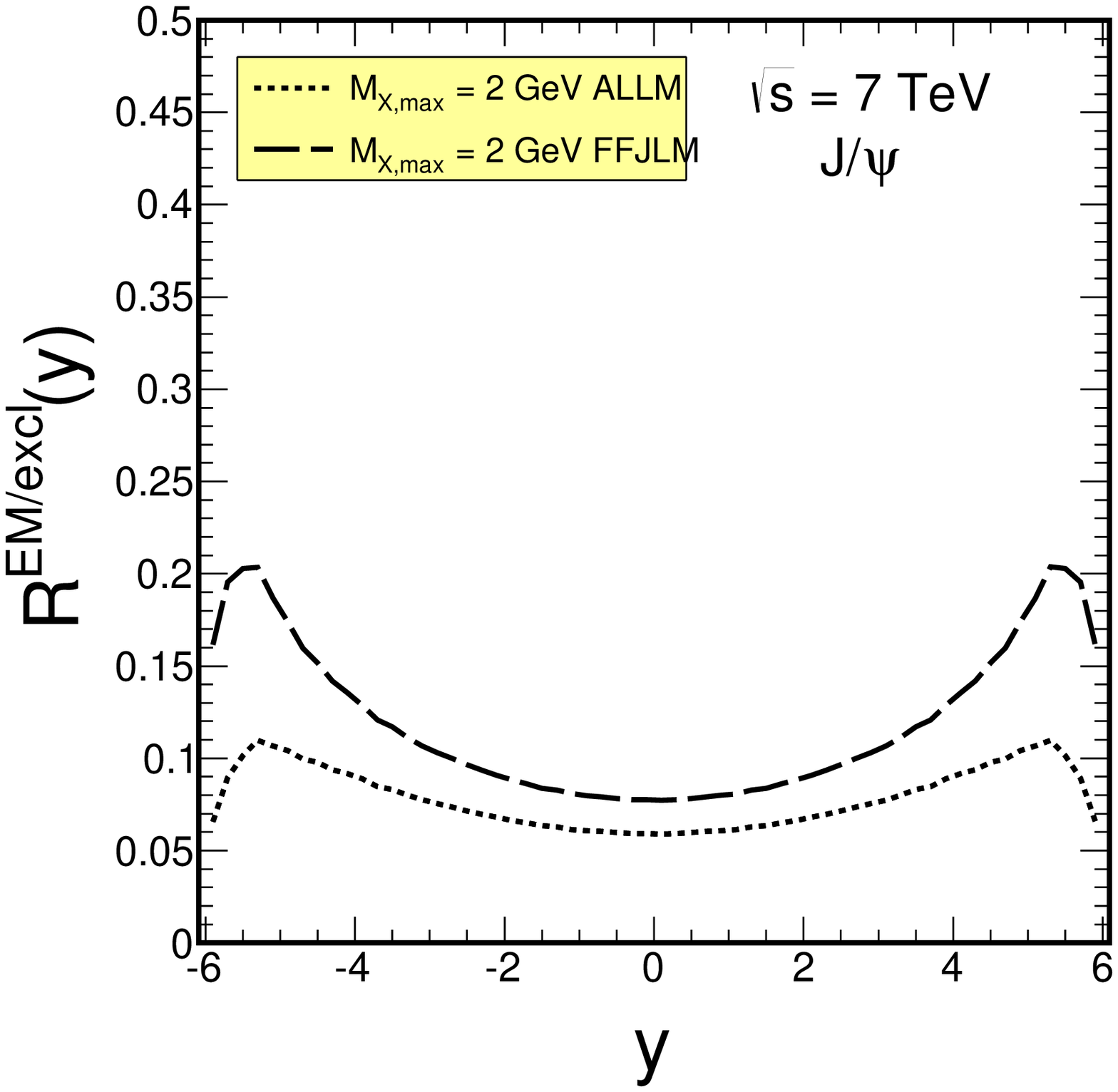}
\includegraphics[height=6.75cm]{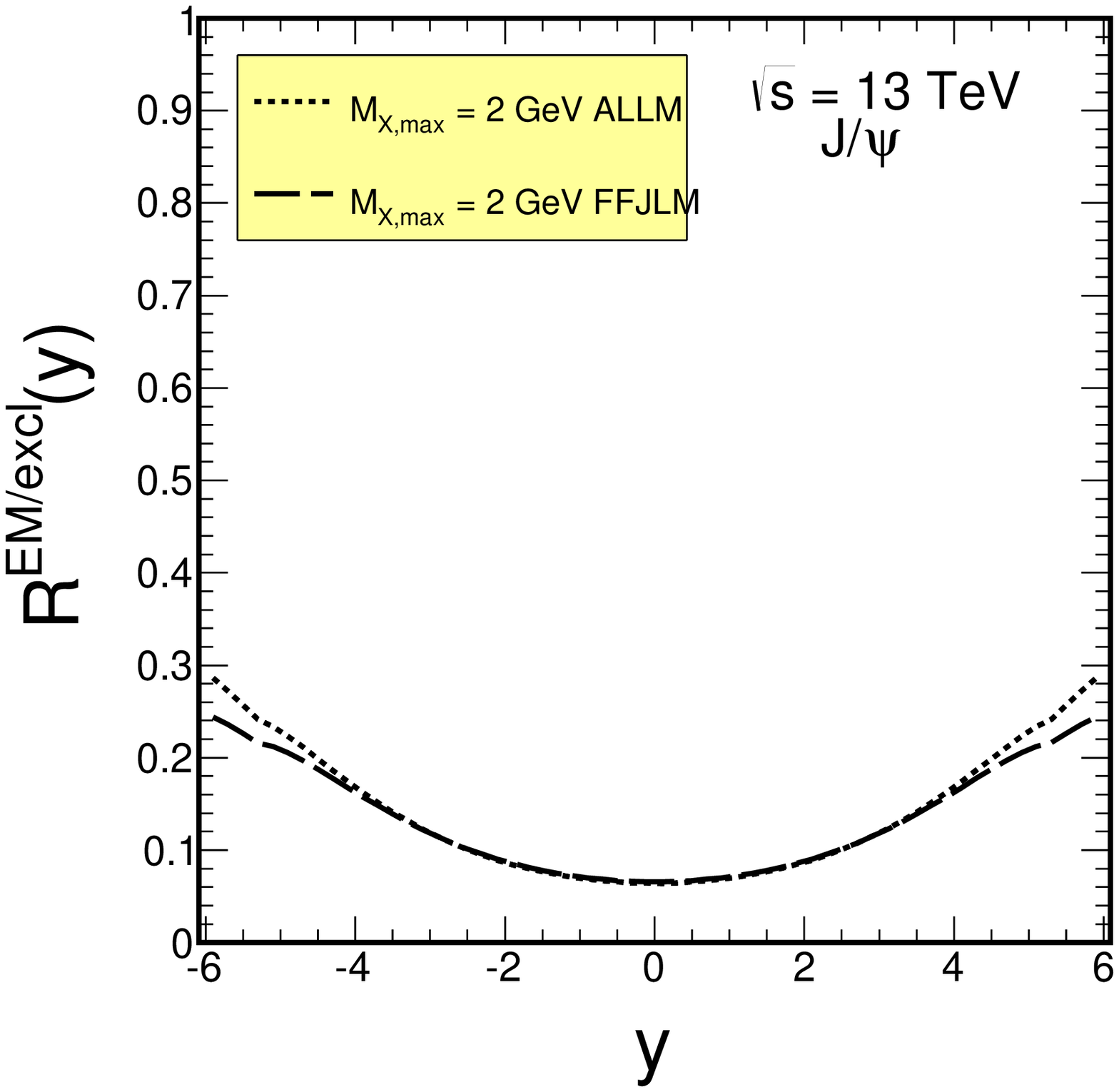}
\includegraphics[height=6.75cm]{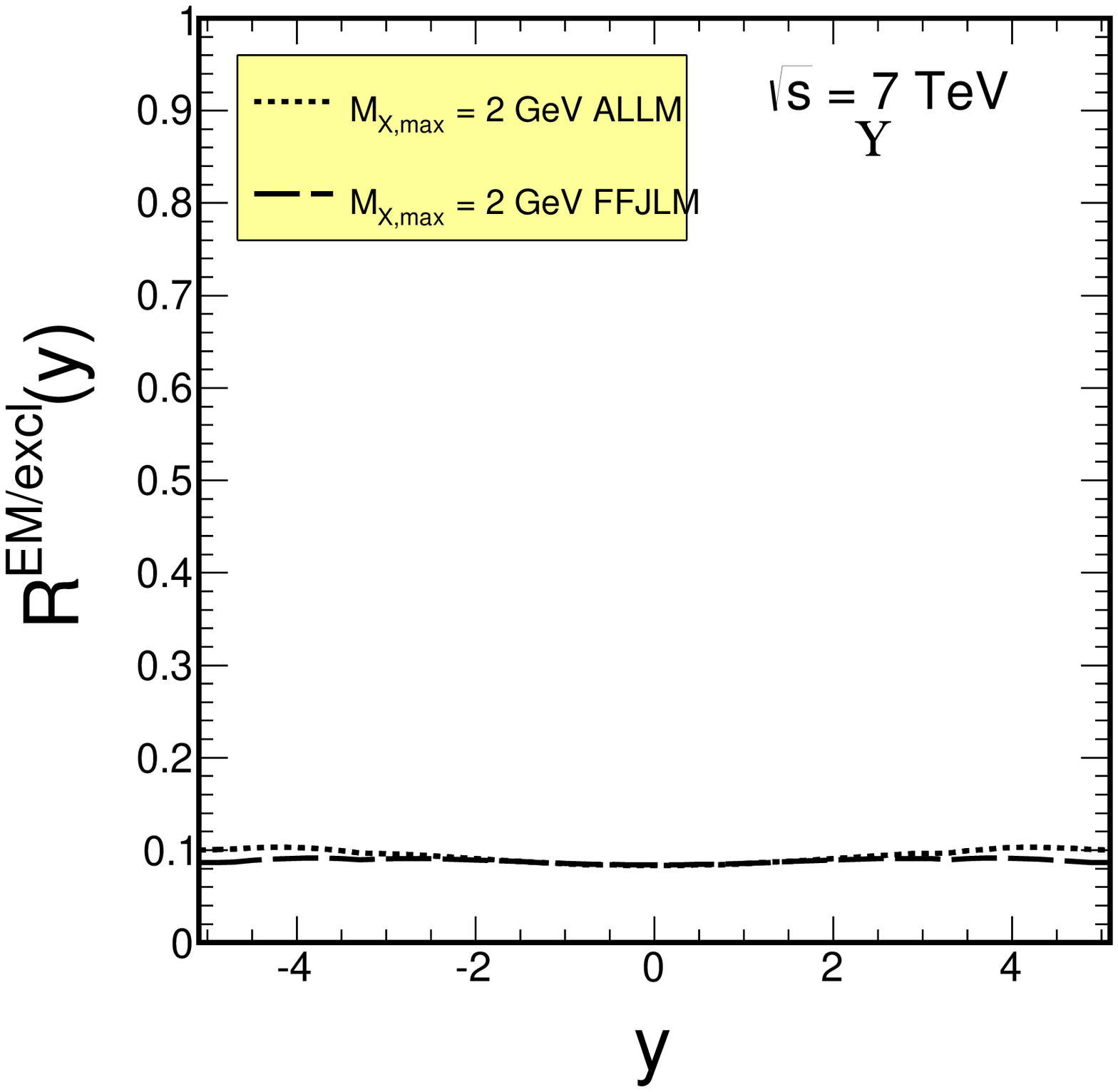}
\includegraphics[height=6.75cm]{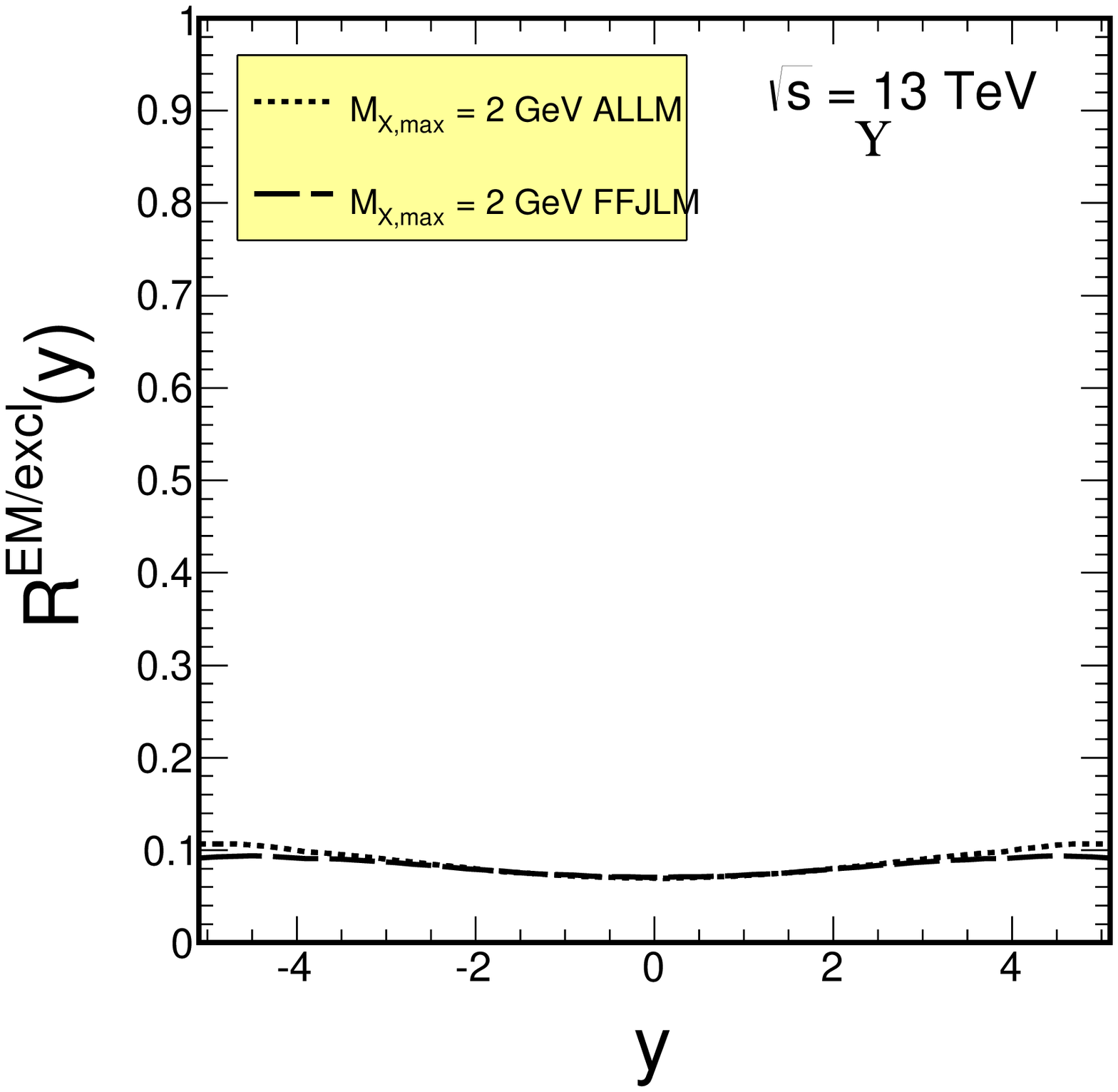}
\caption[*]{Ratio of inelastic diffractive to exclusive vector meson production
as a function of rapidity for low excited masses, integrated up to 
$M_X= 2 \, \rm{GeV}$.}
\label{Ratio_y_low_MX}
\end{figure}

In Figs. \ref{Ratio_y} and \ref{Ratio_y_low_MX} we show the analogous
ratios for the rapidity distribution.



\section{Conclusions}

In this paper we have discussed semiexclusive production of vector
mesons 
in  $p p \to V p X$ processes via photon-pomeron or pomeron-proton fusion, 
where $X$ stands for excited/dissociated proton system and 
$V = \phi, J/\psi, \Upsilon$.
We have investigated the similarities and differences of various
cross sections to the exclusive $p p \to p p V$ process.
Electromagnetic dissociation of protons is calculated using
an inelastic unintegrated photon flux which was calculated based on
modern parametrizations of deep-inelastic proton structure functions.
Different parametrizations from the literature have been used.

A number of differential cross distributions for the vector mesons
(in their rapidity and transverse momentum) as well as for the mass of 
the dissociative system, remnant of the proton, have been calculated.
We have found that in all the considered cases the photon dissociation
cross section is large compared to its purely exclusive counterpart.
The results strongly depend on the parametrization of the structure
function used. 
One should stress, however, in this context that different
parametrizations have quite different status. Some of them are global
fits to the world data, often not ideal in some other regions 
of the phase space.
Some of them focus rather on some corners of the phase space, so are
extremely good there. However, they can be not realistic in other
corners of the phase space.
We have discussed in detail which distributions provide realistic 
estimates of the cross section.

The semiexclusive contributions produce vector mesons with
large transverse momenta. The rapidity distributions of semi-exclusive
and purely exclusive distributions are rather similar.
In the present analysis we have shown also the ratio of the semiexclusive
to the purely exclusive contributions. This ratio depends strongly
on the vector meson transverse momentum and only mildly on rapidity.
In general, a bigger ratio is obtained for heavier quarkonia.

It is obvious that a large fraction of the remnant cannot be seen by
central detectors of different LHC experiments
in the case when protons are not measured using specially dedicated
forward detectors, just installed recently by the CMS-TOTEM or 
ATLAS collaborations.
Without measuring both protons, as is the case of the LHCb experiment,
the so-called exclusive data are not fully exclusive and may contain
the semi-exclusive contributions discussed here.
The LHCb collaboration cuts the large-$p_t$ part of the corresponding
distribution in a purely phenomenological fit of different slope.
It is not completely clear how good is such a procedure.
It would be good to relax requirements on rapidity gap(s) around 
vector mesons and actually measure the semieexclusive contributions.
We encourage experimentalists to perform such analyses.
We note that the semiexclusive contributions were not measure so far,
but are interesting themselves. 
Such measurements would be therefore tests of the method used here.

{\bf Acknowledgments}

This work was partially supported by the Polish National Science Center grant
UMO-2018/31BST2/03537
and by the Centre for Innovation and Transfer of 
Natural Sciences and Engineering Knowledge in Rzesz\'ow.


\end{document}